\documentclass[twocolumn,floatfix,tighten]{aastex62}
\received{}
\revised{}
\accepted{}
\submitjournal{AJ}

\usepackage{listings}
\usepackage{color}

\definecolor{dkgreen}{rgb}{0,0.6,0}
\definecolor{gray}{rgb}{0.5,0.5,0.5}
\definecolor{mauve}{rgb}{0.58,0,0.82}

\newcommand{\DAP}{{\tt DAP}}	
\newcommand{\pPXF}{{\tt pPXF}}
\newcommand{\MAPS}{{\tt MAPS}}
\newcommand{\oi}{\ion{O}{1}}
\newcommand{\oii}{\ion{O}{2}}
\newcommand{\oiii}{\ion{O}{3}}
\newcommand{\nii}{\ion{N}{2}}
\newcommand{\neiii}{\ion{Ne}{3}}
\newcommand{\sii}{\ion{S}{2}}
\newcommand{\siii}{\ion{S}{3}}
\newcommand{\hei}{\ion{He}{1}}
\newcommand{\heii}{\ion{He}{2}}
\newcommand{\kms}{{km$~\!$s$^{-1}$}}
\newcommand{\mileshc}{{\tt MILES-HC}}

\usepackage[T1]{fontenc}
\usepackage{ae,aecompl}
\usepackage{newtxtext,newtxmath}
\shortauthors{Belfiore et al.}
\shorttitle{MaNGA Data Analysis Pipeline: Emission Lines}
\begin{document}

\title{The Data Analysis Pipeline for the SDSS-IV MaNGA IFU Galaxy
Survey: Emission-Line Modeling}

\author[0000-0002-2545-5752]{Francesco Belfiore}
\altaffiliation{ESO fellow, francesco.belfiore@eso.org}
\affiliation{European Southern Observatory, Karl-Schwarzchild-Str. 2, Garching bei M{\"u}nchen, 85748, Germany}
\affiliation{University of California Observatories, University of California Santa Cruz, 1156 High St., Santa Cruz, CA 95064, USA}

\author[0000-0003-1809-6920]{Kyle B. Westfall}
\affiliation{University of California Observatories, University of California Santa Cruz, 1156 High St., Santa Cruz, CA 95064, USA}

\author{Adam Schaefer}
\affiliation{Department of Astronomy, University of Wisconsin-Madison, 475N. Charter St., Madison WI 53703, USA}

\author[0000-0002-1283-8420]{Michele Cappellari}
\affiliation{Sub-department of Astrophysics, Department of Physics, University of Oxford, Denys Wilkinson Building, Keble Road, Oxford, OX1 3RH, UK}

\author{Xihan Ji}
\affiliation{Tsinghua Center of Astrophysics \& Department of Physics, Tsinghua University, Beijing, 100084, China}

\author{Matthew A. Bershady}
\affiliation{Department of Astronomy, University of Wisconsin-Madison, 475N. Charter St., Madison, WI 53703, USA}
\affiliation{South African Astronomical Observatory, P.O. Box 9, Observatory 7935, Cape Town, South Africa}

\author{Christy Tremonti}
\affiliation{Department of Astronomy, University of Wisconsin-Madison, 475N. Charter St., Madison, WI 53703, USA}

\author[0000-0002-9402-186X]{David R. Law}
\affiliation{Space Telescope Science Institute, 3700 San Martin Drive, Baltimore, MD 21218, USA}

\author{Renbin Yan}
\affiliation{Department of Physics and Astronomy, University of Kentucky, 505 Rose Street, Lexington, KY 40506, USA}

\author{Kevin Bundy}
\affiliation{University of California Observatories, University of California Santa Cruz, 1156 High St., Santa Cruz, CA 95064, USA}

\author{Shravan Shetty}
\affiliation{Department of Astronomy, University of Wisconsin-Madison, 475N. Charter St., Madison, WI 53703, USA}


\author{Niv Drory}
\affiliation{McDonald Observatory, The University of Texas at Austin, 1 University Station, Austin, TX 78712, USA}


\author{Daniel Thomas}
\affiliation{Institute of Cosmology \& Gravitation, University of Portsmouth, Dennis Sciama Building, Portsmouth, PO1 3FX, UK}

\author{Eric Emsellem}
\affiliation{European Southern Observatory, Karl-Schwarzchild-Str. 2, Garching bei M{\"u}nchen, 85748, Germany}
\affiliation{Univ Lyon, Univ Lyon1, ENS de Lyon, CNRS, Centre de Recherche Astrophysique de Lyon UMR5574, Saint-Genis-Laval, F-69230, France}


\author{Sebasti\'an F. S\'anchez}
\affiliation{Instituto de Astronomia, Universidad Nacional Aut\'onoma de M\'exico, A.P. 70-264, 04510, Mexico, D.F., Mexico}

\begin{abstract}

SDSS-IV MaNGA (Mapping Nearby Galaxies at Apache Point Observatory) is
the largest integral-field spectroscopy survey to date, aiming to
observe a statistically representative sample of 10,000 low-redshift
galaxies. In this paper we study the reliability of the emission-line
fluxes and kinematic properties derived by the MaNGA Data Analysis
Pipeline (\DAP). We describe the algorithmic choices made in the \DAP\
with regards to measuring emission-line properties, and the effect of
our adopted strategy of simultaneously fitting the continuum and line
emission. The effect of random errors are quantified by studying various
fit-quality metrics, idealized recovery simulations and repeat
observations. This analysis demonstrates that the emission lines are
well-fit in the vast majority of the MaNGA dataset and the derived
fluxes and errors are statistically robust. The systematic uncertainty
on emission-line properties introduced by the choice of continuum
templates is also discussed. In particular, we test the effect of using
different stellar libraries and simple stellar-population models on the
derived emission-line fluxes and the effect of introducing different
tying prescriptions for the emission-line kinematics. We show that these
effects can generate large ($>$ 0.2 dex) discrepancies at low
signal-to-noise and for lines with low equivalent width (EW); however, the
combined effect is noticeable even for H$\alpha$ EW $>$ 6~\AA. We
provide suggestions for optimal use of the data provided by SDSS data
release 15 and propose refinements on the \DAP\ for future MaNGA data
releases.

\end{abstract}
\keywords{methods: data analysis - surveys - techniques: imaging spectroscopy}

\section{Introduction}
\label{sec1}

Advances in our understanding of galaxy evolution are fundamentally
linked to the development of increasingly sophisticated models to derive
physical properties from observables. Integral-field spectroscopy (IFS)
surveys of nearby galaxies, combining large-number statistics with the
information content of resolved spectroscopy, represent some of the
richest datasets currently available to the astronomical community and
pose their own specific data-modeling challenges. Modern IFS surveys of
nearby galaxies -- including ATLAS$^{\rm 3D}$ \citep{Cappellari2011},
CALIFA  \citep{Sanchez2012}, SAMI \citep{Croom2012} and MaNGA
\citep{Bundy2015} -- are designed around a wide variety of science
goals, which often rely on the simultaneous determination of the stellar
and gas kinematics, emission-line ratios and stellar-population
properties (like age and metallicity) via specialized tools.

To provide users with readily available model-independent high-level
data products, the SDSS-IV (Sloan Digital Sky Survey) MaNGA  survey has
developed a data analysis pipeline (\DAP), to process the reduced MaNGA
datacubes in an automated and uniform way, which has now been released
publicly for the first time. In addition to fully reduced data products,
the fifteenth SDSS Data Release \citep[DR15,][]{Arxiv_Aguado2018}
includes the output of the MaNGA \DAP\ for an unprecedented sample of
4688 spatially-resolved galaxies. 

A detailed description of the \DAP\ design workflow, and output is
presented in  \cite{Westfall2019_arxiv}. In short, the MaNGA
\DAP\ is a project-led software effort designed to be both an automated pipeline and a
general-purpose tool. For DR15 the \DAP\ provides stellar kinematics,
emission-line properties, and assessments of stellar-continuum features
as measured by spectral indices, such as the Lick indices and D4000.

\cite{Westfall2019_arxiv} present a detailed assessment of
the stellar kinematics provided by the \DAP. Here we provide a
complementary analysis of the emission-line properties, focusing on
fluxes and kinematics. We validate the measurements as well as the
statistical fidelity of the \DAP-produced uncertainties. Our approach
follows both the classical perspective of adding noise to mock data as
well as making use of repeat observations specifically obtained for
testing the repeatability of the MaNGA survey output.

Importantly, we also recognize that the derivation of emission-line
fluxes and kinematics suffer from a certain amount of model-dependent
systematics. In this work, therefore, we explore several sources of
systematic error - e.g. the use of different stellar-continuum
templates, how one ties the kinematic parameters of different emission
lines, and the simultaneous  or sequential optimization of fits of the
emission lines and underlying continuum. Although these issues are not
new to the literature, we discuss them here in a coherent framework,
which we hope will constitute a useful reference for the
spectral-fitting community beyond the users of the MaNGA data itself.
 
A key aspect of accurately measuring the nebular emission lines is
properly accounting for the stellar continuum. This is particularly
important for the Balmer lines, where underlying stellar absorption can
reduce the H$\beta$ emission-line equivalent width by up to 10~\AA\ at
low spectral resolution \citep{Groves2012}.  In early work focused on
\ion{H}{2} regions, it was common to assume a constant 2~\AA\ correction
\citep{Mccall1985}.  The development of more sophisticated
stellar-population models enabled a more rigorous approach whereby the
stellar continuum is fit using a linear combination of simple
stellar-population models (SSPs) with reddening treated as an additional
free parameter.  This approach was first applied on a large scale to the
SDSS data by the MPA-JHU group in their analysis of the SDSS-I spectra
\citep{Tremonti2004, Brinchmann2004, Aihara2011}.   They carried out the
fitting in two stages: first, they masked the emission lines and modeled
the stellar continuum using a linear combination of \cite{Bruzual2003}
SSP models modified by a \cite{Charlot2000} dust law with the velocity
dispersion and redshift constrained a priori.  Next, the stellar
continuum was subtracted, low-order residuals were removed using a
sliding median, and the emission lines were simultaneously fit with
Gaussian functions. 

A downside of treating the stellar-population modeling and emission-line
fitting as separate steps is that valuable regions of the spectrum are
masked during the continuum fit and uncertainties in the continuum fit
are not propagated forward into the emission-line fits.  To circumvent
these issues and accurately measure very weak lines in early-type
galaxies, \cite{Sarzi2006} introduced a routine called {\tt GANDALF}
(gas and absorption-line fitting algorithm), based on an early version
of \pPXF\ \citep{Cappellari2004}, that {\it simultaneously} fits the
stellar continuum and nebular emission lines, given a
previously determined stellar kinematics solution for the continuum.
This code was subsequently applied to the SDSS data by \cite{Oh2011}.  

IFS data poses a particular  challenge to analyze because the outer
regions of galaxies often have low signal-to-noise (S/N) in the
continuum (i.e., S/N per pixel of 3--5).  This makes it difficult to
accurately constrain the stellar continuum, especially when stellar
kinematics are determined simultaneously with stellar-population ages
and metallicities.  The penalized pixel-fitting (\pPXF) software
 \citep{Cappellari2004, Cappellari2017}, developed for use on the SAURON
data \citep{Emsellem2004}, pioneered a robust pixel-fitting method,
particularly optimized for determining template mixes and robust
kinematics from data with moderate S/N ($\sim10-20$) and resolution.
This technique can be coupled with adaptive Voronoi binning
\citep{Cappellari2003} to achieve the S/N needed to accurately fit the
stellar continuum in IFS data. 

The CALIFA survey has led the way in terms of the development of
spectral-fitting pipelines suitable for a wide range of galaxy types
\citep[e.g.][]{CidFernandes2013a}. The {\tt FIT3D} pipeline
\citep{Sanchez2006} and its newer implementation {\tt Pipe3D}
\citep{Sanchez2016a} have developed a detailed procedure for employing
different binning schemes for the stellar and emission-line properties.
For example, {\tt Pipe3D} performs an initial spatial binning based on
continuum signal-to-noise ratio and analyzes the binned spectra to
determine the properties of the stellar continuum. It then re-scales the
best-fit continuum model to match the flux in each individual spaxel in
the bin, subtracts the re-scaled continuum, and fits the nebular lines \citep{Sanchez2016b}.
An emission-line-free spectrum is then created, and the process is
iterated without the emission-line masks. 

Other emission-line fitting codes have been developed to optimize the
information extracted from different data sets. For example, the SAMI
IFS data has higher spectral resolution in its red-wavelength arm than
either MaNGA or CALIFA, and, as a consequence, many of the emission
lines show complex line profiles that are not well-fit by a single
Gaussian profile \citep{Hampton2017, Green2018a}.  The {\tt LZIFU}
\citep{Ho2016a} code constrains the stellar continuum in individual
spaxels using \pPXF, and then it fits the emission lines with multiple
Gaussians where needed. 

In the MaNGA \DAP{}, we employ a `hybrid' binning scheme (Voronoi for
the continuum and individual spaxels for the emission lines; see Section
\ref{sec2.2}) and simultaneously fit the continuum and emission lines,
which is made possible by the latest version ($>6.0$ in \texttt{python})
of the \pPXF\ software package \citep{Cappellari2017}.\footnote{Available here 
\url{https://pypi.org/project/ppxf/}.} 

Although the \DAP\ fits the stellar continuum with the aim of deriving
accurate emission-line fluxes, the code does not provide
stellar-population properties (age, metallicity, etc.), because the MaNGA team 
considered these quantities to be too model-dependent to be provided by a general-purpose tool. 
Stellar population analysis of MaNGA galaxies presented in DR15 are therefore 
released as value-added catalogs (VACs). In the context of SDSS, a VAC is a 
product which is not generated by the SDSS project team, but instead contributed 
by specific members of the collaboration.

Two teams have released catalogs of stellar population properties 
for the MaNGA galaxies in DR15, constituting the {\tt FIREFLY} and {\tt Pipe3D} VACs.
{\tt FIREFLY} \citep{Wilkinson2015, Wilkinson2017} is a specialised full-spectral 
fitting code, which uses the output from the \DAP\ for binning, 
determination of the stellar kinematics and subtraction of nebular line
emission and computes stellar population properties 
(mean age, metallicity and dust extinction). 

The MaNGA data has been independently analysed with the {\tt Pipe3D}
\citep{Sanchez2016a, Sanchez2016b} code. Unlike {\tt FIREFLY}, {\tt Pipe3D} performs independent measurements 
of the stellar kinematics and emission lines, in addition to providing stellar population properties.  
A brief overview of these two VACs can be found in 
\cite{Arxiv_Aguado2018}.

Finally, we warn the potential user that the \DAP{} remains limited in
how well-suited its output is to certain emission-line-related science
goals. For example, studies of chemical composition of the interstellar
medium (ISM) within galaxies \citep{Sanchez2014,
Belfiore2017, Barrera-Ballesteros2017, Poetrodjojo2018a} rely on
accurate derivation of emission-line fluxes and their ratios within
H\textsc{ii} regions, and such studies benefit from the highest spatial
resolution allowed by the data to avoid contamination from the diffuse
ISM. The study of the ISM in early-type galaxies \citep{Sarzi2010,
Belfiore2016}, on the other hand, requires careful modeling of the
stellar continuum in order to recover the fluxes of faint
low-equivalent-width lines and may benefit from ad-hoc spatial binning.
Studies of the diffuse ionized ISM \citep{Zhang2017} and extra-planar
gas \citep{Jones2017} also crucially rely on binning and stacking of
low-surface-brightness emission, while galactic outflows can be
dissected by careful analysis of asymmetries in the emission-line
profiles \citep{Gallagher2018}. We anticipate that several users will
use the \DAP\ output as a reference and starting point towards more
complex and tailored analysis. 

In this paper, we describe the algorithmic choices made in the \DAP{}
with regards to measuring emission-line properties (Section \ref{sec2}).
The effects of random errors are quantified by studying various
fit-quality metrics, idealized recovery simulations and repeat
observations in Section \ref{sec3}. The systematic uncertainty in
emission-line properties introduced by the choice of continuum templates
is analyzed in Section \ref{sec4}. In particular, we test the effect of
using different stellar libraries and simple-stellar-population models
on the derived emission-line fluxes.  In Section \ref{sec5} we consider
other systematics introduced by our algorithmic choices, such as the
adopted strategy of simultaneously fitting the continuum with the
emission lines.  In Section \ref{sec6} we summarize our recommendations
for optimal use of the data provided by DR15 and some ideas for future
\DAP\ development. A brief summary is given in Section \ref{sec7}.
Throughout this paper, and in all data released by SDSS, wavelengths are
given in vacuum.

\section{The MaNGA \DAP{} algorithm}
\label{sec2}

\subsection{The input MaNGA data}
\label{sec2.1}

The MaNGA survey is one of the three key components of SDSS-IV
\citep{Blanton2017}, and aims to obtain IFS data for a representative
sample of 10~000 galaxies in the redshift range 0.01 $<$ z $<$ 0.15 by
2020. The MaNGA instrument operates on the SDSS 2.5m telescope at Apache
Point Observatory \citep{Gunn2006} and consists of a set of 17 hexagonal
fiber bundles of different sizes, plus a set of mini-bundles and sky
fibers used for flux calibration and sky subtraction respectively
\citep{Drory2015, Law2015, Yan2016, Yan2016a}. All fibers are fed into
the dual-beam BOSS spectrographs covering the wavelength range from
3600~\AA\ to 10300~\AA\ with a spectral resolution R $\sim$ 2000
\citep{Smee2013}. 

MaNGA galaxies are selected from an extended version of the NASA-Sloan
Atlas (NSA) and are observed out to 1.5 $\rm R_e$ (primary sample,
comprising 2/3 of the total sample) or 2.5 $\rm R_e$ (secondary sample,
comprising 1/3 of the total sample). Targets are selected to be
representative of the overall galaxy population at each stellar mass in
the range $\rm 9.0 < log(M_\star/M_\odot) < 11.0$. In practice the
absolute $i$-band magnitude is used for sample selection to avoid the
systematic uncertainty intrinsic in deriving stellar masses
\citep{Wake2017}.

The starting point of this paper are the datacubes produced for DR15, by
the MaNGA data-reduction pipeline (DRP; \citealt{Law2016}), with
additional modifications described in \cite{Arxiv_Aguado2018}. The MaNGA
\DAP\ takes as input the reduced MaNGA datacubes generated with
logarithmic wavelength sampling.

\begin{figure*}
	\centering
 	\includegraphics[width=0.95\textwidth, trim=0 60 0 20, clip]{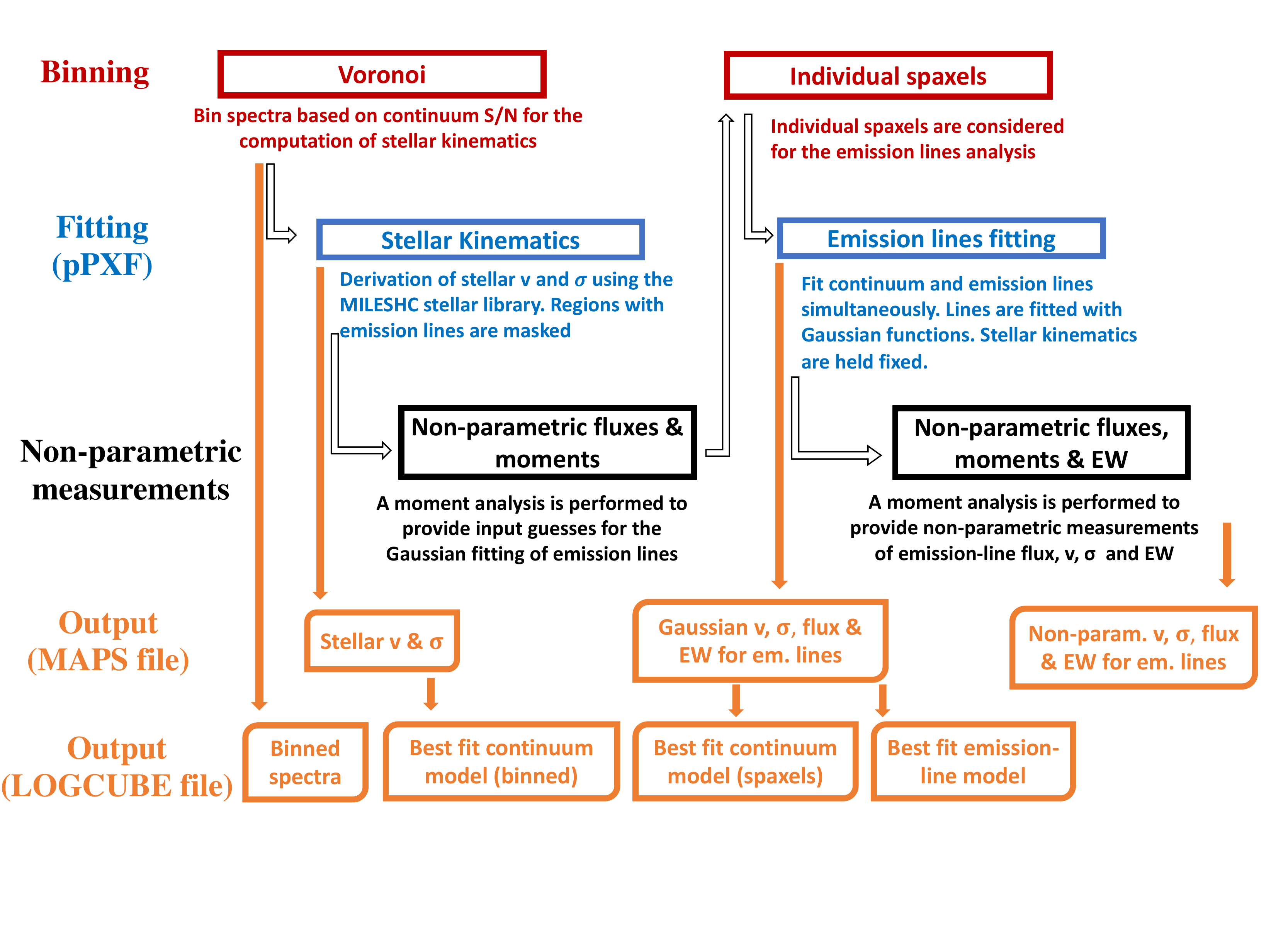}
     \caption{A graphical representation of the \DAP\ workflow. The
     figure highlights the interplay between different binning schemes
     (red), fitting steps performed with \pPXF\ (blue), non-parametric
     measurements (black) and output quantities (orange).  The algorithm
     can be roughly divided into two sections: the first step is
     dedicated to the extraction of the stellar kinematics (left side of
     the figure), while the second one performs the simultaneous fitting
     of gas and stars (right side of the figure). The main outputs
     produced by the \DAP\ are the \MAPS\ and {\tt LOGCUBE} files,
     containing respectively the 2D maps of derived parameters (e.g.
     velocities, fluxes, etc.) and the 3D best-fit models (in datacube
     format).}
     \label{dap_flow}   
\end{figure*}
  
An overview of the \DAP\ workflow is presented in \cite{Westfall2019_arxiv}. Here, we summarize the aspects of the algorithm that are
most relevant to the derivation of emission-line properties, in order to
motivate the tests performed in the rest of the paper. A graphical
overview of the relevant components of the \DAP\ workflow is presented
in Figure \ref{dap_flow} and discussed in the following sections.

\begin{figure*}
 	\includegraphics[width=1\textwidth, trim=0 0 0 0, clip]{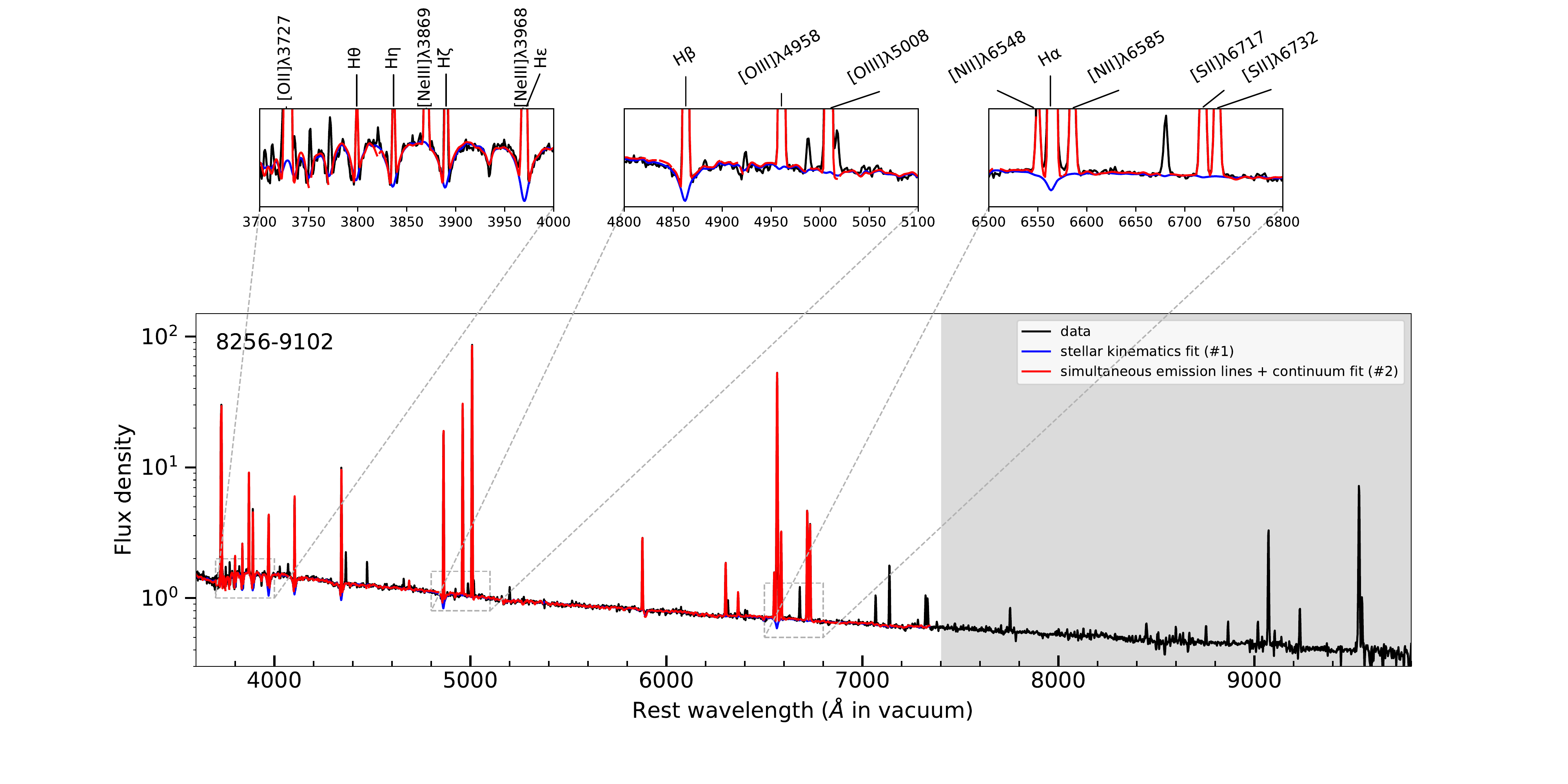}
 	\includegraphics[width=1\textwidth, trim=0 0 0 0, clip]{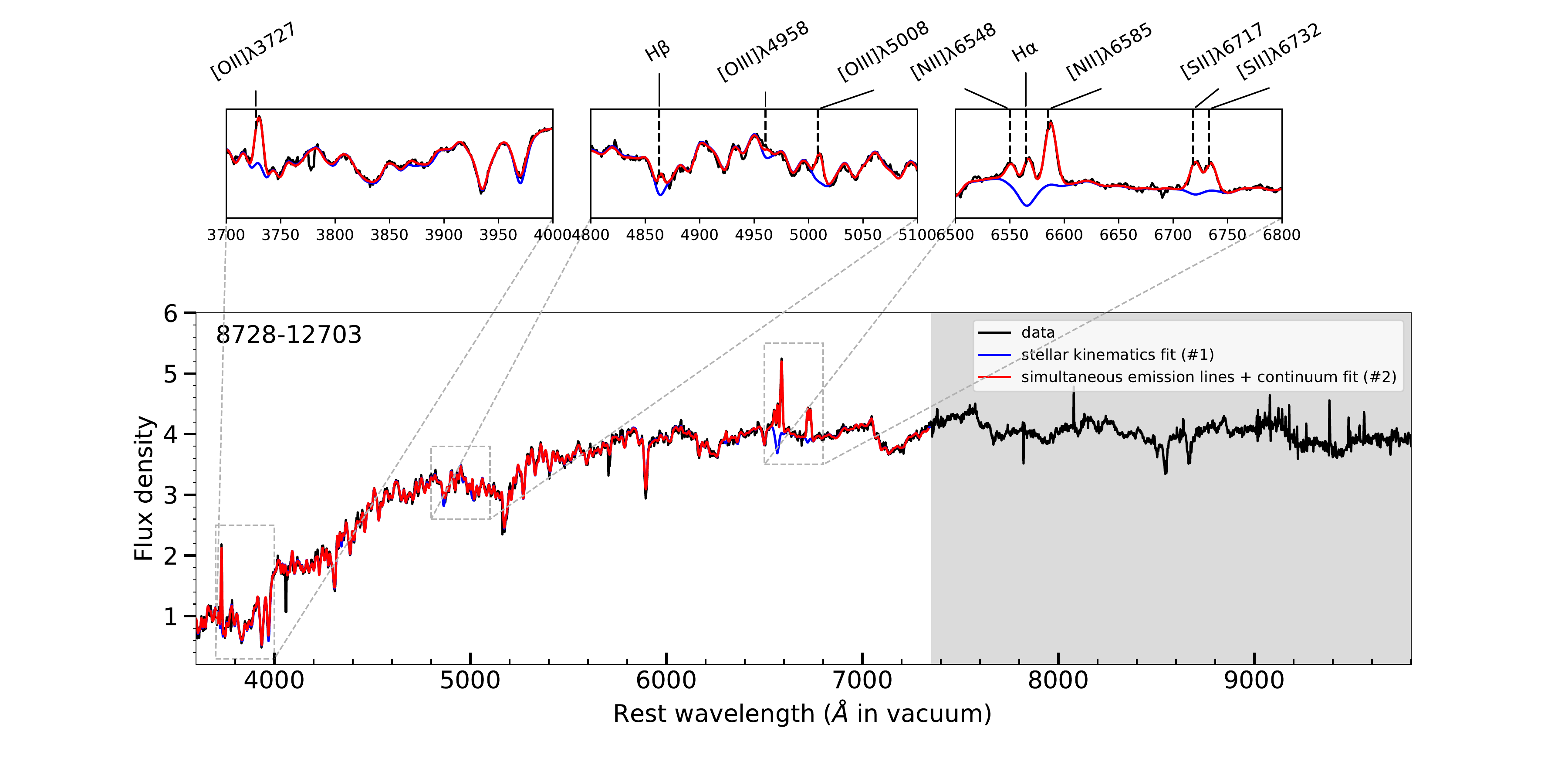}
     \caption{Two spectra fitted by the MaNGA \DAP. The top spectrum
     belongs to the central regions of a star-forming galaxy with very
     bright emission lines (8256-9102), while the bottom spectrum is
     taken from the central regions of an early-type galaxy with low-EW
     line emission (8728-12703). Zoomed views are provided for the
     wavelength regions around  [\oii]$\lambda\lambda$3727,29 and the
     4000~\AA\ break, H$\beta$ and [\oiii]$\lambda$5007 and H$\alpha$ and
     [\nii]$\lambda\lambda$ 6548,84. The figures show the data in black,
     the results of the first fit (optimized for the determination of
     stellar kinematics) in blue and the second fit (optimized for
     emission-line parameters) in red. During the first fit the spectrum
     is masked around the expected positions of the main emission lines.
     The region redder than $\sim$ 7400~\AA\ is not fitted in DR15
     because of the limited wavelength coverage of the \mileshc\
     templates.}
     \label{example_spectra}   
\end{figure*}

\subsection{Overview of the MaNGA \DAP\ with regards to emission lines}
\label{sec2.2}

The \DAP\ currently performs two full-spectrum fits, both using \pPXF.
The first fit primarily determines the stellar kinematics and the second
models the emission lines. Two example spectra fitted by the MaNGA \DAP\
can be inspected in Figure \ref{example_spectra}. For DR15, the stellar
kinematics are determined only for spectra binned to $g$-band S/N$>$10,
as produced by applying the Voronoi binning algorithm implemented in the
\texttt{python} language\footnote{Available here
\url{https://pypi.org/project/vorbin/}} by \citet{Cappellari2003}.  The
subsequent emission-line modeling is done for both the binned spectra
and after deconstructing the bins into the individuals spaxels; a full
description of the emission-line-fitting module of the \DAP\ is provided
in Section 8 of \cite{Westfall2019_arxiv}. As it is most relevant
to emission-line science, we focus on the results provided by the latter
approach, which we term the `hybrid' binning scheme (see below).

\subsubsection{The stellar-kinematics fit (first fitting stage)}

When using \pPXF\ to determine the stellar kinematics (\citealt{Westfall2019_arxiv}, Section 7), spectral regions potentially affected by
line emission are masked.  The adopted masks extend $\pm$750 \kms\
around the expected emission-line wavelength at the galaxy systemic
redshift.  In DR15, we use a set of templates determined using a
hierarchical-clustering analysis of the MILES stellar library
\citep{Sanchez-Blazquez2006, Falcon-Barroso2011}, which we refer to as the \mileshc\ library (see Section 5 of \cite{Westfall2019_arxiv}, and Section \ref{sec4.1} below).  We also
include an eighth-order additive Legendre polynomial, motivated by the
experience from the SAURON \citep{Emsellem2004} and ATLAS$\rm ^{3D}$
\citep{Cappellari2011} teams, to improve the quality of the derived
kinematics by providing a closer match between data and spectral
templates.  However, it is important to note that the additive
polynomials modify the absorption-line depth of the stellar templates,
which becomes important to our discussion in Section \ref{sec5.2}.

\subsubsection{The hybrid binning approach}

Since the line emission surface-brightness can be very different from
the continuum surface-brightness, the relevant binning scale is not
necessarily the same for continuum and emission-line science.  Indeed,
the emission-line fitting can be performed by optimally rebinning the
data for the extraction of the emission-line properties
\citep{CidFernandes2013a, Belfiore2016, Sanchez2016}; however, this
strategy has not yet been implemented within the \DAP.  Instead, we
model the emission lines in each spaxel as follows: we first remap the
best-fitting stellar kinematics determined for the binned spectra to the
individual spaxels and then, keeping the stellar kinematics fixed, we
simultaneously optimize the stellar-continuum and emission-line
templates to determine the best-fit model spectrum.  We refer to this as
the `hybrid' binning approach because the stellar kinematics uses the
Voronoi binned data, whereas the emission-line results are for
individual spaxels.  Importantly, during the emission-line modeling, the
algorithm re-optimizes the continuum templates to fit each spaxel,
rather than simply rescaling the best-fit stellar continuum from the
Voronoi-binned fit to each spaxel (as done, for example, by {\tt
Pipe3D}).  {\it The output of this scheme} ({\tt HYB10-GAU-MILESHC}; see
below) {\it is the recommended data product in DR15 for users interested
in emission-line properties of MaNGA galaxies.} A fit to the
emission lines on the same Voronoi bins as the stellar continuum is also
provided for users whose science goals require, e.g., the stellar and
gas kinematics to be computed over the exact same spatial scales.

We considered it important in the hybrid binning scheme to fix the stellar kinematics 
based on the binned spectra when fitting the individual spaxels because non-linear 
parameters (such as velocity and $\sigma$) may suffer from biases when derived in low-S/N single-spaxel spectra.

\begin{table*}
	
	\centering

    \caption{Wavelengths and ionization potential of the relevant ion
    for each emission line fit for DR15, subdivided in the groups
    defined to study the different tying schemes described in Section
    \ref{sec5.3}.  Ritz wavelengths in vacuum are taken from the
    National Institute of Standards and Technology (NIST;
    \url{http://physics.nist.gov/PhysRefData/ASD/Html/help.html}). The
    \DAP\ string name is reported in the header of the \MAPS\ files for
    the emission-line extensions and allows the users to associate each
    map with the correct line (see Section \ref{sec6.1}).  Ionization
    potentials are taken from \cite{Draine2011}.  Lines redder than
    $\sim$ 7400~\AA\, corresponding to the red cutoff of the \mileshc\
    stellar templates, are not fit in DR15.}
	
	\begin{tabular}{c c c c c}
		\hline
		line name & wavelength (vacuum) [\AA] & \DAP\ string name & Ionization potential [eV] & Fixed ratio \\
		\hline
		\multicolumn{5}{c}{Hydrogen Balmer lines} \\
		\hline
		H$\theta$ (H10) & 3798.983  & Hthe-3798 & 13.60 & no \\
		H$\eta$ (H9) & 3836.479 & Heta-3836 & 13.60  & no\\
		H$\zeta$ (H8) & 3890.158 & Hzet-3890 & 13.60 & no\\
		H$\epsilon$ (H7) & 3971.202 &Heps-3971 & 13.60 & no \\
		H$\delta$ &  4102.899 &Hdel-4102 & 13.60 & no\\
		H$\gamma$ & 4341.691 &Hgam-4341 & 13.60  &no \\
		H$\beta$ & 4862.691 &Hb-4862& 13.60 & no \\
		H$\alpha$ & 6564.632 & Ha-6564& 13.60  & no\\
		\hline
		\multicolumn{5}{c}{Low ionization lines} \\
		\hline
		\mbox{[\oii]}$\lambda$3727 & 3727.092 & OII-3727 & 13.61 & no\\
		\mbox{[\oii]}$\lambda$3729 & 3729.875 & OII-3729 & 13.61 & no\\
		\mbox{[\oi]}$\lambda$6300 & 6302.04 & OI-6302 & 0.0 & no\\
		\mbox{[\oi]}$\lambda$6364 & 6365.535 & OI-6365 & 0.0 & 0.328 [\oi]$\lambda$6300 \\
		\mbox{[\nii]}$\lambda$6548 &  6549.86 &NII-6549 & 14.53 & 0.327 [\nii]$\lambda$6584\\
		\mbox{[\nii]}$\lambda$6584 &  6585.271 &NII-6585 & 14.53 & no \\
		\mbox{[\sii]}$\lambda$6717 & 6718.294 &SII-6718 & 10.36 & no \\
		\mbox{[\sii]}$\lambda$6731 & 6732.674 &SII-6732 & 10.36 & no\\
		\hline
		\multicolumn{5}{c}{High ionization lines} \\
		\hline
		\mbox{[\neiii]}$\lambda$3869 & 3869.86  & NeIII-3869& 40.96 & no\\
		\mbox{[\neiii]}$\lambda$3968 & 3968.59  & NeIII-3968 & 40.96 & no\\
		\heii$\lambda$4687 & 4687.015  & HeII-4687 & 54.41 & no\\
		\mbox{[\oiii]}$\lambda$4959 & 4960.295  & OIII-4960& 35.12 & 0.340 [\oiii]$\lambda$5007 \\ 
		\mbox{[\oiii]}$\lambda$5007 & 5008.240  & OIII-5008 &35.12 & no\\ 
		\hei$\lambda$5876 		& 5877.243  & HeI-5877 & 24.58  & no\\
	\end{tabular}
	
	\label{emission_line_groups}
\end{table*}

\subsubsection{Simultaneous fit of gas and stars (second fitting stage)}

To simultaneously optimize the fit to the stellar continuum and the
emission lines, we use \pPXF\ with Gaussian emission-line templates
associated to kinematic parameters that are independent of those used
for the stellar templates \citep{Cappellari2017}.  Simultaneous fits of
emission-line and continuum templates has been introduced and
recommended in previous work \citep{Sarzi2006, Oh2011} as a way of
minimizing the bias resulting from masking of the wings of the Balmer
absorption profiles and other stellar features on the recovered
emission-line fluxes. This approach allows one to enforce the physical
constraint that emission lines cannot be negative, while optimizing the
fit to the stellar continuum. \pPXF\ adopted the same idea for the gas
fitting, but implemented this strategy in a different way as described
in \cite{Cappellari2017}. An example of the results of this fitting
algorithm can be seen in Figure \ref{example_spectra} for both a highly
star-forming and an early-type galaxy. We assess the difference in the
best-fit models derived by simultaneous versus subsequent fitting of the
emission lines in Section \ref{sec5.1}.

The 22 emission lines fit by the \DAP\ in DR15 is presented in Table
\ref{emission_line_groups}. The flux ratios of doublets are fixed when
such ratios are determined by atomic physics \citep[Table
\ref{emission_line_groups};][]{Osterbrock2006}.  For DR15 we tie the
velocity of all the fitted emission lines, but do not tie the velocity
dispersions with the exception of most of the doublets.  Specifically,
the line doublets with fixed flux ratios and the
[\oii]$\lambda\lambda$3727,29 doublet have their velocity dispersions
tied between the doublet lines, but each doublet and all other lines
have independent velocity dispersions.\footnote{Tying the
[\oii]$\lambda\lambda$3727,29 doublet is particularly important because
it is unresolved at the MaNGA spectral resolution and large degeneracies
in the fit would result otherwise.}  While tying the kinematics of
different emission lines may prove advantageous to recover the flux of
weak lines, we do note tie all velocity dispersions in DR15 to allow for
modest inaccuracies (generally a few percent; Law et al., {\it in
preparation}) in the wavelength-dependent line-spread function (LSF)
determined by the MaNGA DRP; further discussion of the current
line-tying strategy is presented in Section \ref{sec5.2}.  Given the
limited spectral range of the adopted \mileshc\ stellar library, lines
redder than $\sim$7400~\AA\ are not fit for DR15.  In future data
releases, however, we plan to use templates with a larger wavelength
range to allow for a determination of the continuum under the emission
lines redder than 7400~\AA\ (see Section \ref{sec4}).

During the fitting procedure, the \DAP\ does not constrain the
emission-line fluxes to follow a specific attenuation law resulting from
dust present in the host galaxy (as generally done in {\tt GANDALF},
e.g., \citealt{Oh2011}, and optionally available in \pPXF).  All line
fluxes are, however, corrected for Galactic foreground extinction using
the maps of \cite{Schlegel1998} and the reddening law of
\cite{O'Donnell1994}.  Users comparing the \DAP\ output with the output
generated by other spectral-fitting pipelines may need to take this
factor into account.

An eighth-order multiplicative Legendre polynomial is used in DR15 to
match the overall spectral shape of the data, which can deviate from
that of the models both because of dust extinction and small
inaccuracies in the spectrophotometric calibration. A physically
motivated extinction law may be used in the MaNGA \DAP\ instead of
multiplicative polynomials, but this was found to produce worse fits to
the stellar continuum, especially at the blue end of the spectrum (see
Section \ref{sec5.1}). Additive polynomials are not, and indeed ought
not to be, used in this stage as they modify the depth of stellar
absorption lines in the templates, therefore potentially leading to
degeneracies with emission-line strengths. This point is further
elucidated in Section \ref{sec5.1}. Neither polynomials nor extinction
corrections are applied to the emission-line templates by the fit, but
only to the stellar continuum. 

\subsubsection{Non-parametric emission-line properties and EW}

The \DAP\ also calculates non-parametric emission-line moments (zeroth,
first, and second), both before and after the emission-line modeling
(see Figure \ref{dap_flow}).  Both iterations subtract a best-fit
stellar-continuum model before calculating the moments; additional
detail is provided in Section 9 and Table 2 of \cite{Westfall2019_arxiv}.  The first iteration subtracts the best-fit stellar
continuum used to determine the stellar kinematics, and the first
moments are used as initial guesses for the ionized-gas velocities in
the emission-line modeling.  The second iteration subtracts the best-fit
stellar continuum determined during the emission-line modeling to
account for the re-optimization of the continuum fit.  The integrated
fluxes (zeroth moments) from the second iteration are provided in the
\DAP\ output ({\tt SFLUX} extensions in the output file), in addition to
the values derived from the Gaussian fitting ({\tt GFLUX} extensions).
Equivalent widths (EW) for each line are obtained by dividing the flux
in the line by local pseudo-continua. Both summed and Gaussian-fit
fluxes are used, leading to two computations of the EW in the final
output ({\tt SEW} and {\tt GEW}, respectively).

\subsection{Output files}

A full description of the \DAP\ output data model is provided by
\cite{Westfall2019_arxiv}, particularly in Sections 2 and 11 and
Appendix A.

Emission-line fluxes (from both Gaussian fitting and the moments
analysis), velocities, velocity dispersion and associated errors and
masks are consolidated into the main \DAP\ output file, the \MAPS\ file,
for each analyzed datacube.  The \MAPS\ file is a multi-extension fits
file, where each extension provides a set of 2D maps of \DAP\
measurements.\footnote{For the datamodel  of the \MAPS\ file see
\url{https://data.sdss.org/datamodel/files/MANGA_SPECTRO_ANALYSIS/DRPVER/DAPVER/DAPTYPE/PLATE/IFU/manga-MAPS-DAPTYPE.html}
and Section 11.1 and Table 4 of \cite{Westfall2019_arxiv}.}

The best-fit continuum and emission-line models are given as extensions
in the \DAP\ model {\tt LOGCUBE} file.  Most extensions in this file are
three-dimensional datacubes, presented on the same world coordinate
frame as the input MaNGA datacube.  The model {\tt LOGCUBE} provides the
results of both the continuum-only fit used to determine the stellar
kinematics, and the combined fit used to simultaneously model the
continuum and emission.\footnote{For the model {\tt LOGCUBE} datamodel see
\url{https://data.sdss.org/datamodel/files/MANGA_SPECTRO_ANALYSIS/DRPVER/DAPVER/DAPTYPE/PLATE/IFU/manga-LOGCUBE-DAPTYPE.html}
and Section 11.2 and Table 5 of \cite{Westfall2019_arxiv}.}  We
highlight here, however, that the continuum from the stellar-kinematics
fit should \textit{not} be used to recompute emission-line parameters.  

\subsection{Known limitations}

While the vast majority of spaxels are successfully fit by the MaNGA
\DAP, users should be aware of some known failure modes, discussed in
Section 10.2.2 of \cite{Westfall2019_arxiv}. In the context of
line emission, it is particularly important to note that broad-line AGN
are generally not well fit.  All current runs of the \DAP\ assume a
single Gaussian component per emission line, meaning that it is not
currently possible to recover the broad and narrow components present in
Type I AGN.  This represents a notable limitation of the current \DAP,
and we therefore recommend users interested in Type I AGN -- roughly 1\%
of MaNGA galaxies \citep{Sanchez2017} -- to perform their own spectral
fitting.  Further recommendations and descriptions of known bugs in the
DR15 \DAP\ run are presented in Section \ref{sec6.1}.

\section{Quality assessment and estimate of random errors}
\label{sec3}

In the section we address the question of whether the \DAP\ performs a
successful fit to the emission lines found in the MaNGA data. We further
address the robustness of the errors provided by the \DAP\ by making use
of both idealized recovery simulations and analysis of repeat
observations.

\subsection{Emission line quality fit metrics}
\label{sec3.1}

\paragraph{Defining line S/N and A/N.}

\begin{figure*}
 	\includegraphics[width=0.48\textwidth, trim=0 0 0 0, clip]{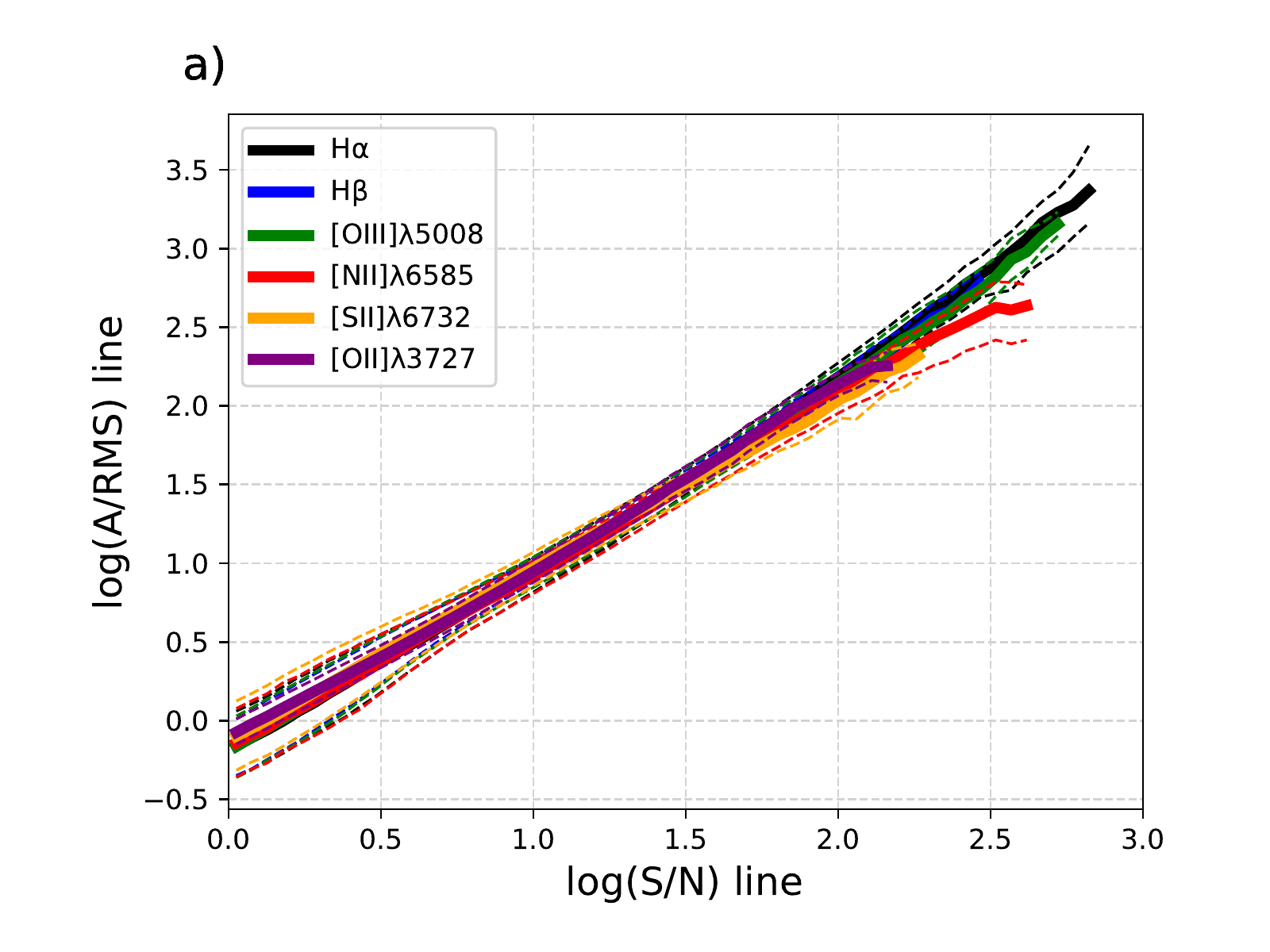}
     \includegraphics[width=0.48\textwidth, trim=0 0 0 0, clip]{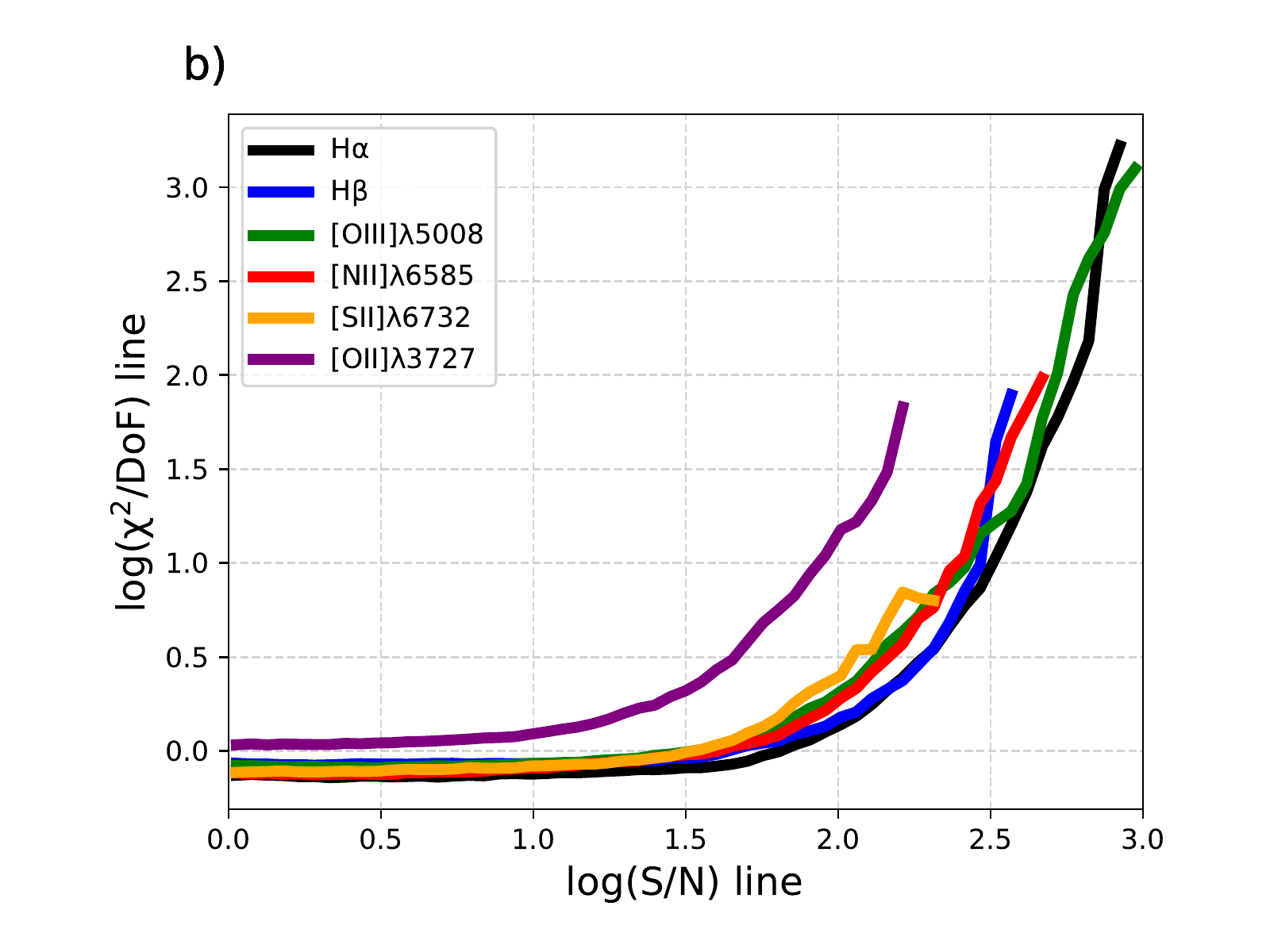}
     \caption{\textbf{a)}: The relation between $\rm S/N_{line} $
     (Equation \ref{s_n_def}) and  $\rm A/N_{line}$ (Equation
     \ref{a_n_def}) for a sample of spaxels taken from 300 random
     MaNGA galaxies in DR15. Different colors represent the median
     relation obtained for different strong emission lines, while the
     dashed curves represent the 16$\rm ^{th}$ and 84$\rm ^{th}$
     percentiles. This panel demonstrated that the A/N closely tracks the S/N as measured by the \DAP.
      \textbf{b)} The relation between the $\chi^2$ per
     degree of freedom and $\rm S/N_{line} $ for the same sample of
     spaxels and set of strong emission lines as in panel (a). The $\rm \chi^2/DoF$ increases for high S/N as a consequence of template mismatch (i.e. a Gaussian is not a perfect model for a high-S/N emission line). We argue in the text that even in this regime the line fluxes measured by the MaNGA DAP are accurate.}
     \label{fit_metrics1}   
 \end{figure*}

Several fit quality measures have been employed in the literature to
describe the reliability of measured parameters for an emission line.
The most used characterization of fit quality is the fractional error on
the recovered line flux $\rm Err_{Flux} / Flux$, which is often quoted
in terms of the `signal-to-noise' of a line, defined as
\begin{equation}
\rm S/N_{line} \equiv  Flux/ Err_{Flux}.  \label{s_n_def}
\end{equation}

A more empirical way of assessing line detection relies on quantifying
how much the line protrudes above the noise level in the spectrum. This
is usually measured by the amplitude over noise ratio,
\begin{equation}
\rm A/N_{line} \equiv Amplitude/ RMS,
\label{a_n_def}
\end{equation}
where the amplitude refers to the best-fit Gaussian amplitude and the
root-mean-square (RMS) is calculated from the residuals between the data
and the model in small regions on either side of the line.

\paragraph{Line S/N as a good measure of fit quality.}

In Figure \ref{fit_metrics1} we plot the $\rm S/N_{line}$ obtained from
the \DAP\ \MAPS\ file versus the $\rm A/N_{line}$ obtained using the
\DAP\ fit residuals around the position of the line for a sample of 300
random galaxies in DR15 ($\sim 4 \cdot 10^4$ spaxels). The RMS is
computed as the mean RMS in sidebands bluewards and redwards of the
position of each line.\footnote{The same sidebands are used to determine
the continuum term in the computation of EW, and are listed in Table 3
of \cite{Westfall2019_arxiv}.}

The figure highlights the tight linear relation in log space between the
two quantities across almost three dex in $\rm S/N_{line}$ for strong
lines spanning a large fraction of the MaNGA wavelength range.  This
tight relation, which shows an increase in scatter only for $\rm
S/N_{line}<3$, implies that the S/N computed by the \DAP\ is equivalent
to the more empirical A/N to very good accuracy. This scaling is indeed
expected, since most emission lines in MaNGA are unresolved and their
velocity dispersion is roughly comparable to one pixel in the spectral
direction. 

To check whether the fit residuals at the position of different emission
lines are comparable with the error spectrum, we computed the $\chi^2$
per degree of freedom (dof) of the Gaussian fit in 15-pixel windows
around the fitted position of each line center. In the right panel of
Figure \ref{fit_metrics1} we show the median relation between the $\rm
\chi^2/dof$ for each line and $\rm S/N_{line}$. All strong lines
considered follow a similar relation, except [\oii]$\lambda3727$, which
suffers from worse $\rm \chi^2$ at fixed $\rm S/N_{line}$, possibly as a
result of the difficulty in correctly fitting the unresolved doublet.
For $\rm S/N_{line} < 30$, the other strong lines show a roughly
constant $\rm \chi^2/dof \sim 0.8 $. At higher $\rm S/N_{line}$, the
$\rm \chi^2/dof$ increases sharply, up to 3 orders of magnitude. We
interpret this as ``template mismatch'', in the sense that our Gaussian
model represents an increasingly worse representation of the observed
line profiles at high S/N. In this regime the $\rm \chi^2$ could be
lowered by fitting each emission line with more than one Gaussian
component (e.g. \citealt{Gallagher2018}), but this goes beyond the scope
of the current MaNGA \DAP.  We note that a similar behavior in the
$\rm \chi^2/dof$ over the full spectrum in both full-spectrum-fitting
modules of the \DAP, as shown in Figure 19 of \cite{Westfall2019_arxiv}; however, the discrepancies between model and data are orders
of magnitude stronger in these small windows near each line.

\begin{figure}
	\includegraphics[width=0.5\textwidth, trim=0 0 0 0, clip]{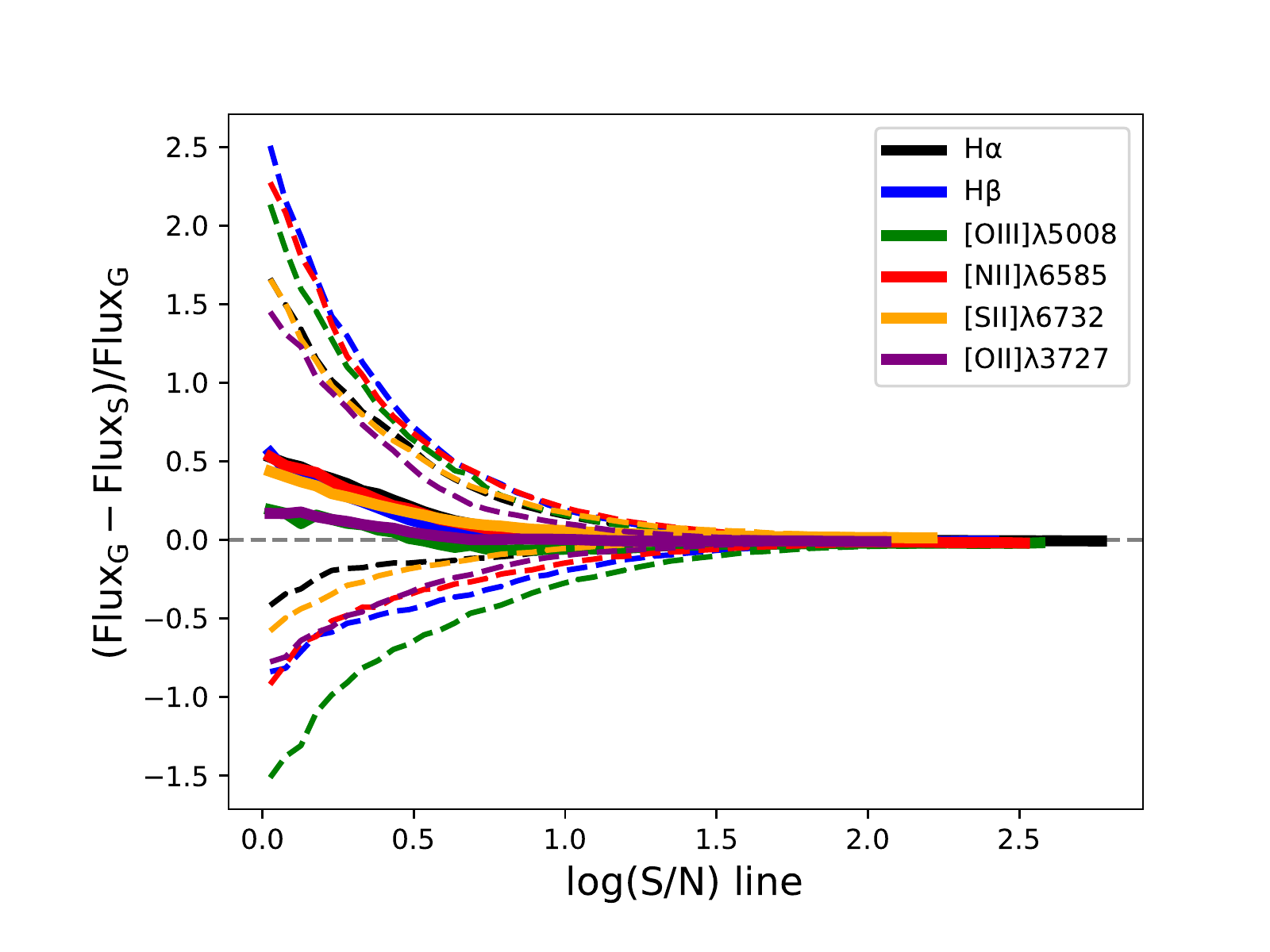}
	\caption{The relative difference between the flux derived from Gaussian fitting (Flux$_{G}$) and that derived from zeroth-moment analysis (also referred to as summed flux, Flux$_{S}$) as a function of line S/N (estimated from the Gaussian flux). In the case of the $\rm [OII]\lambda\lambda 3737,29$ both Gaussian and summed fluxes refer to both components of the doublet. The comparison demonstrates excellent agreement between Gaussian and summed fluxes for all lines in the high S/N regime. At low S/N Gaussian fluxes then to be higher than summed fluxes because we constrain Gaussian models to have positive amplitude. }
	\label{gauss_vs_summed}
\end{figure}

We note that this increase in $\rm \chi^2/dof$ does not mean that the fluxes obtained via
Gaussian fitting are unreliable. In fact we have checked that the fluxes obtained for Gaussian fitting
agree exceedingly well with line fluxes obtained by simply summing the flux around 
the position of the line ({\tt SFLUX} extension of the DAP MAPS files), as can be seen in Fig. \ref{gauss_vs_summed}. For S/N $<$ 3, we start to see a discrepancy between the two flux measurements, with Gaussian fluxes being higher on average. This effect is partially due to the fact that we fit Gaussian with positive amplitude, while we allow negative summed fluxes. Exclusion of the negative summed fluxes from the comparison improves the median agreements at low S/N (not shown).

In light of this discussion, we conclude that emission lines are
statistically well-fit by a single Gaussian model at low S/N. 
The increasing $\rm \chi^2/dof$ does not imply that line fluxes from Gaussian 
fitting are unreliable in this regime, as can be demonstrated by a comparison with the
non-parametric summed fluxes. 
Since $\rm S/N_{line}$ correlates very well with the empirically derived
A/N, we suggest that, despite its simplicity, $\rm S/N_{line}$ is an
excellent metric for the uncertainty in the fit. Henceforth, in this
paper, $\rm S/N_{line}$ is used instead of A/N. 

\begin{figure*}
    \includegraphics[width=\textwidth, trim=0 0 0 0, clip]{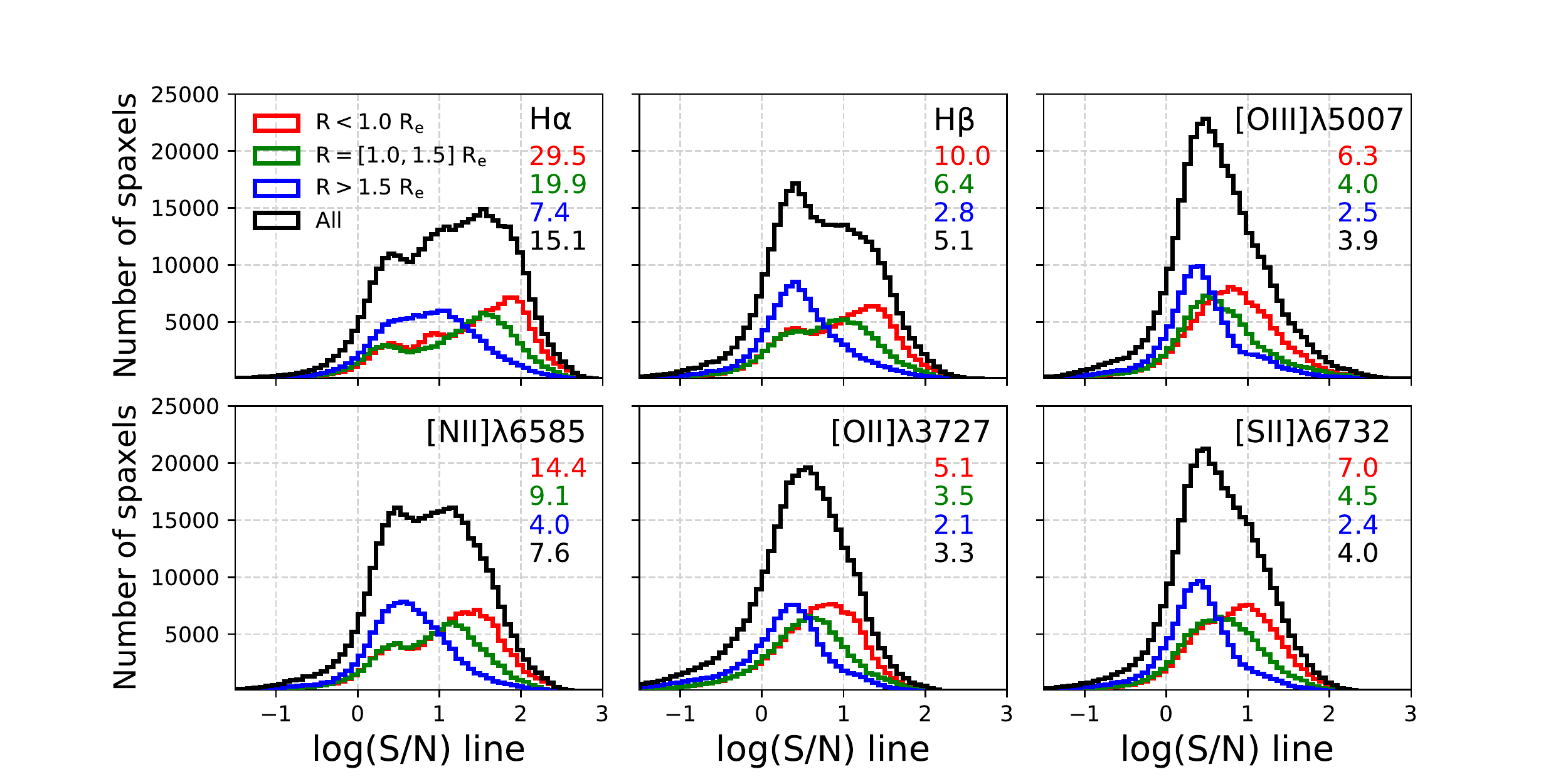}
     \caption{Distributions of emission-line signal-to-noise ratios for
     a large sample of spaxels (drawn from a random sample of 300 MaNGA
     galaxies in DR15). The black histogram includes all spaxels, while
     other histograms represent spaxels at different galactocentric
     distances ($\rm < 1 R_e$: red; [1.0-1.5] $\rm R_e$: green; $\rm >
     1.5 R_e$:blue). In color, on the right-hand side of each plot, we
     list the average S/N ratio for each line in the associated radial
     range.}
     \label{fit_metrics2}
\end{figure*}

\paragraph{The typical line S/N of MaNGA data} To conclude this section,
we present in Figure  \ref{fit_metrics2} the S/N distribution of the
same sample of spaxels used in Figure \ref{fit_metrics1}, which is
representative of the line S/N distribution in the MaNGA data. Only
spaxels with S/N $>$ 0 are plotted; i.e., we do not plot the large
number of spaxels that have no detected line emission according to our
fitting procedure.  Colored lines show the S/N distribution in three
different radial bins.  The radial variation of these S/N distributions
highlight the decrease in S/N even in the strong nebular lines for $\rm
R> 1.5 \ R_e$.  We note that the MaNGA sample includes both star-forming
and passive galaxies that are characterized by low-S/N line emission.
In particular, the bimodality between star-forming and low-ionisation
emission-line regions (LIERs, \citealt{Belfiore2016}) is evident as a
bimodality in the S/N of the Balmer lines, especially at small
galactocentric radii.

\subsection{Idealized recovery simulations}
\label{sec3.2}

In order to test the the presence of possible systematic errors in the
recovery of emission-line parameters and the statistical correctness of
the errors produced by the \DAP\ we have carried out a set of idealized
recovery simulations. Four test galaxies were selected to span a wide
range of stellar continuum and emission-line properties (two
star-forming blue galaxies and two red LIER galaxies). Considering all
four galaxies, our mock dataset consists of $\sim$ 5000 spaxels with S/N
> 1 in H$\alpha$.

The MaNGA datacube for each galaxy was fit using the DR15 version of the \DAP\ and the best-fit model cube, including both continuum and
emission lines, was used as a template for generating `mock' datacubes.
For each spaxel in the model cube, Gaussian noise was added to the model
spectrum, with a standard deviation given by the error vector in the
input MaNGA data. Assuming the MaNGA {\tt DRP} errors are accurate, this
procedure generates mock cubes with the same noise level as the original
data. Mock cubes with twice and half the noise level of the original
data were also created. 

All the mock cubes were run through the MaNGA \DAP\ in the same way as
real MaNGA data. In particular, the same \mileshc\ stellar templates
were used to generate and then fit the mock cubes. In Section
\ref{sec4.4} we repeat this exercise using a different template set to
fit the simulated data, and discuss the effect of template mismatch.

\begin{figure*}
	\centering
 	\includegraphics[width=0.48\textwidth, trim=0 0 0 0, clip]{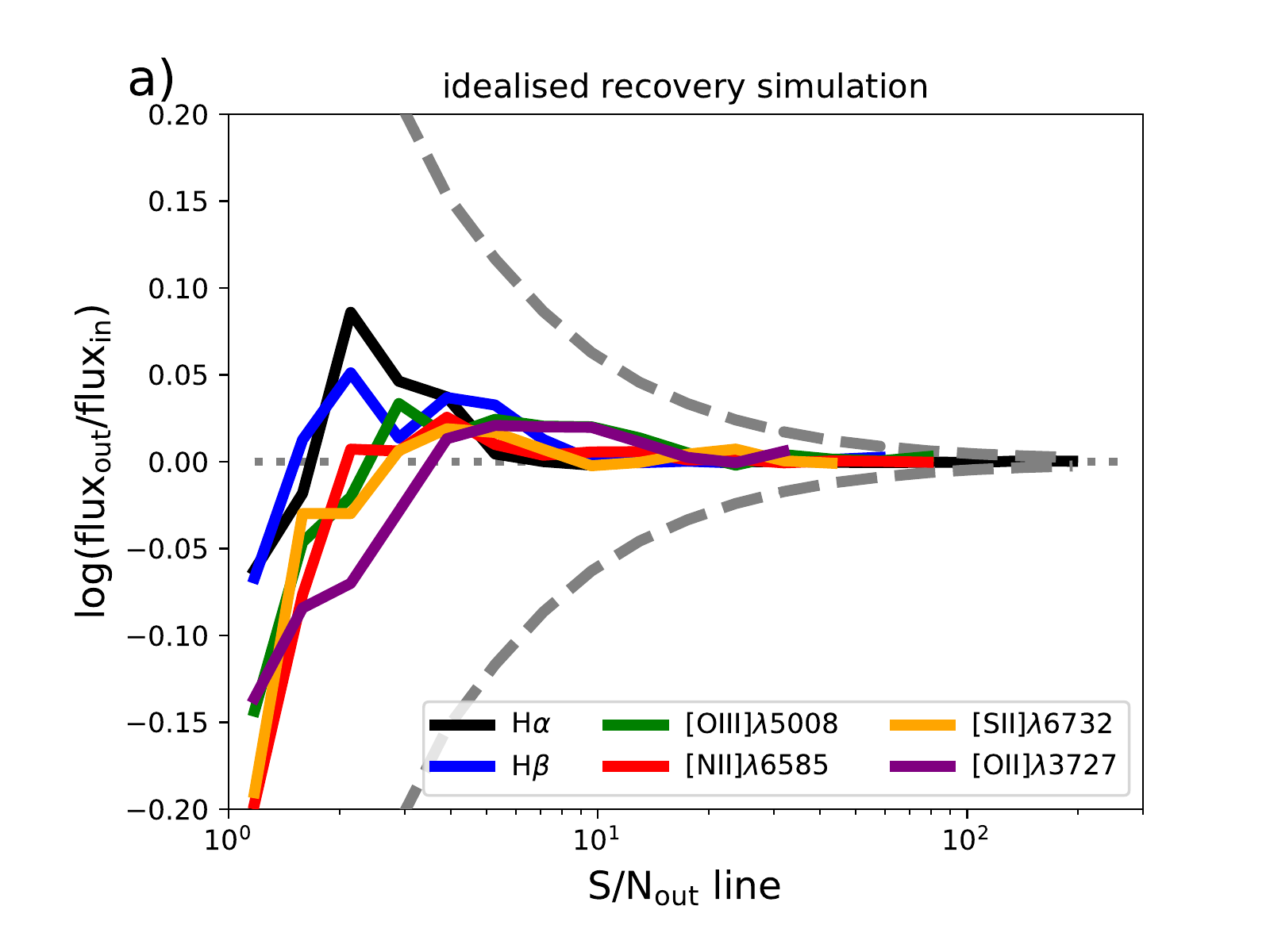}
 	\includegraphics[width=0.48\textwidth, trim=0 0 0 0, clip]{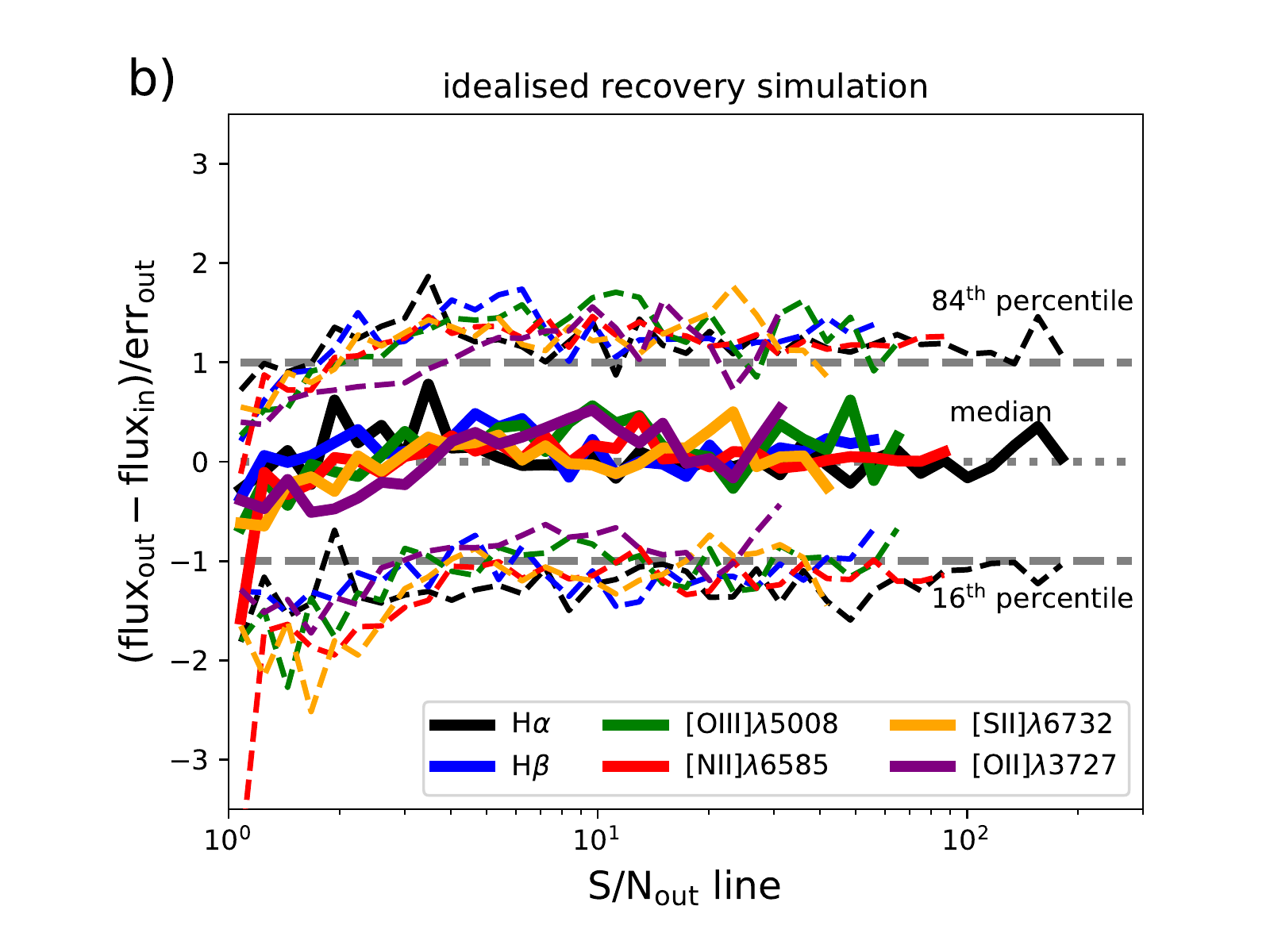}
    \includegraphics[width=0.48\textwidth, trim=0 0 0 0, clip]{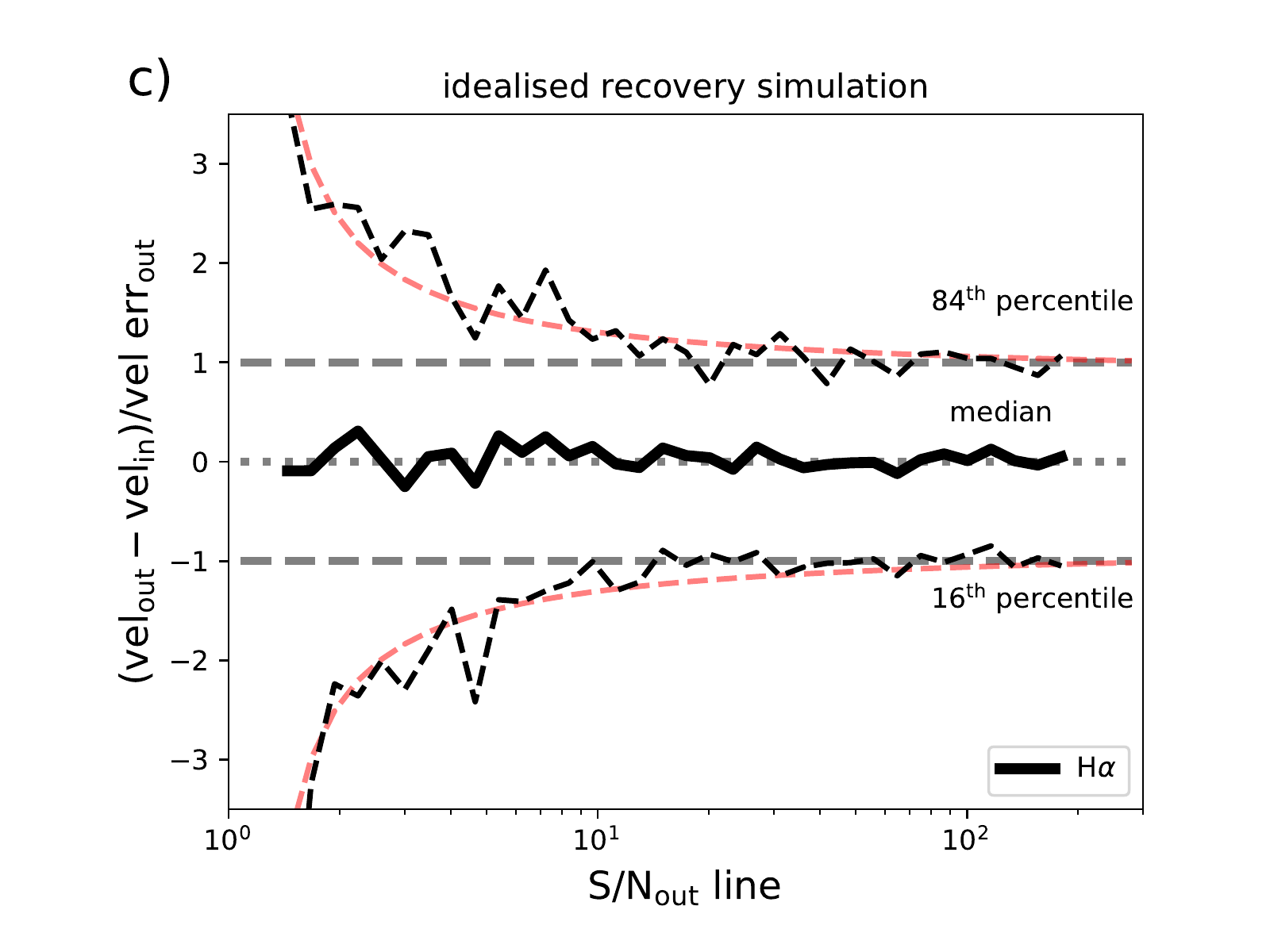}
    \includegraphics[width=0.48\textwidth, trim=0 0 0 0, clip]{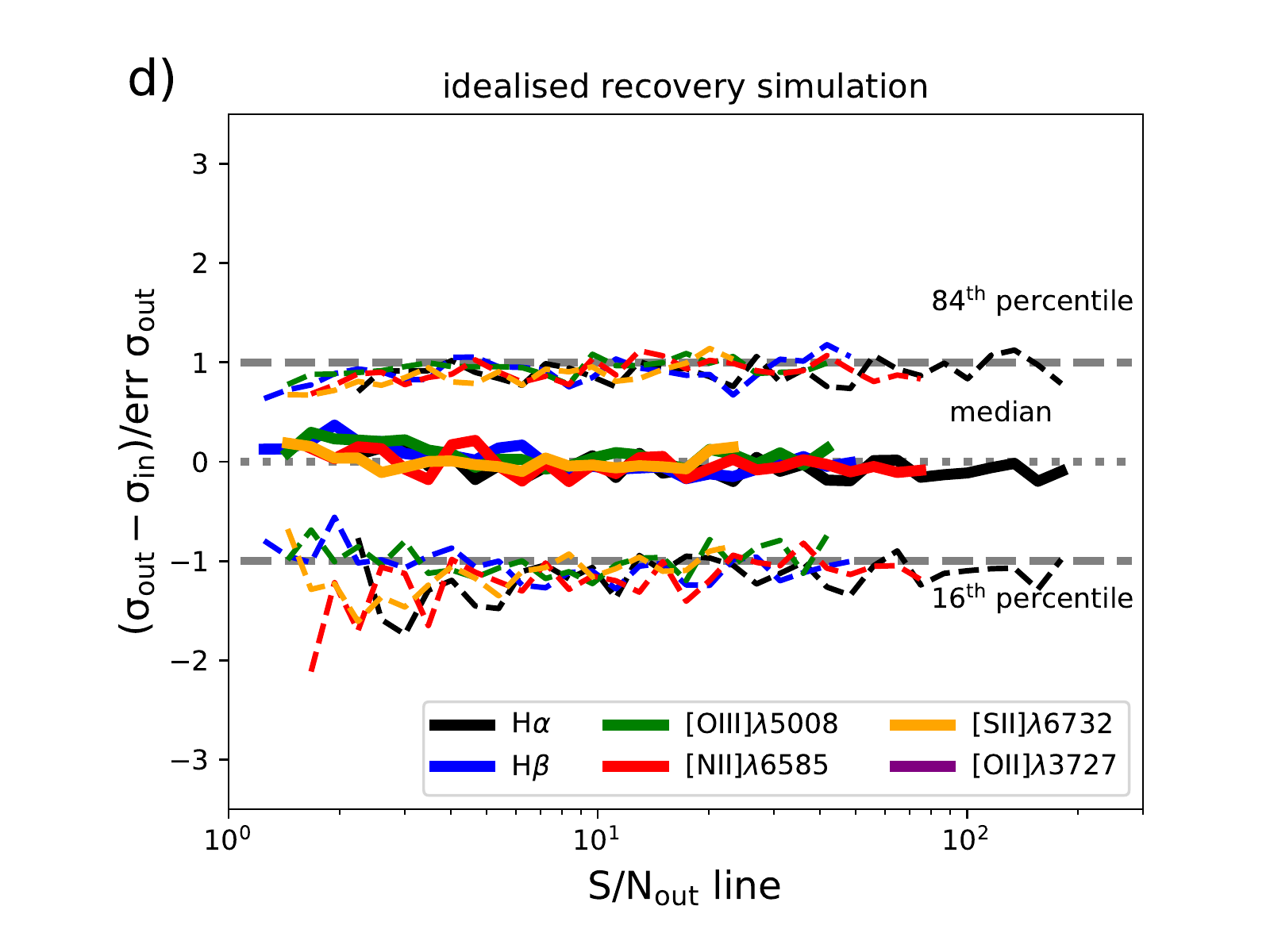}
    \caption{\textbf{a)} The median offset (in dex) between the input
    and output fluxes of a parameter-recovery simulation. Different
    strong emission lines are represented by different colors, as noted
    in the legend. The dashed gray lines correspond to the level of
    offset consistent with random errors. \textbf{b)}The median flux
    difference between the input and output flux, normalized by error in
    the output flux ($\rm (Flux_{out} - Flux_{in}) / err_{out}$), as a
    function of output signal-to-noise ratio (S/N=Flux/Error) of each
    line for an idealized simulation.  The 16$\rm ^{th}$ and 84$\rm
    ^{th}$ percentiles of the distribution as a function of S/N are
    shown as dashed lines. In the case of perfect recovery the median
    values should be zero and the 16$\rm ^{th}$ and 84$\rm ^{th}$
    percentiles should be $\pm$ 1.  \textbf{c)} Same as panel b) but
    showing the velocity normalized residuals versus the S/N for the
    H$\alpha$ line. Only H$\alpha$ is shown since the velocities of all
    lines are tied. The red lines represent fits to the 16$\rm ^{th}$
    and 84$\rm ^{th}$ percentiles of the distribution given by Equation
    \ref{correct_velocity} which could be used to correct the errors.
    \textbf{d)} Same as panel b) but for the velocity dispersion. 
Overall the results of these idealised simulations demonstrate that we are capable of recovering accurate fluxes down to low S/N and that the errors given by the MaNGA \DAP\ are realistic for flux and velocity dispersion. In equation \ref{correct_velocity}  we provide a correction formula to obtain the correct velocity errors at low S/N.}
    \label{sims}   
\end{figure*}

In the ideal case, the fits to the mock datacubes would recover the
input values of flux, velocity and velocity dispersion for all emission
lines with no bias and the (1$\sigma$) errors for these quantities
reported by the \DAP\ should be equal to the standard deviation of the
residuals between the output and input values. In other words, we expect
$\rm <q_{in} - q_{out}> = 0$ and $\rm std((q_{in} -
q_{out})/err_{out})=1$, where angle brackets denote averaging, std is
the standard deviation, $\rm q_{in}$ and $\rm q_{out}$ are input and
output values for a physical quantity and $\rm err_{out}$ is the
\DAP-provided error in $\rm q_{out}$. In this section we will refer to
$\rm (q_{in} - q_{out})/err_{out}$ as the normalized residual for
quantity q.

In Figure \ref{sims}a we plot the offset between input and output fluxes
(in dex) as a function of measured (output) S/N for six strong emission
lines ([\oii]$\lambda$3727, H$\beta$, [\oiii]$\lambda$5007, H$\alpha$,
[\nii]$\lambda$6584 and [\sii]$\lambda$6731). The plot demonstrates the
existence of a small positive bias in the recovered flux for S/N $<6$,
which then becomes a sizable decrease in the recovered flux for the
lowest S/N levels ($\sim$ 2-3). Overall, the ability of the code to
recover the input fluxes is better than 0.05 dex (12 \%) for S/N $>$6. 

In Figure \ref{sims}b,c,d we plot the normalized residuals for the flux,
velocity and velocity dispersion as a function of the output S/N. The
solid colored lines correspond to the median values of the normalized
residuals in logarithmic bins of S/N, while the dashed lines represent
the 16$\rm ^{th}$ and 84$\rm ^{th}$ percentiles. In all panels, in the
case of perfect recovery, the median lines would lie at zero, and,
assuming Gaussian errors, the 16$\rm ^{th}$ and 84$\rm ^{th}$
percentiles would follow horizontal lines at $\pm 1$. 

From Figure \ref{sims}b we observe that fluxes of all the emission lines
considered are recovered with negligible bias down to S/N $\sim$ 1.5.
More notably, the errors are also correctly estimated, since the 16$\rm
^{th}$ and 84$\rm ^{th}$ percentiles closely follow the $\pm1$ lines in
normalized residuals.  For S/N $<$ 1.5 the flux is systematically
underestimated. At these low S/N the distribution of normalized
residuals also deviates from a Gaussian, showing a long tail at low
normalized residuals. We note that, while different lines cover
different ranges in S/N, the behavior of different lines are remarkably
similar.

In Figure \ref{sims}c we show the normalized residuals for the
emission-line velocities as a function of H$\alpha$ S/N. Only the $\rm
H\alpha$ line is plotted in this panel, since all the emission lines are
fit with the same velocity. The figure shows that the input
emission-line velocities are recovered with no bias down to S/N $\sim$
1.5. However, the formal errors calculated by the \DAP\ are
underestimated for S/N $<$ 10. In particular, at S/N $\sim$ 2, the
output error is a factor of $\sim$ 3 lower than expected. At high S/N,
on the other hand, the output errors are consistent with the scatter in
the normalized residuals. The source of this underestimation likely lies
in the fact that the formal error provided by \pPXF\ for the gas fluxes
are computed, for computational efficiency, from the covariance matrix
of the gas emission templates alone. This implies that the uncertainties
currently ignore the covariance between the fluxes (which are a linear
parameters in the fit) and the gas kinematics (which are non-linear
parameters). Proper uncertainties could be computed via bootstrapping at
the expense of a significantly larger computation time, or by
re-computing the covariance matrix with respect to all variables at the
best fitting solution. 

In order to quantify this deviation we have fit
the observed relation with a simple functional form
\begin{equation}
\rm (v_{in} - v_{out})/err_{v out} = (0.8 \pm 0.1) + (0.49 \pm 0.07) \
log(S/N)^{-1}.
\label{correct_velocity}
\end{equation}
The resulting fit provides a very good representation of the data and is
shown in light red in Figure \ref{sims}b. This correction is not applied to DR15 \DAP\
output and needs to be taken into account by the user. We anticipate that users 
interested in fitting detailed kinematical models to the emission line velocity field may need to take this correction factor into account.

Figure \ref{sims}d shows the normalized residuals for the velocity
dispersions of the different emission lines considered. The
velocity-dispersion trends are similar to those observed for the flux,
and are indeed their likely cause, since the flux is positively
correlated with dispersion. We observe remarkably good agreement in both
the median and 16$\rm ^{th}$ and 84$\rm ^{th}$ percentiles values down
to S/N $\sim$ 1.5. Below that value the dispersion shows a larger tail
of negative normalized residuals.

Overall, idealized recovery simulations with no template mismatch
demonstrate that the values and errors of flux and velocity dispersion
can be recovered accurately with negligible bias down to S/N $\sim$ 1.5.
The velocities can also be recovered reliably down to low S/N, but their
associated errors appear to be underestimated for S/N $<$ 10.

\begin{figure*}
	\centering
	\includegraphics[width=0.62\textwidth, trim=0 0 0 0, clip]{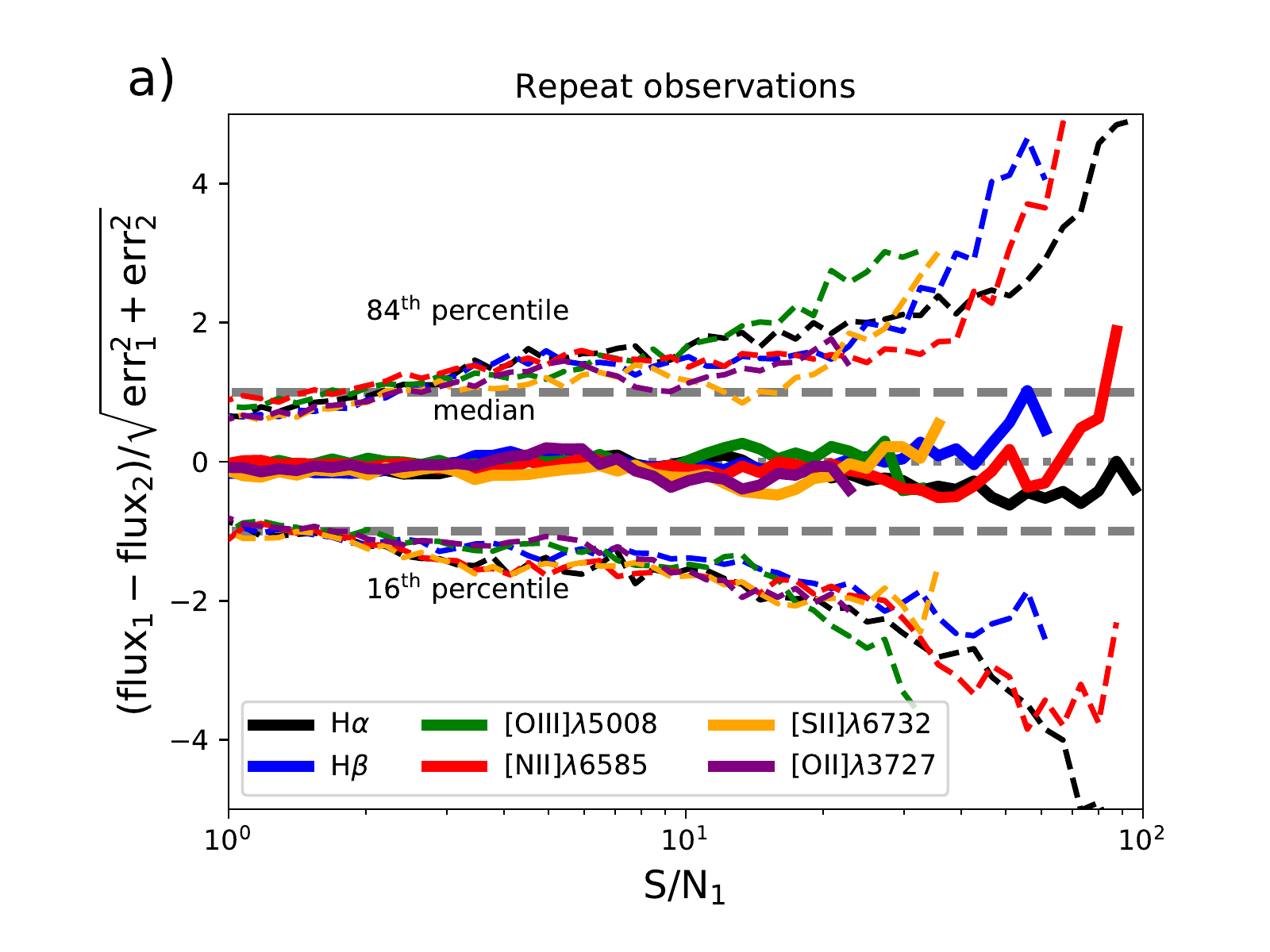}
	\includegraphics[width=0.42\textwidth, trim=0 0 0 0, clip]{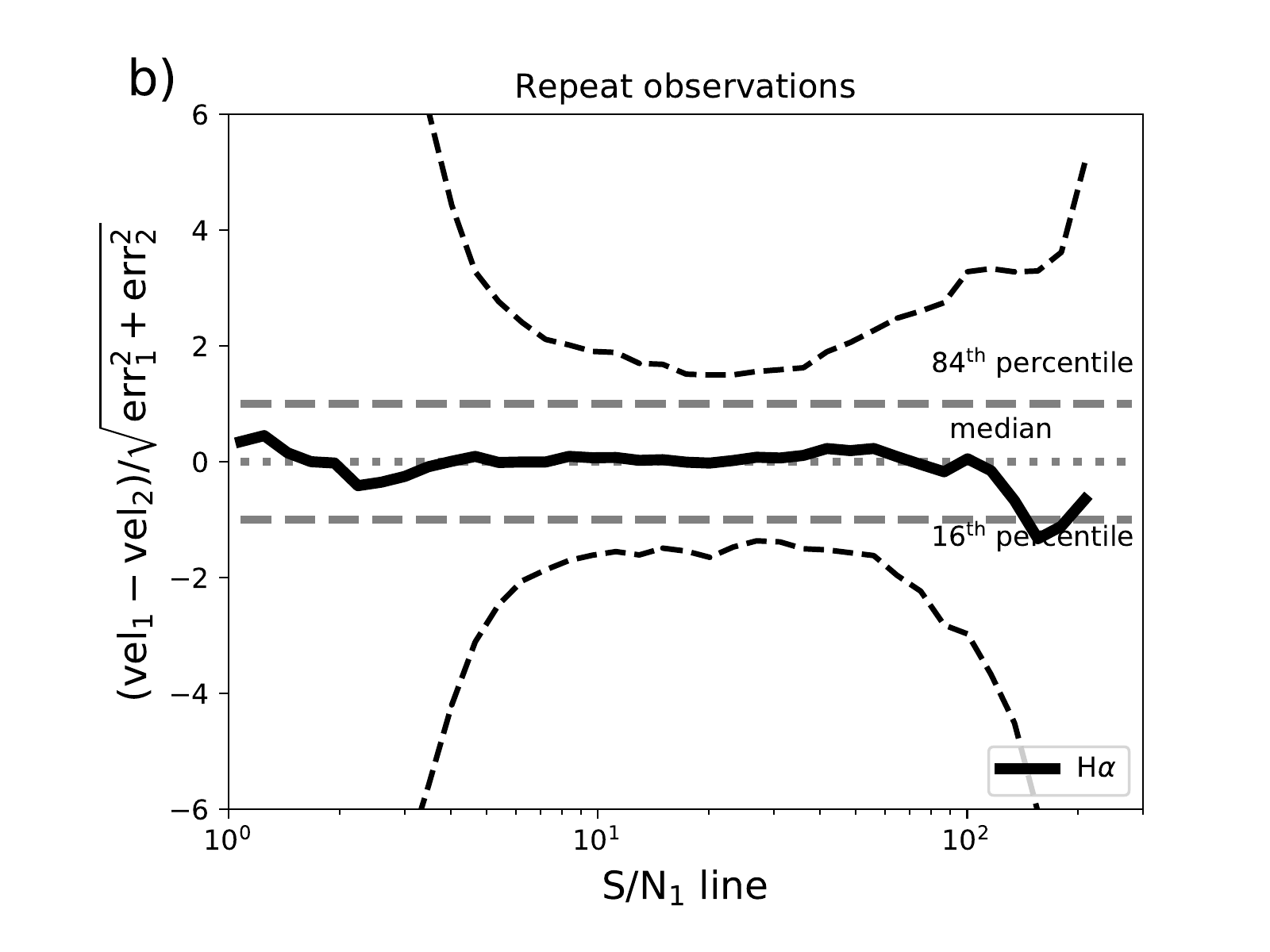}
	\includegraphics[width=0.42\textwidth, trim=0 0 0 0, clip]{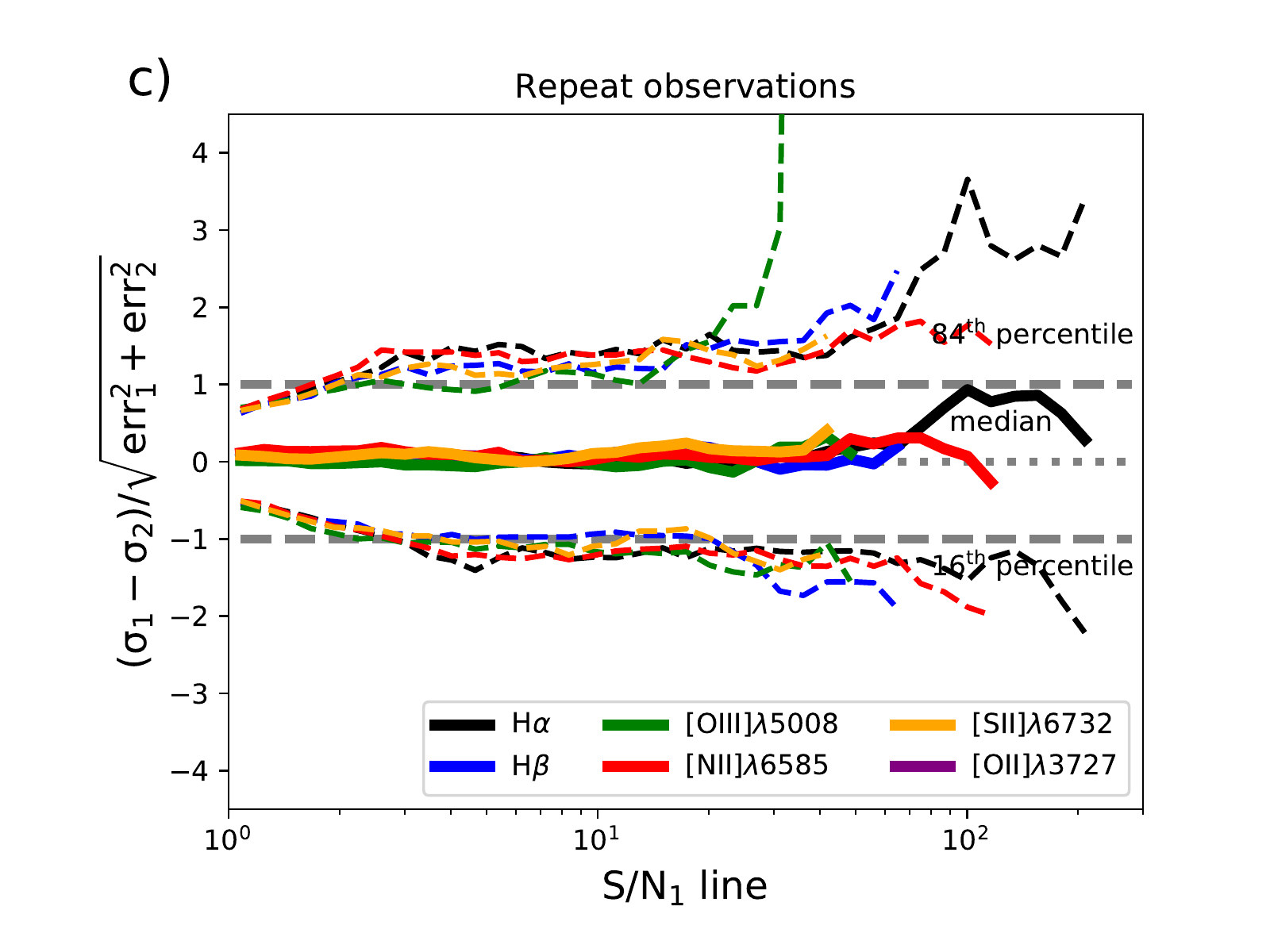}
	\includegraphics[width=0.42\textwidth, trim=0 0 0 0, clip]{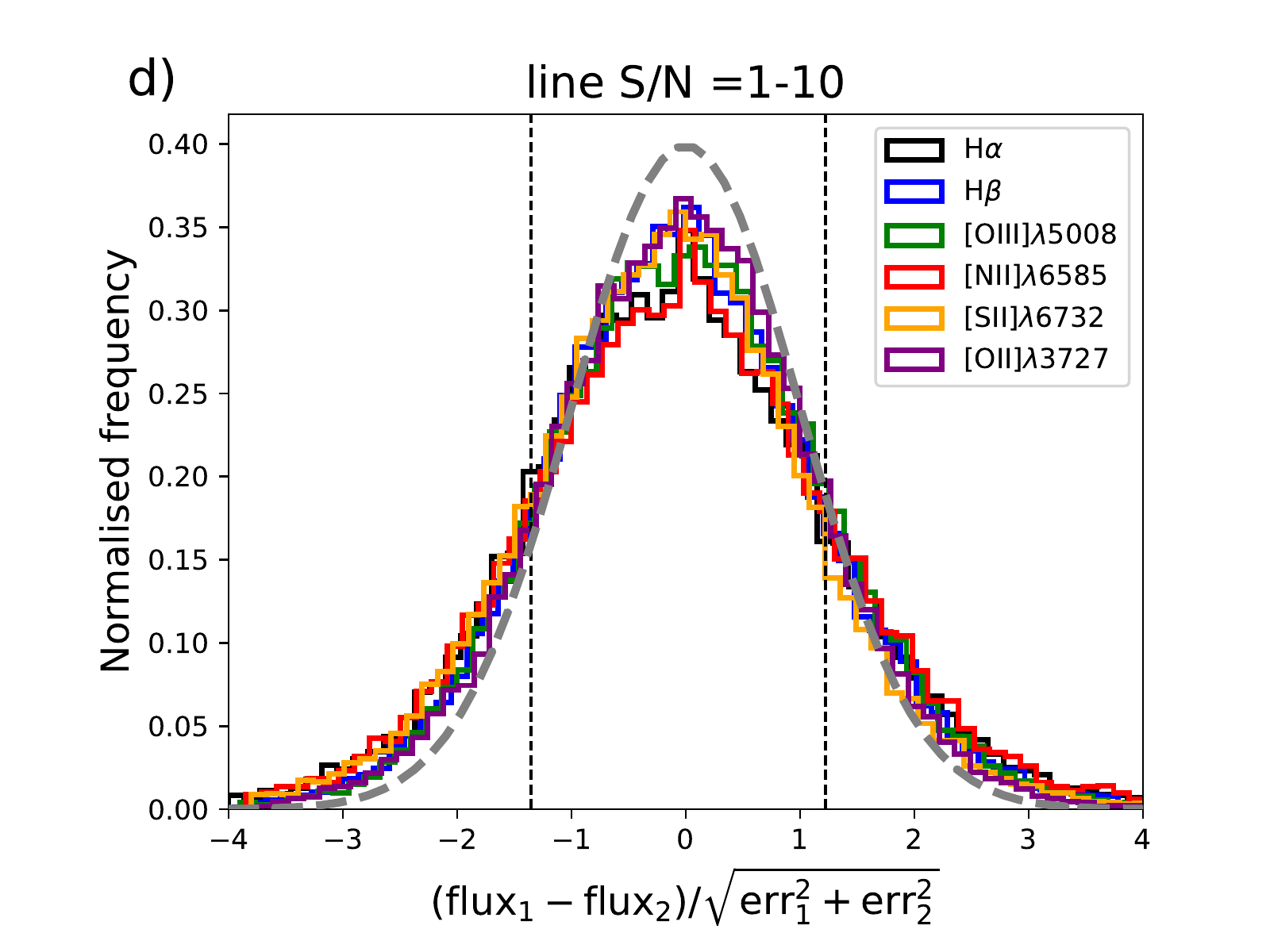}
	\includegraphics[width=0.42\textwidth, trim=0 0 0 0, clip]{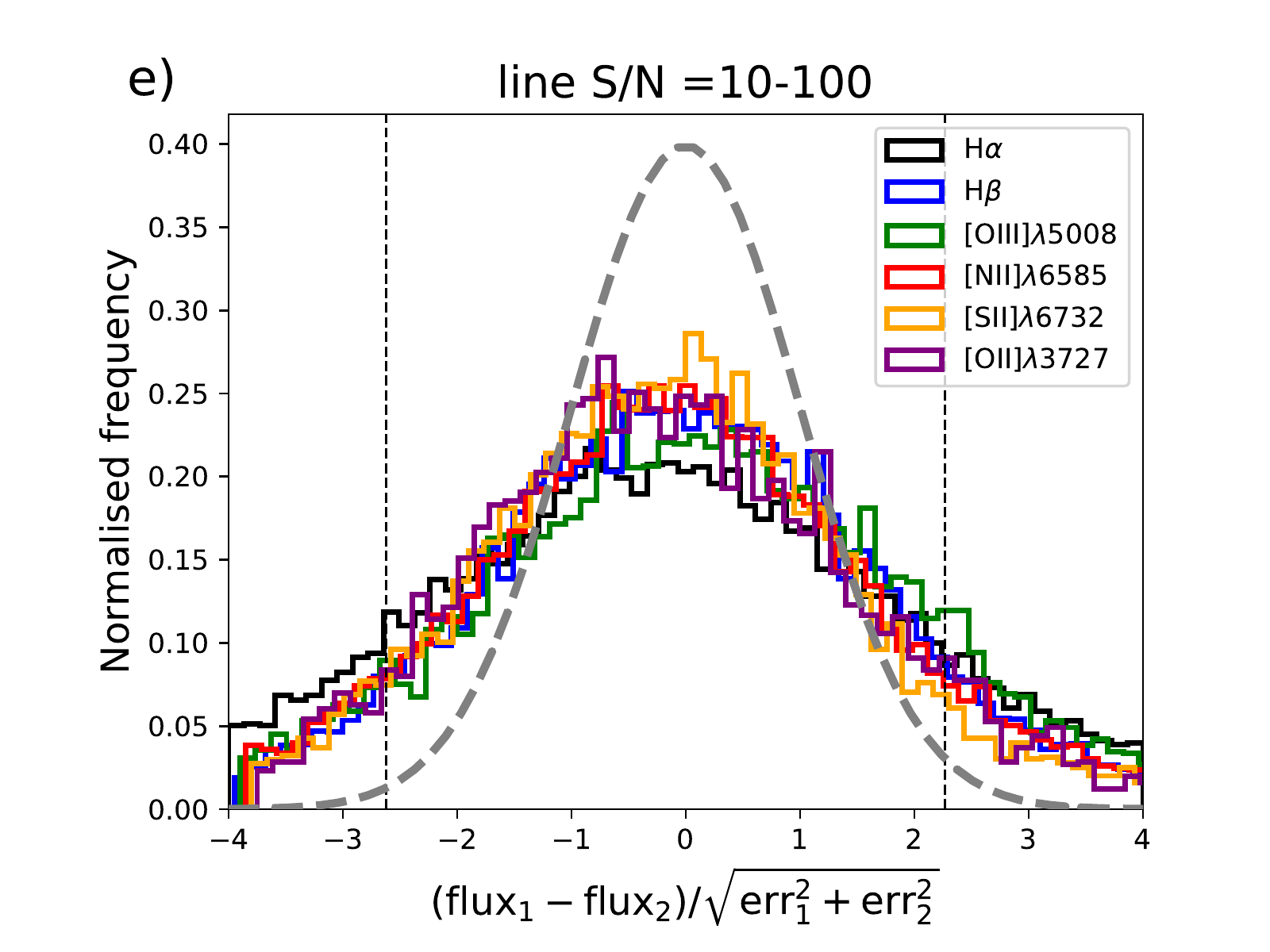}
    \caption{ \textbf{a)} The median flux difference between pairs of
    repeat observations normalized by root-mean-square error, i.e. $\rm
    (Flux_{1} - Flux_{1}) / (err_{1}^2 +err_2^2)^{1/2} $ as a function
    of signal-to-noise ratio (S/N=Flux/Error) of each line. Different
    strong emission lines are represented by different colors, as noted
    in the legend. The 16$\rm ^{th}$ and 84$\rm ^{th}$ percentiles of
    the distribution as a function of S/N are shown as dashed lines. If
    the output errors are correct, the 16$\rm ^{th}$ and 84$\rm ^{th}$
    percentiles of the normalized residual distribution should be $\pm$
    1.  \textbf{b)} Same as panel a) but showing the velocity normalized
    residuals versus the S/N for the H$\alpha$ line. Only H$\alpha$ is
    shown since the velocities of all lines are tied.  \textbf{c)} Same
    as panel a) but for the velocity dispersion.  \textbf{d)} The
    distribution of normalized flux residuals in the S/N range [1-10]
    for six strong emission lines, listed in the legend. The gray dashed
    line represents the expected distribution assuming Gaussian errors.
    The black vertical dashed lines are the 16$\rm ^{th}$ and 84$\rm
    ^{th}$ percentiles of the H$\alpha$ normalized residual
    distribution. If the errors are correct, these values should lie at
    $\pm$ 1.  \textbf{e)} Same as b), but for the S/N range [10-100].
    Note that in this regime the normalized residuals distributions are broader and the
    errors appear to be underestimated. }
	\label{repeat}   
\end{figure*}

\subsection{Error statistics from repeat observations}
\label{sec3.3}

In this section, we further analyze the error statistics for the
emission-line measurements provided by the \DAP\ by using repeat
observations. In DR15, 56 galaxies have been observed more than once,
mainly for the purpose of testing random and systematic errors (\citealt{Westfall2019_arxiv}, Table 1).\footnote{Forty-three galaxies have
been observed twice, 12 have been observed three times and one has been
observed four times, for a total of 70 pairs of galaxies with repeat
observations.}. After processing these galaxies through the \DAP, their
\MAPS\ files were transposed into the same world coordinate system, in
order to account for small shifts in the IFU bundle positions and
orientations between observations. The world coordinate system is
derived by the MaNGA {\tt DRP} by matching the MaNGA cubes to
pre-existing SDSS photometry in the advanced astrometry module
\citep[][Section 8]{Law2016}, and therefore takes into account small
shifts and rotations of the IFU fiber bundles. In comparing repeat
observations we do not, however, take into account possible changes in
the seeing conditions.

Similar to the procedure adopted for analyzing the recovery
simulations in the previous section, we calculate the normalized
residuals as a function of S/N. Since for repeat observations we do not
know the true value of any physical quantity, we define the normalized
residual as $\rm (q_1 - q_2) / (err_1^2 +err_2^2)^{1/2}$,  where 1 and 2
refer to a pair of repeat galaxies and err is the error in quantity q.
Considering all repeat galaxies, we obtain a sample of $\sim$ 5$\cdot
10^4$ pairs of spaxels with two independent measurements with S/N > 1
for H$\alpha$.

In Figure \ref{repeat}a we show the normalized flux residual as a
function of S/N of the first galaxy in the pair. Following the same
graphical conventions as in Figure \ref{sims}, the solid colored lines
represent the median while the dashed colored lines represent the 16$\rm
^{th}$ and 84$\rm ^{th}$ percentiles as a function of S/N. The median
residual is found to be close to zero. The  16$\rm ^{th}$ and 84$\rm
^{th}$ percentiles, on the other hand, are found to be close to $\pm 1$
at S/N $\sim$ 2 and show a systematic deviation towards larger values at
higher S/N. This deviation is particularly evident for S/N $>$ 10.

In Figures \ref{repeat}b and \ref{repeat}c we show the normalized
residuals as a function of S/N for the velocity and velocity dispersion.
For velocity, the errors are underestimated at all S/N, with the worst
discrepancy at low S/N, similarly to what was found in our study of
idealized simulations in the last section. Different from what was
seen in the previous section, the errors also diverge from expectations
at high S/N (S/N > 20-30), while they appear to be underestimated by a
factor less than 2 in the range S/N = [6, 50]. The velocity dispersion
appears to be much better behaved, with no evidence for large error
underestimation until S/N$>$100. Interestingly, for $\rm S/N =[1,2]$ the
errors appear to be overestimated.

Figure \ref{repeat}d shows the distribution of normalized residuals for
six strong lines (see legend) in the S/N range [1-10]. The gray dashed
line shows a normalized Gaussian of unit standard deviation, which
represents the theoretical expectation in case of ideal error
measurements. In Figure \ref{repeat}d we show as vertical black dashed
lines the  16$\rm ^{th}$ and 84$\rm ^{th}$ percentiles of the observed
distribution for H$\alpha$ (1.22 and $-$1.35 respectively). We note
that, although the data presents slightly non-Gaussian tails, it is
well-fit to first order by a Gaussian with standard deviation 1.25 (fit
not shown).

Figure \ref{repeat}e is the same as Figure \ref{repeat}d, but represents
the S/N range [10-100]. As already evident in Figure \ref{repeat}a, at
this S/N level the errors are either underestimated by a factor of 2-3,
or some other systematic enters in the repeat observations comparison. 

Since this large error underestimation is not seen in the idealized
recovery simulation we consider possible systematic effects which could
cause this. First, as already seen in Section \ref{sec3.1}, at S/N >
20-30, our Gaussian model may be insufficient to accurately fit the line
profiles in real data, leading to higher normalized residuals and,
possibly, underestimated errors. Secondly, regions of bright line
emission tend to be clumpy, and the measured fluxes are therefore
particularly sensitive to differences in PSF. In this case the increased
error in flux is due to intrinsic scatter in the amplitude, and not to
larger errors in the recovered velocity dispersion, which would be in
agreement with the findings from Figure \ref{repeat}c. The same effect
would be caused by small astrometric misalignments between repeat
observations. 

We gained some insight into these issues by visually inspecting
difference and normalized residuals maps for different pairs of repeat
observations. This exercise clearly revealed that the largest normalized
residuals are indeed associated with bright and clumpy line emission. We
have therefore performed a simple test to quantify the effect of
astrometric offsets and PSF differences. One of the galaxies showing the
largest differences in normalized residuals was selected as an example.
For this galaxy we considered the output of the MaNGA astrometry module,
which matches the MaNGA IFU data to the underlying SDSS photometry in
order to correct for small deviations of the rotation and centroid
position of the MaNGA IFU ferrules in a given exposure due to the
mechanical tolerance of the ferrule and rotational clocking pin holes.
We artificially added random error to the best-fit astrometric solution,
consistent with the uncertainty calculated by the astrometry module
(typically about 0.25$^{\circ}$ and 0.1'' for the rotational and
translational components respectively). A datacube was then produced,
following the usual MaNGA reduction recipies, and fit using the \DAP. We
compared the output produced by this datacube with additional
astrometric error to reference datacube generated for DR15. At low S/N
the dispersion in line fluxes between the two datacubes is negligible,
but it flares at high S/N in a fashion consistent with that observed in
Figure \ref{repeat}a. In particular we find that astrometric errors
consistent with those expected by registering MaNGA data to SDSS
photometry are sufficient to explain the observed increase in the error
budget in repeat observations. 

\subsection{Summary and recommendations with regards to errors}

In summary, recovery simulations demonstrate that the errors for flux
and velocity dispersion of different emission lines behave in a
statistically correct fashion down to S/N $\sim 1.5 $. Errors in the
velocity are underestimated for S/N < 10, and the source of this
discrepancy is not know at the time of writing.  Equation
\ref{correct_velocity} quantifies this underestimation and can be
utilized to rescale the errors based on the outcome of the recovery
simulation. In DR15 we leave it to the user to apply this correction if deemed necessary to their science goal.

These trends are largely confirmed by the analysis of repeat
observations. Repeat observations, however, also show underestimation of
the errors in the high S/N regime. We have demonstrated that this trend
can be entirely explained by small astrometric errors in individual
exposures, which are consistent with the uncertainties derived by the
MaNGA DRP astrometric registration routine.  In light of this
discussion, we leave it up to the user to consider whether adding this
extra error contribution is advisable for their specific science goal.

\section{Systematic errors from the modeling of the continuum }
\label{sec4}

In the section we address the systematics on emission-line properties
that arise from the modeling of the continuum. We present \mileshc, the
stellar library used to fit the MaNGA data in DR15, and discuss the
differences in the recovered emission-line properties obtained using
several different SSPs.

\subsection{Stellar and SSP template libraries}
\label{sec4.1}

In the following section we briefly outline the characteristics of the
spectral libraries that we will discuss and compare in this paper.

\paragraph{The hierarchically clustered MILES templates (\mileshc)}

As discussed in Section 5 of \cite{Westfall2019_arxiv}, we have
applied a hierarchical clustering algorithm to the MILES stellar library
spectra, which consists of 985 stars covering the wavelength range 3525
- 7500~\AA\ \citep{Sanchez-Blazquez2006, Falcon-Barroso2011}. The
clustering algorithm subdivides the stars in the MILES library into a
number of groups that are defined to be maximally different from each
other. Forty-nine such groups are generated and a composite spectrum is
obtained for each group as the average of the spectra of the
contributing stars. These 49 spectra were visually inspected and 7 of
them were removed due to artifacts and/or the presence of emission lines
(in flaring late-type stars), leading to a total of 42 stellar
templates. The resolution of the MILES library has been independently
derived by \cite{Beifiori2011} and \citet{Falcon-Barroso2011} and is
2.54~\AA\ (FWHM). This library is used in generating all the \DAP\ DR15
data products.

\paragraph{The Maraston 2011 SSP models based on MILES (\textit{M11-MILES})}

These SSP are generated using the MILES stellar library by the Maraston
stellar-population-synthesis code
\citep{Maraston2011}.\footnote{\url{www.maraston.eu/M11}} One hundred
ten models are used with ages ranging from 6.5 Myr to 15 Gyr, 3
metallicities (Z$=$0.01, 0.02, 0.04) and a Salpeter IMF. The spectral
resolution is the same as that of the MILES library (2.54~\AA\ FWHM).

\paragraph{The Vazdekis MIUSCAT SSP models (\textit{MIUSCAT})} 

This is a set of 72 SSP models generated according to
\cite{Vazdekis2012}\footnote{
\url{http://miles.iac.es/pages/webtools/tune-ssp-models.php}}
with a Salpeter IMF (unimodal IMF with slope$=$1.3) and a set of 24 ages
(0.0631, 0.0794, 0.1000, 0.1259, 0.1585, 0.1995, 0.2512,  0.3162,
0.3981, 0.5012, 0.6310, 0.7943, 1.0000, 1.2589, 1.5849, 1.9953, 2.5119,
3.1623, 3.9811, 5.0119, 6.3096,7.9433, 10.0000, 12.5893 Gyr) and 3
metallicities ([M/H] $=$ -0.4, 0.0, 0.22). MIUSCAT extends the
wavelength range of MILES to cover the full range 3465 - 9469~\AA\
making use of the near-IR CaT library of \cite{Cenarro2001}. Stellar
spectra from the Indo-US stellar library are used to fill in the gap
left between the MILES and and  CaT spectral ranges and also to extend
towards the blue and red the wavelength coverage of the MILES and CaT
libraries respectively. The spectral resolution is the same as the MILES
library, as the higher-resolution CaT and Indo-US libraries are
convolved to the MILES spectral resolution.  The MaNGA VAC generated by
the {\tt Pipe3D} team also uses a subset of MIUSCAT templates for
spectral fitting, although the exact set of templates differs from the
ones described above.

\paragraph{The Bruzual and Charlot SSP models based on STELIB (\textit{BC03})}

This is the set of 40 \cite{Bruzual2003} models used in the MPA-JHU
catalog for the SDSS
DR4.\footnote{\url{https://www.mpa.mpa-garching.mpg.de/SDSS/DR4/}} The
SSP templates cover a range of 10 ages (0.005, 0.025, 0.101, 0.286,
0.640, 0.904, 1.434, 2.500, 5.000,	10.000 Gyr) and 4 metallicities (Z
$=$ 0.008, 0.004, 0.02, 0.05) using a Chabrier IMF. The SSP models are
based on the STELIB stellar library \citep{LeBorgne2003a} and have
nominal spectral resolution of 2.3~\AA. These SSP models are different
from the BC03 SSP based on MILES used to produce the MPA-JHU value-added
catalogue for later data releases, such as the latest DR8
version.\footnote{\url{www.sdss.org/dr12/spectro/galaxy_mpajhu}}

\subsection{Choosing between stars and SSPs}
\label{sec4.2}

\begin{figure*}[htb]
	\includegraphics[width=\textwidth, trim=0 0 0 0, clip]{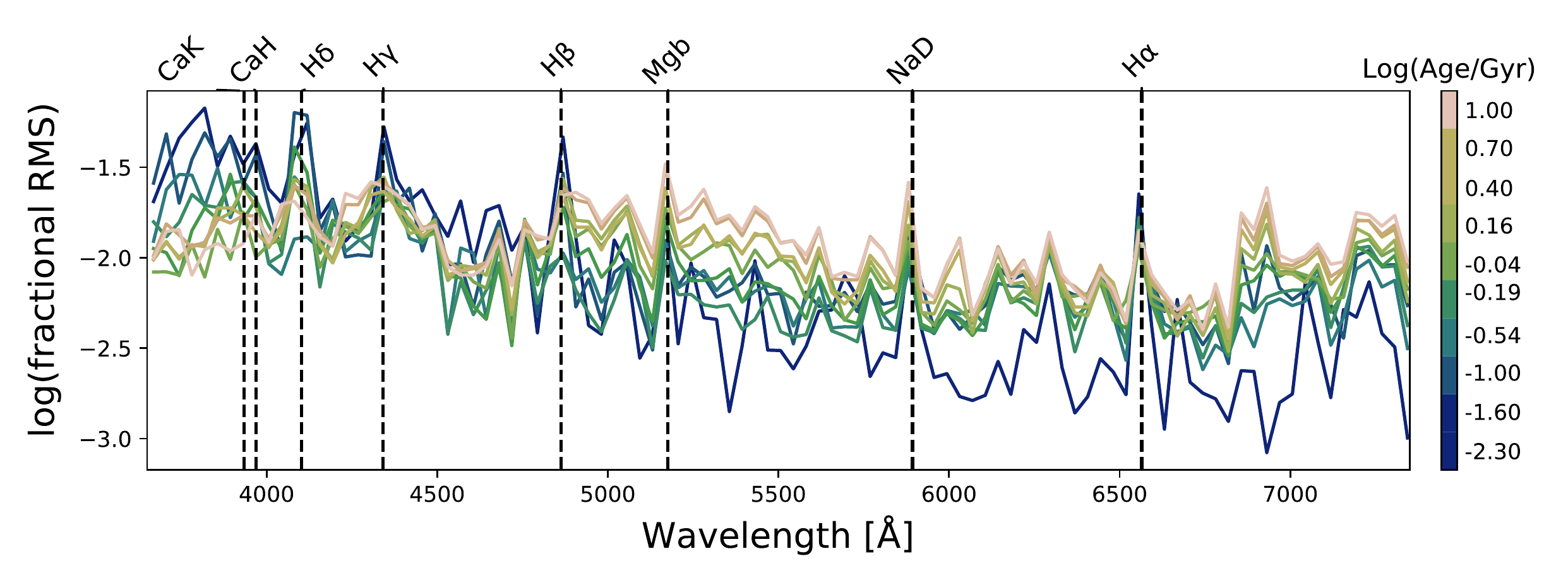}
    \caption{The fractional RMS of the residuals obtained when fitting BC03 SSP
    models of different ages (and solar metallicity) with \mileshc\
    stellar templates (and no polynomials). The residuals are computed in 40~\AA\ bins
    spanning 3650 -- 7400~\AA\ and the ages of the templates considered as given in the colourbar. Prominent metal absorption lines (e.g. Mgb and NaD) lead to localised increases in the RMS especially for the older SSPs.
    An increase in RMS is also evident at the positions of Balmer series lines, especially at young ages.}
	\label{fit_SSP_with_MILESHC}
\end{figure*}

\begin{figure*}[htb]
	\centering
	\includegraphics[width=0.24\textwidth, trim=10 20 20 20, clip]{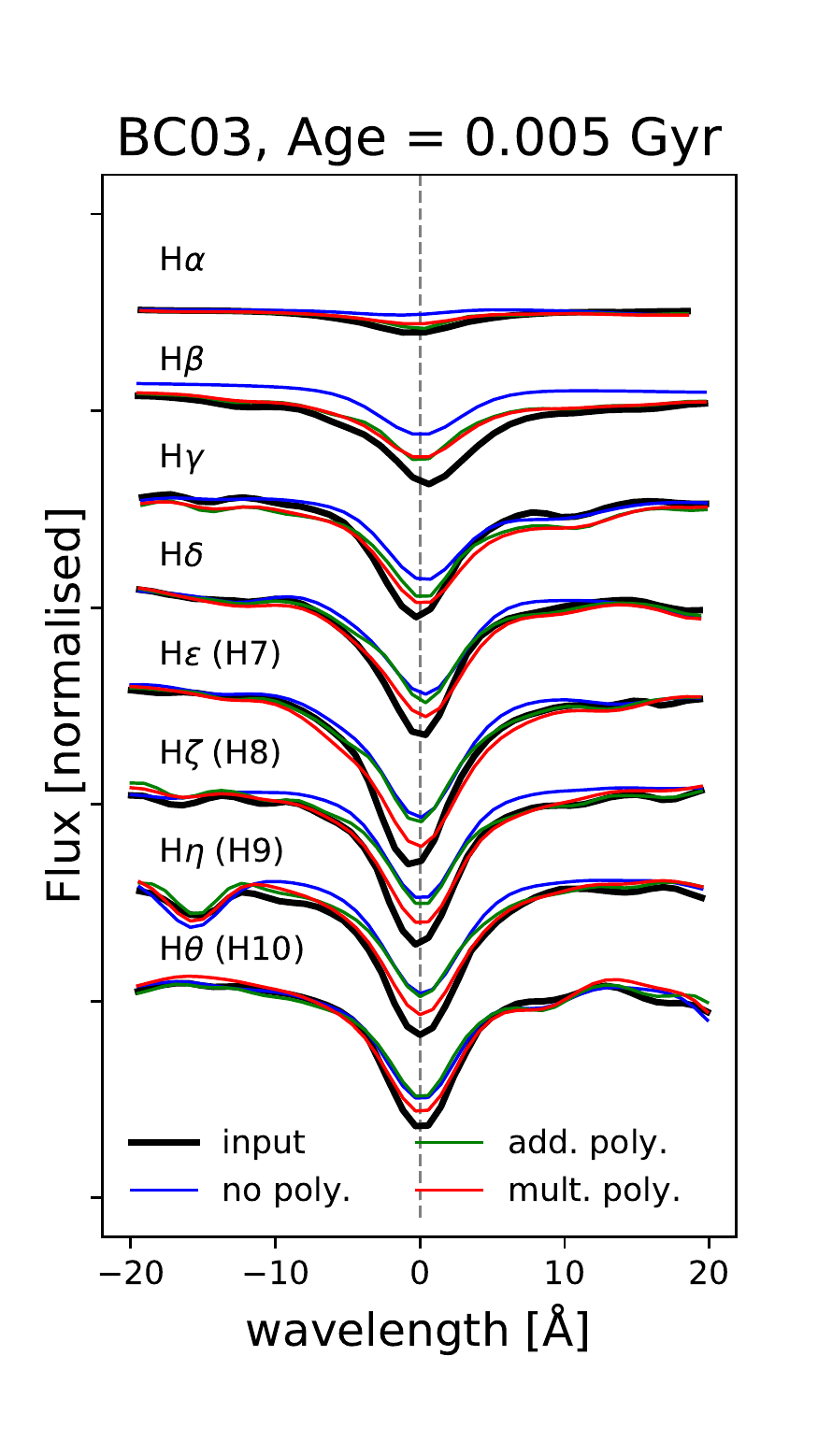}
	\includegraphics[width=0.24\textwidth, trim=10 20 20 20, clip]{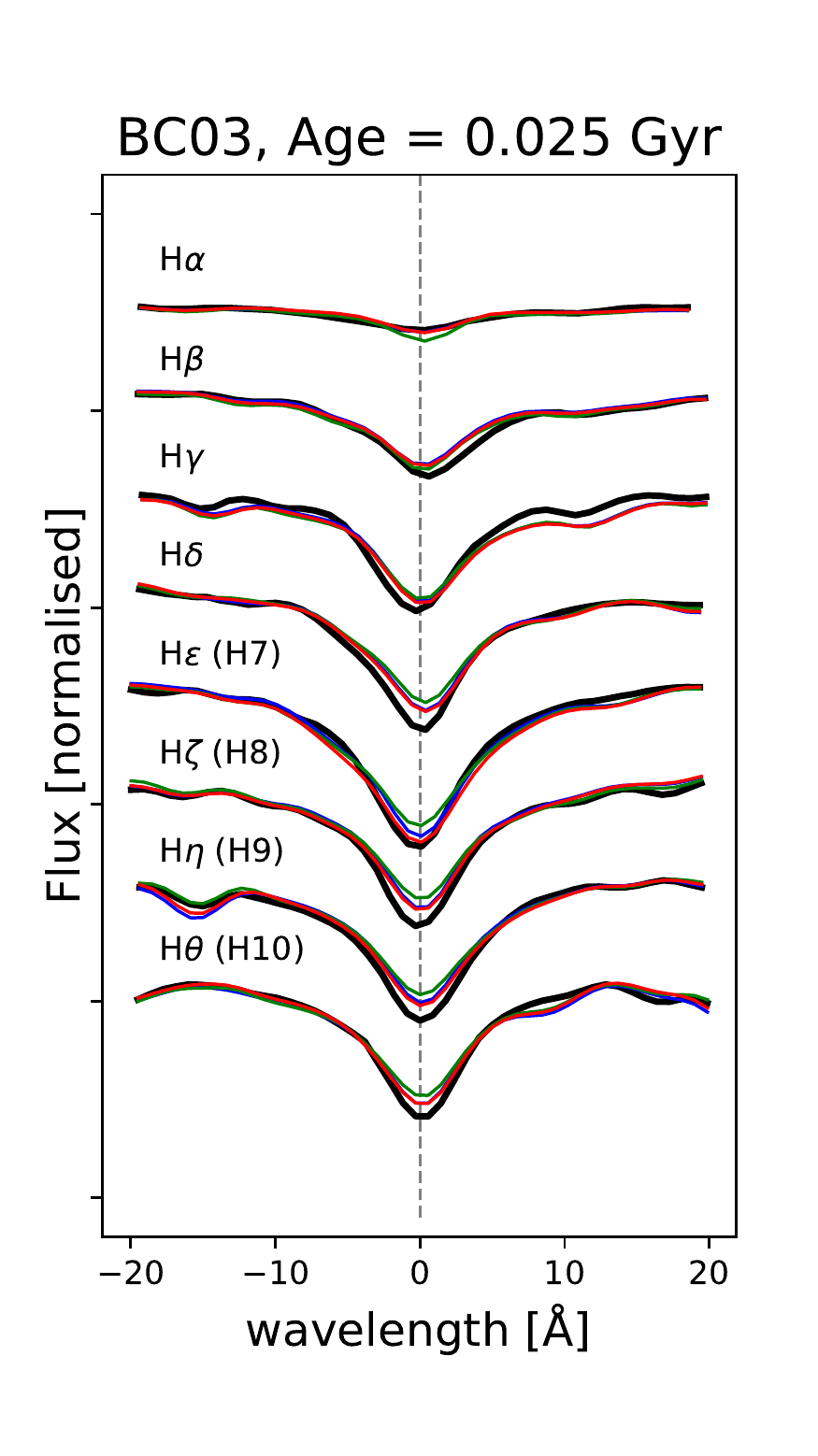}
	\includegraphics[width=0.24\textwidth, trim=10 20 20 20, clip]{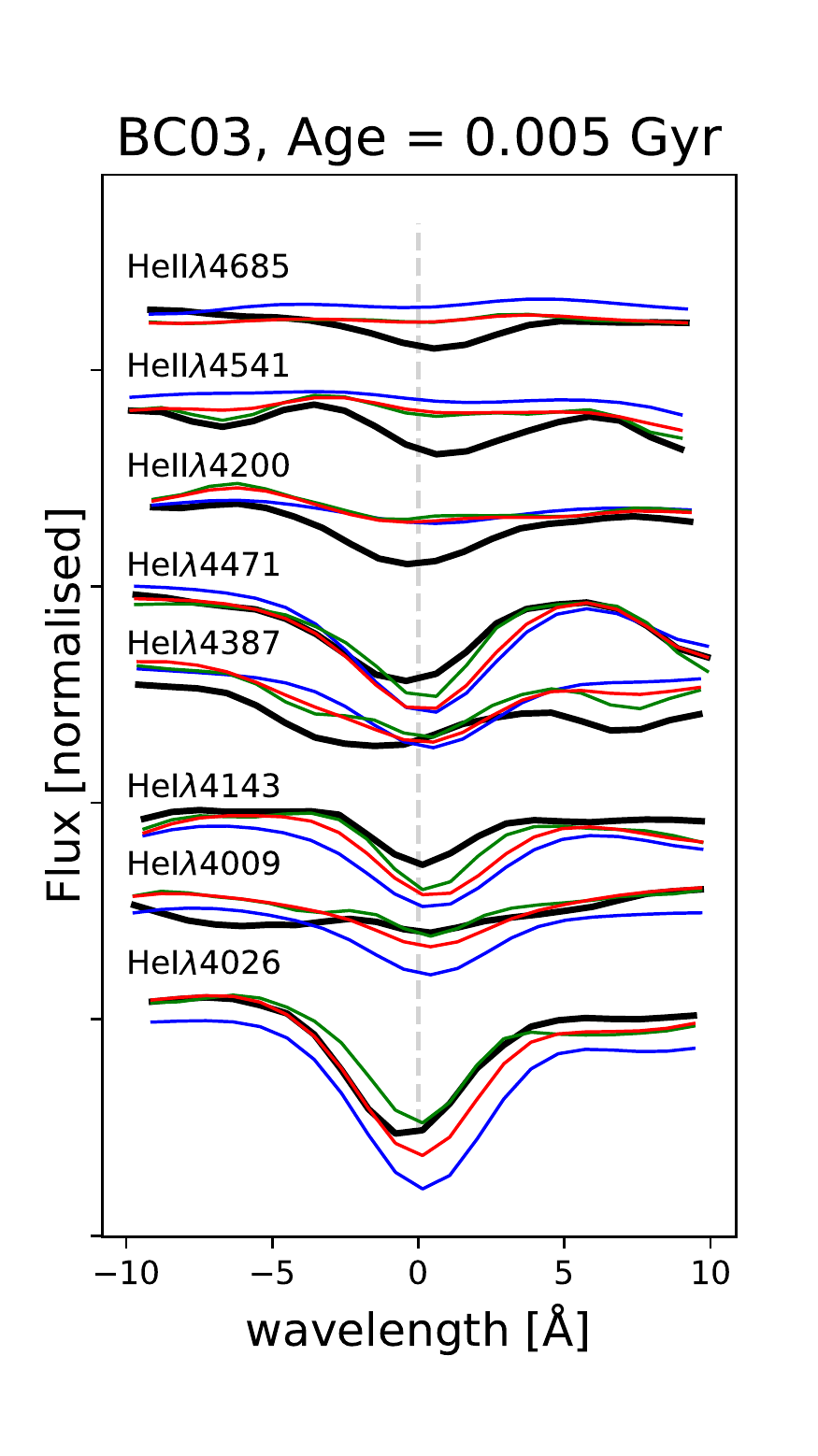}
	\includegraphics[width=0.24\textwidth, trim=10 20 20 20, clip]{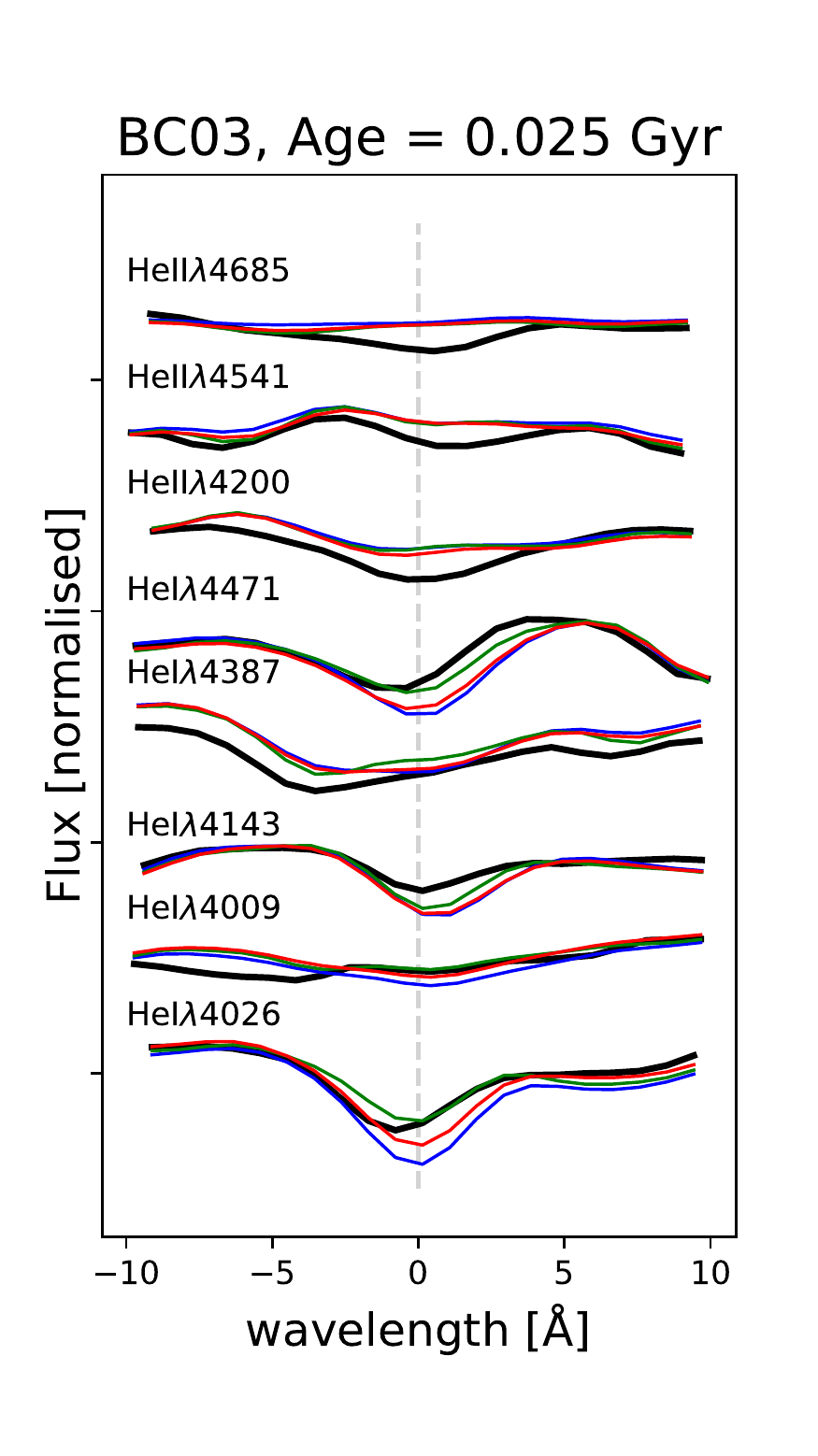}
    \caption{Zoom-in around the Hydrogen Balmer (left) and Helium
    (right) lines. The black spectrum corresponds to the input SSP (from
    BC03) being fit using the \mileshc\ library and the colored lines
    represent the best fits (blue: with no polynomials, green: with
    8$\rm ^{th}$-order additive polynomials, red: with 8$\rm
    ^{th}$-order multiplicative polynomials). SSPs of two different ages
    (5 and 25 Myr) are shown.}
	\label{Balmer_lines}   
\end{figure*}

In this section we investigate the consistency between the \mileshc\
stellar library used in DR15 and the SSP models described in the
subsection above. SSP models prescribe parameterized combinations or
interpolations between elements of stellar libraries based on physical
models of stellar interiors, atmospheres, and their evolution
(isochrones). To the extent that these physical models are incomplete or
simply inaccurate, one may worry about whether SSP and stellar spectra
are consistent in both overall shape and in the details of their
absorption features. SSP models based on observed stellar libraries
mitigate, e.g., incomplete line-lists in synthetic spectra, but suffer
problems due to library incompleteness and stellar misclassification.
Library incompleteness is a general concern for all continuum-fitting
methods, and it is the specific concern motivating the work in this
section. 

In particular, the lack of hot (O-type) stars in the MILES library
potentially undermines our ability to fit young stellar populations with
\mileshc. The hierarchical clustering performed on the spectra may
further dilute the blue continua of the few B-stars present in MILES.
While O-star spectra are largely featureless, their very blue continua
cannot be reproduced by a linear sum of other stellar spectra. A common
solution to this problem, within the \pPXF\ framework, is to allow the
inclusion of additive and/or multiplicative polynomials in the fit.
These polynomials are generally of low order to avoid any effect of the
polynomial on the spectral features of individual absorption lines. 

However, hot stars do have some critical features, notably in their
hydrogen and helium lines that have distinct and systematic changes with
temperature. These changes in stellar absorption lines include both the
equivalent width, the core width (due to rotation and winds), and the
relative strength and shape of the wings (due to the Stark effect).
None of these are modified by multiplicative polynomials, and only the
equivalent width (not the shape) can be modulated by additive polynomials.
Further, because of the different and non-linear temperature dependence
of these features it is not possible to accurately simulate spectra
containing O-stars with a library that does not include these stars. 
For a library with such a deficiency, we would expect to see Balmer lines that are too narrow
and a deficiency of He absorption. Since this will
lead to systematics in the continuum model at the location of key
emission lines, the amplitude of such systematics is important to
assess.

We therefore test the ability of the \mileshc\ library to reproduce
different stellar populations by fitting the BC03 SSP models of
different ages (and solar metallicity) with \mileshc\ spectra. The fit
has been performed in three ways: (1) with no polynomials; (2) with an
8th-order additive polynomial; and (3) with an 8th-order multiplicative
polynomial. Aside from the polynomial type, we perform this fit in the
same way as the first (i.e., the stellar kinematics) fitting stage of
the \DAP. In each case, we compute the residual between the best-fit
model and the input SSP, and the resulting RMS over the wavelength range
3700-7400~\AA. 

The fractional RMS (i.e. the standard deviation of output - input/ input) calculated over $\sim 40$~\AA\ windows is shown in
Figure \ref{fit_SSP_with_MILESHC} for BC03 templates of different ages fitted with no polynomials.

The figure demonstrates the overall
shape of the BC03 SSPs are well-fit by \mileshc, with median residual RMS values of $10^{-2.0}$. The largest RMS values are seen both at the blue end of the spectrum at very young ages and in very localised wavelength regions, generally corresponding to notable absorption lines. Balmer series lines are particularly problematic and increasingly so at young ages. Metal lines (such as Mgb and NaD), on the other hand, are fit worse at older ages. The inclusion of polynomials does not lead to an overall improvement of the fit quality, although it does have an effect on the fit around the Balmer and helium lines. 

Figure \ref{Balmer_lines} compares the fits of the \mileshc\
library to the BC03 SSPs for ages of 5 and 25 Myr with and without
polynomials. This figure is worth careful scrutiny. Inspection reveals \mileshc\ under-predicts the hydrogen line-depths,
increasingly for lower-order lines. This is mostly ameliorated by either
additive or multiplicative polynomials, which do a good job at matching
the wings but fail to match the core.  The same relative statements are
true for the 25 Myr SSP, but the amplitudes of the differences are
decreased, i.e., the \mileshc\ fit is substantially better on its own
without polynomials. In contrast, the \mileshc\ fits over-predict
\ion{He}{1} and under predict \ion{He}{2} lines for BC03 for both ages.
Polynomials do little to help remedy the mismatch in equivalent width
and often degrade the quality of the fit. 


We can interpret the over-predicting of the \ion{He}{1} and the under-predicting of 
\ion{He}{2} lines as due to the lack of very hot O stars in \mileshc. B stars are 
the hottest stars in the library, and they do not have \ion{He}{2} lines. If \mileshc\
lacks templates with strong \ion{He}{2} lines, then the fit will use more B stars to
compensate for the spectral shape and and up over-fitting \ion{He}{1}, while still not
producing the  \ion{He}{2} features.

A detailed accounting of the stellar templates (and their
weights) that go into the specific SSPs would be one way to make
progress on this question, but this critical `deconstruction' of SSPs is
beyond the scope of this paper.

We conclude that even with additive and/or multiplicative polynomials
\mileshc\ is likely to have small systematic residuals in the cores of
the hydrogen lines that lead to emission-line overestimates at the very
youngest ages. The systematics for helium lines are more significant and
varied, and in some cases are minimized without including polynomials. 
Polynomials do not significantly improve the overall match of the spectral between SSPs and stellar templates, which is excellent except for the youngest ages for $\rm \lambda < 4000 \AA$.

\subsection{The effects of the continuum model on line fluxes}
\label{sec4.3}

\begin{figure*}
	\centering
	\includegraphics[width=0.8\textwidth, trim=40 20 40 20, clip]{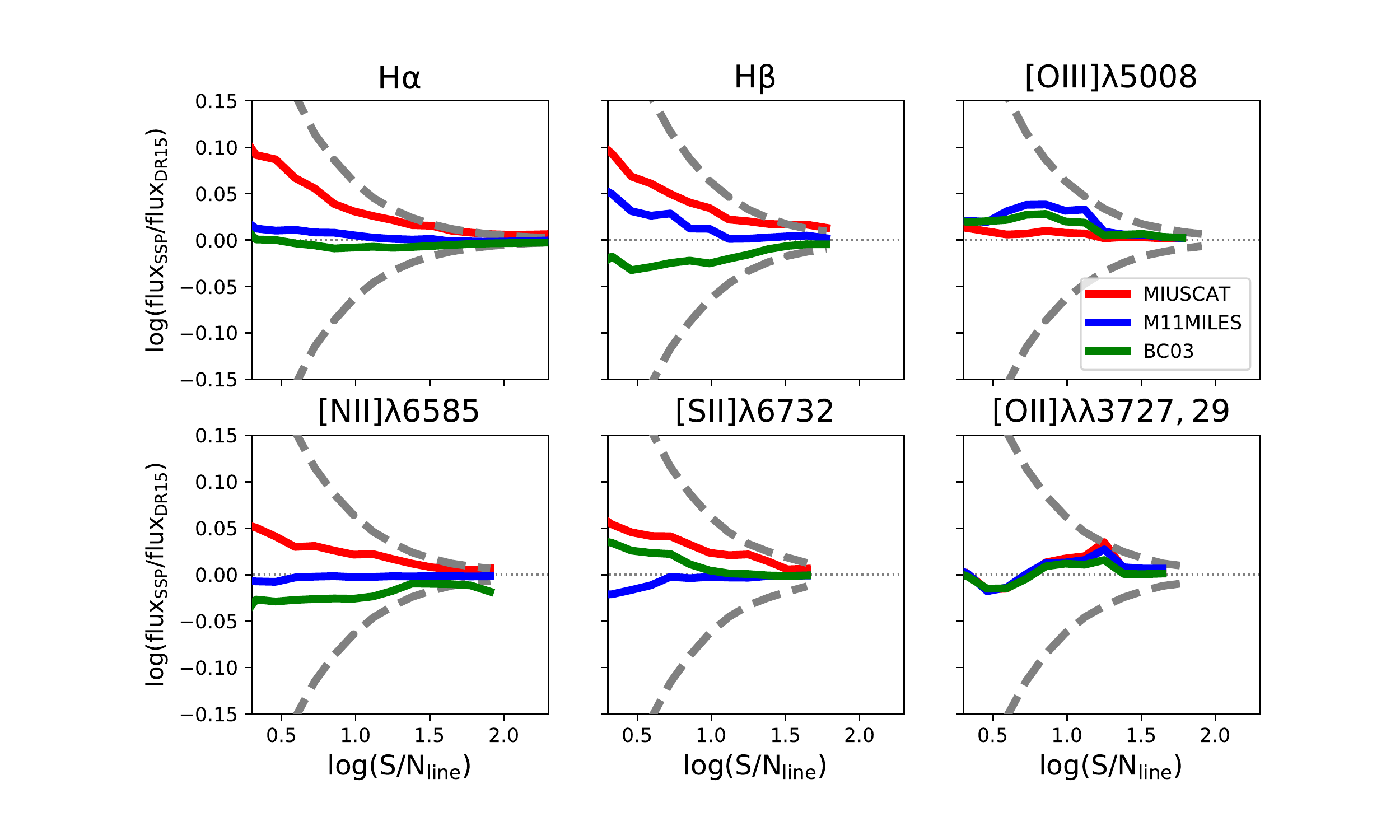}
    \caption{The ratio in the emission-line flux (in dex) obtained
    when modeling the continuum with various SSP templates (MIUSCAT: red,
    M11-MILES: blue, BC03: green) with respect to the fluxes obtained in
    DR15 (which uses \mileshc\ templates) as a function of line S/N.
    The dashed gray lines correspond to the level where the flux
    discrepancy is equal to the random error. The largest systematic
    discrepancies are found for Balmer lines at low S/N. Low-S/N line
    emission in MaNGA is generally associated with low-EW line
    emission, so this figure  looks equivalent with the flux discrepancy
    plotted as a function of EW instead of S/N.}
	\label{template_comp_SNR}
\end{figure*}

\begin{figure*}
	\includegraphics[width=\textwidth, trim=20 0 0 0, clip]{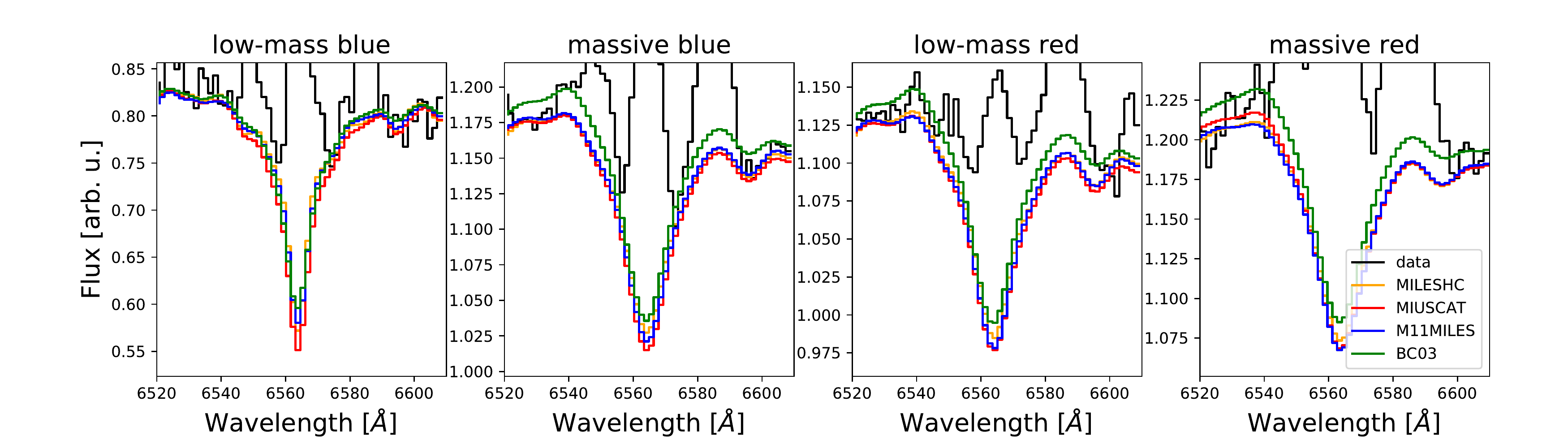}
    \caption{Continuum fits around the H$\alpha$ and
    [\nii]$\lambda\lambda$6548,84 for the central spaxels of four
    example galaxies spanning different regions of the $NUV -r$ versus
    $\log(M_\star/M_\odot)$ plane using different
    stellar (\mileshc) and SSP (M11-MILES, MIUSCAT, BC03) templates.  It
    should be noted how MIUSCAT and M11-MILES prefer deeper Balmer
    absorption-line cores and how BC03 displays significantly
    different line wings, which affect the continuum under the
    [\nii]$\lambda\lambda$6548,84 doublet.  The galaxies in figure are:
    low-mass blue: 7815-6101; massive blue: 8138-12704; low-mass red:
    8329-1901; massive red: 8258-6102.}
	\label{ha_models_new} 
\end{figure*}

In order to explicitly test the effect on measured emission-line fluxes
caused by the use of different stellar-continuum models we fit a subset
of DR15 MaNGA datacubes with the three sets of SSPs discussed in Section
\ref{sec4.1}, in addition to the \mileshc\ fit performed in DR15.  In
particular, we used SSPs to perform the second fitting stage in the
\DAP, but not for the extraction of the stellar kinematics, for which
stars are generally recommended. When necessary, the difference between
the intrinsic spectral resolution of the MILES stars and that of the SSP
templates used for the second fitting stage has been taken into account.

The SSP fits were carried out for a sample of 15 galaxies, evenly
sampling the $NUV -r$ versus $\log(M_\star/M_\odot)$
plane, in order to have access to a wide variety of stellar populations.
We only considered galaxies with line emission (including extended LIER
galaxies on the red sequence). Considering the entire galaxy subsample,
we obtain $\sim 2 \cdot 10^4$ spaxels with H$\alpha$ S/N $>$ 1.

In Figure \ref{template_comp_SNR} we compare the emission-line fluxes
obtained using \mileshc\ for both fitting stages (i.e., the DR15 data
products) with the fluxes obtained after switching to an SSP template
for the second fitting stage. The flux ratios are presented as a
function of line SNR for different strong lines. The dashed gray lines
represent the level at which the flux difference is comparable to the
random error. 

Figure \ref{template_comp_SNR} shows a number of interesting features.
For the Balmer lines (H$\alpha$ and H$\beta$ are shown in the figure),
different templates give systematically different line fluxes because of
the different best-fit stellar Balmer absorptions, especially at low
S/N. MIUSCAT prefers deeper Balmer absorption, leading to larger
Balmer-line fluxes than \mileshc.  BC03 and M11-MILES lead to better
agreement with DR15 for H$\alpha$ but show significant differences in
H$\beta$. It is interesting to note that the systematic discrepancies
are substantial (up to 0.1 dex at S/N $=$ 2 between DR15 and MIUSCAT).
They are, however, smaller than the random errors at low S/N, while they
become comparable (or larger) than the random error at high S/N. We have
checked the behavior of the flux differences as a function of EW of the
lines, and the resulting plot is very similar to Figure
\ref{template_comp_SNR}, especially for the Balmer lines, since S/N
largely tracks the EW.

On the other hand, for metal lines, such as [\oiii]$\lambda$5007,
[\sii]$\lambda$6732 and [\oii]$\lambda\lambda$3727,29, the discrepancies
between fluxes obtained with different templates are less extreme and do
not correlate as well with line S/N. [\nii]$\lambda$6585 and
[\sii]$\lambda$6732 stand out from the other metal lines for showing
comparatively larger discrepancies.  In Figure \ref{ha_models_new} we
show some example fits to the spectral regions around H$\alpha$ and the
[\nii] doublet for the central spaxel of four galaxies (low-mass blue:
7815-6101; massive blue: 8138-12704; low-mass red: 8329-1901; high-mass
red: 8258-6102) spanning a range of properties in the $NUV -r$ versus
$\log(M_\star/M_\odot)$ plane. These example fits
highlight the previously discussed differences in the core of the
H$\alpha$ line, but also the resulting effect on the nearby [\nii]
lines, which are the outer edge of the Balmer absorption wings. For
example, the BC03 templates generate best-fit models that have
substantially different line wings from those of other template sets,
therefore affecting the [\nii] flux in addition to H$\alpha$. 

\begin{figure*}    
	\centering
	\includegraphics[width=0.8\textwidth, trim=40 20 40 20, clip]{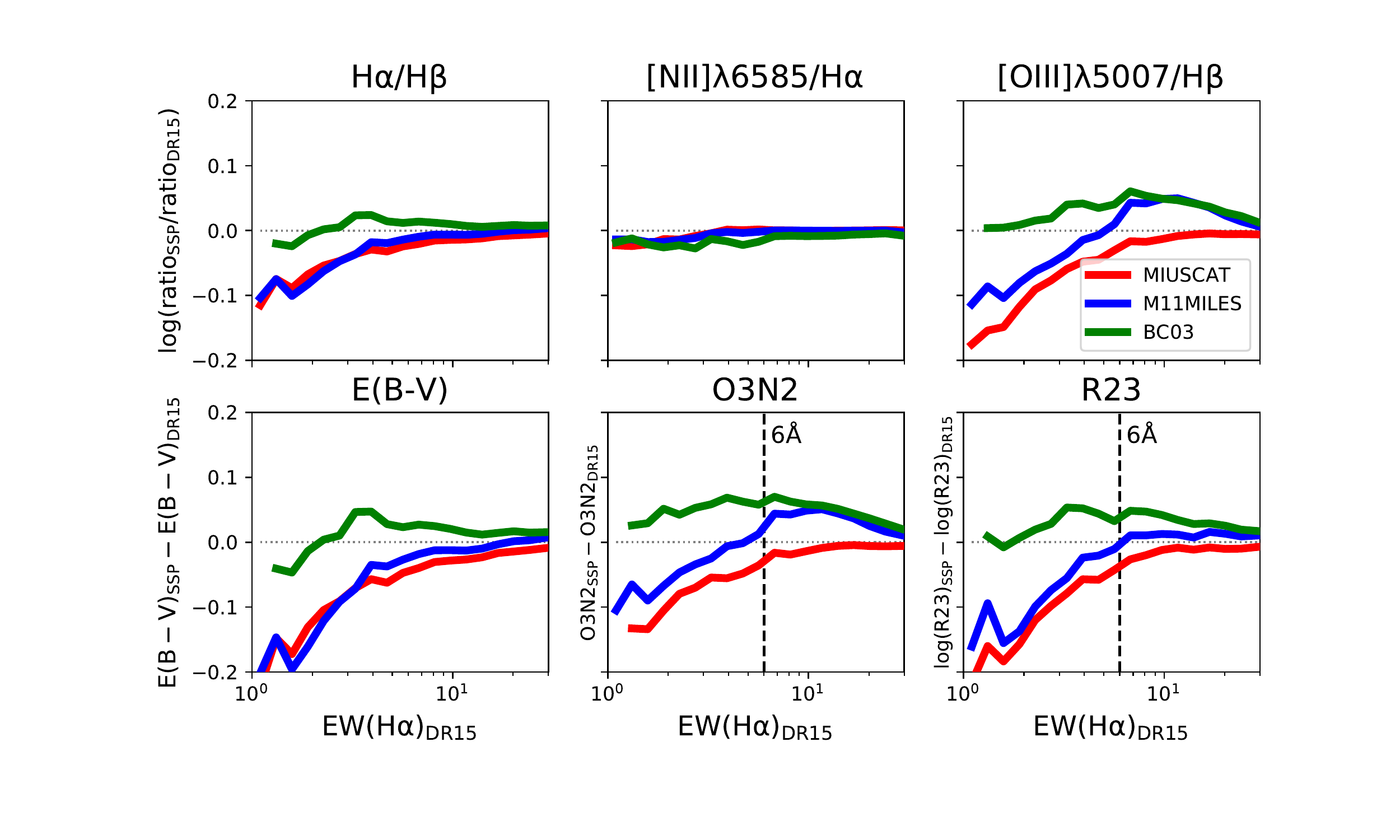}
    \caption{The difference in dex between specific line ratios when the
    spectral fitting is preformed with various SSP templates (MIUSCAT:
    red, M11-MILES: blue, BC03: green) with respect to the line ratios
    obtained in DR15 (which uses \mileshc\ templates) as a function of
    line EW. $E(B-V)$ is computed from the Balmer decrement assuming and
    intrinsic ratio $\rm H\alpha/H\beta$ = 2.86 and a
    \protect\cite{Calzetti2001} extinction curve. In the case of
    $E(B-V)$ the plotted difference is in magnitudes. The R23 index is
    computed after correcting for dust attenuation, while the other
    indices are not corrected.}
	\label{template_comp_EW}
\end{figure*}

Although these flux discrepancies are smaller than the random error,
they are systematic and behave differently for the different lines
considered, therefore leading to biases in the derived line ratios. In
Figure \ref{template_comp_EW} we show the differences in dex for several
line ratios and other derived quantities between the cases fit with SSPs
and DR15 as a function of EW(H$\alpha$). In the first row we show the
Balmer decrement (H$\alpha$/H$\beta$) and two classical BPT
(Baldwin-Phillips-Terlevich, \citealt{Baldwin1981}) line ratios ([\nii]/H$\alpha$ and [\oiii]/H$\beta$).
At low S/N, the Balmer decrement measured with MIUSCAT and M11-MILES
differs substantially from that inferred in DR15 or using BC03.
Estimating $E(B-V)$ using an intrinsic ratio $\rm H\alpha/H\beta$ = 2.86
and a \cite{Calzetti2001} extinction curve, deviations up to 0.1 dex are
evident at EW(H$\alpha$) $\sim$ 2~\AA, growing worse at even lower EW.
Regarding the BPT line ratios, [\nii]/H$\alpha$ is relatively unaffected
by template choice, possibly because of the vicinity of the two lines
means they are affected by the best-fit continuum shape in a correlated
way.  [\oiii]/H$\beta$, on the other hand, displays significant
differences for low EW lines (there is a 0.2 dex difference between
MIUSCAT and DR15 at EW(H$\alpha$) $\sim$ 2~\AA). These biases will have
a measurable impact on the BPT diagram positions of low EW regions,
which tend to be associated with LIER emission and diffuse ionized gas.

In the bottom row of Figure \ref{template_comp_EW} we also show two
metallicity-sensitive indices often employed in the literature, O3N2 =
([\oiii]$\lambda$5007/H$\beta$)/([\nii]$\lambda$6583/H$\alpha$)
\citep{Pettini2004} and  R23=([\oii]$\lambda\lambda$3727,29 +
[\oiii]$\lambda\lambda$4959,5007) / H$\beta$ \citep{Pagel1979}. While
O3N2 is relatively insensitive to dust extinction, we correct the
measured line fluxes for extinction when computing R23. It is evident
from the figure that discrepancies larger than a tenth of a dex are
present at low EW for both indicators. A cut on EW(H$\alpha$) $>$ 6 is
sometimes performed in studies of ISM metallicity in order to minimize
the contamination from gas not directly associated with \ion{H}{2}
regions \citep{Sanchez2014}. Here we show this threshold as a dashed
black line for these two indicators in Figure \ref{template_comp_EW},
demonstrating for larger EWs the systematic effects from the choice of
continuum templates are non-negligible. This exercise demonstrates that
care is needed when comparing results from IFU surveys calculating
emission-line fluxes with different underlying stellar or SSP models.

In the Appendix we perform a similar comparison on the line fluxes measured 
by the \DAP\ and the {\tt Pipe3D} VAC for DR15. The \DAP\ and 
 {\tt Pipe3D} differ in many fundamental aspects beside the choice of continuum templates, so 
 it is more difficult to attribute discrepancies to just one factor. However, for some line ratios (like H$\alpha$/H$\beta$ and  [\oiii]/H$\beta$) we find discrepancies between the results of the two pipelines which are comparable,
 at least qualitatively, with those between the \DAP\ DR15 run and the \DAP\ run utilising MIUSCAT templates. These differences may therefore be attributed at least partly to the different choice of continuum templates in the two pipelines.
 
 However, while the results in this section show no strong impact of template choices on the [\nii]/H$\alpha$, we do find a significant discrepancies for this line ratio between the  \DAP\ and {\tt Pipe3D}. This fact is discussed further in Section \ref{appendix_ratios}.

\subsection{Simulating the effect of template mismatch}
\label{sec4.4}

The aim of this section is not to select the `correct' set of
templates, but simply to quantify the effect of different templates on
the resulting emission-line fluxes. In general, we cannot determine
which set of templates is the correct one for our galaxy data, so we
devise an artificial exercise similar to the recovery simulation
presented in Section \ref{sec3.2} to study the effect of using the
`wrong templates' in the presence of noise. 

In particular, we take the best-fit model from the previous section
based on a set of SSP templates and add noise in the same way as was
done in Section \ref{sec3.2}. Here we discuss the results of using mock
datacubes generated using the MIUSCAT best-fit models which are then fit
using the standard DR15 approach (i.e., using the \mileshc\ library). In
light of the results of the previous section, we expect the recovered
Balmer line fluxes to be lower than the input ones on average, given the
preference for MIUSCAT to fit deeper Balmer absorption features.

\begin{figure}		
	\includegraphics[width=0.5\textwidth, trim=20 0 20 0, clip]{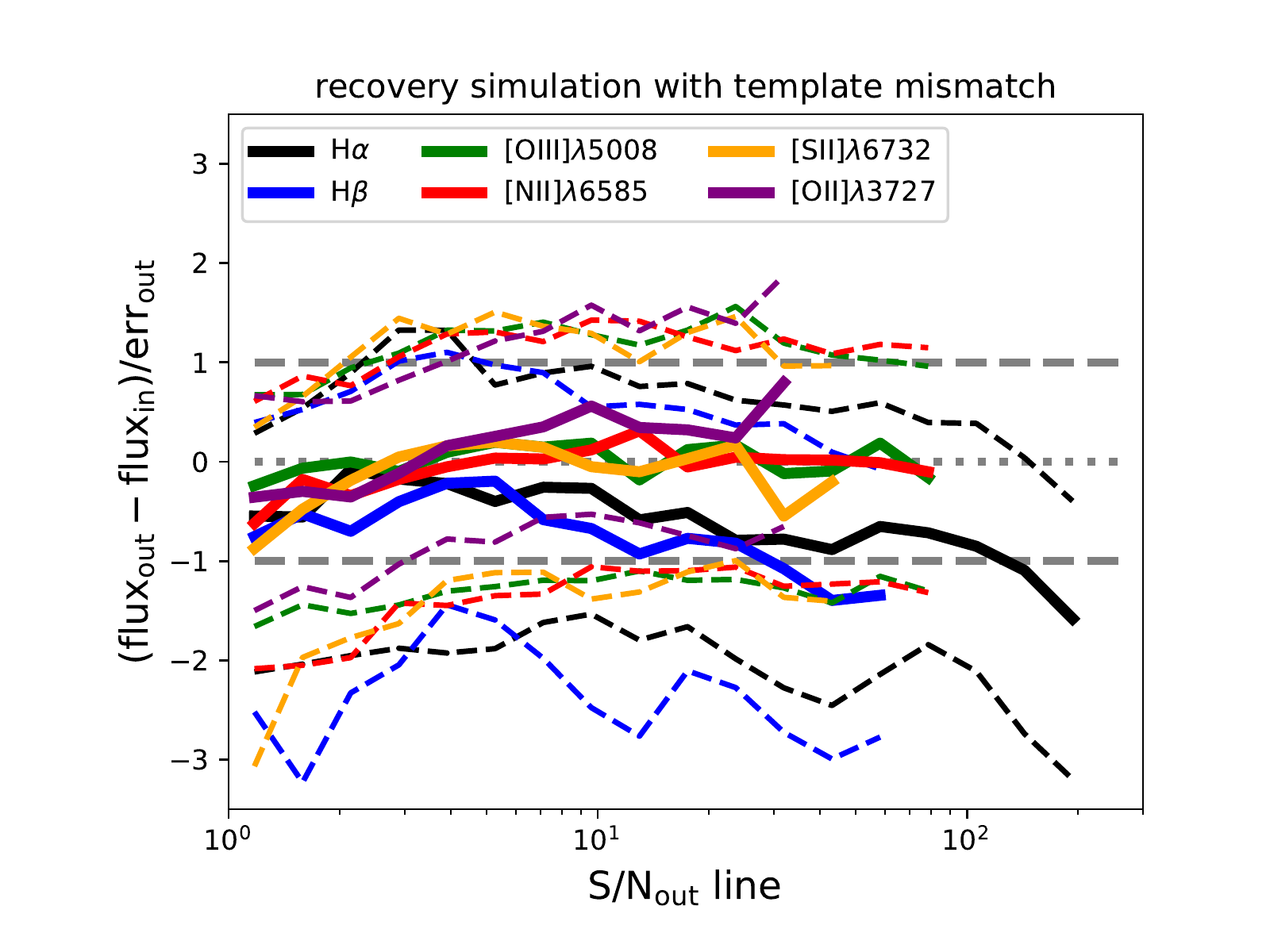}
    \caption{The difference between the input and output fluxes,
    normalized to the error, for a parameter-recovery simulation that
    includes template mismatch (see text). The setup of the figure is
    the same as Figure \ref{sims}a. The solid colored lines represent
    the median and the dashed colored lines the 16th and 84th
    percentiles of the normalized residual distributions for the
    different strong lines considered.  The Balmer lines (H$\alpha$ in
    black and H$\beta$ in blue) display the expected systematic offset
    at high S/N.} \label{sim_temp_mis}
\end{figure}

In Figure \ref{sim_temp_mis} demonstrates this effect. The recovered
H$\alpha$ and H$\beta$ fluxes are indeed systematically lower than the
input ones. It is interesting to note, however, that while the median
flux is systematically biased, the 16th and 84th percentiles still
approximately correspond to $\pm 1$ with respect to the median,
indicating that template mismatch does not dramatically affect the
statistical validity of the emission-line flux errors.  At low S/N, we
also observe a bias in the recovered fluxes, in the sense that the flux
tends to be under-estimated, as already noted in Section \ref{sec3.2}
(see Figure \ref{sims}a).

\section{Systematics from algorithmic choices}
\label{sec5}

In the section we address the systematic errors on emission-line
parameters that may result from specific algorithmic choices. In
particular, we study the effect and importance of the polynomial
corrections adopted in DR15 and critically assess the strategy of
simultaneously fitting the continuum and emission lines in the second
fitting stage of the \DAP. We also explore different schemes for tying
the velocity and velocity dispersion of different emission lines and
compare them to the approach we have followed in DR15.

\begin{figure}
	\includegraphics[width=0.5\textwidth, trim=0 0 0 0, clip]{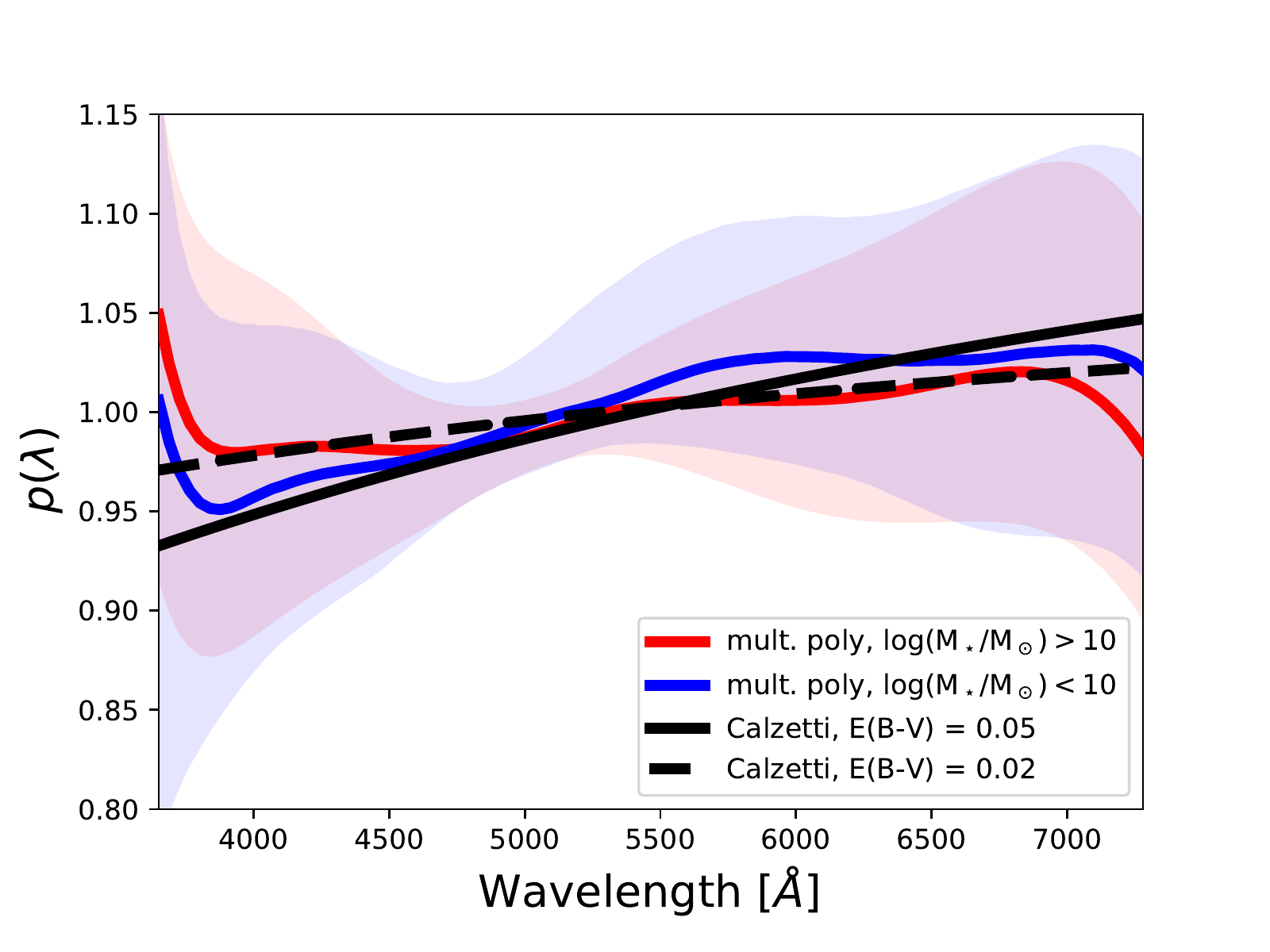}
    \caption{The median multiplicative polynomial correction applied to
    the best-fit continuum templates calculated for all spaxels in a
    sample of 100 random galaxies from DR15. Red and blue solid lines
    correspond to the median correction for
    $\log(M_\star/M_\odot) >10$ and
    $\log(M_\star/M_\odot) <10$ galaxies,
    respectively. The shaded regions correspond to the 16th and 84th
    percentiles. The black curves represent a
    \protect\cite{Calzetti2001} extinction curve with different values
    of $E(B-V)$.  }
	\label{mult_poly}
\end{figure}	

\begin{figure*}
	\centering
	\includegraphics[width=0.8\textwidth, trim=40 0 0 0, clip]{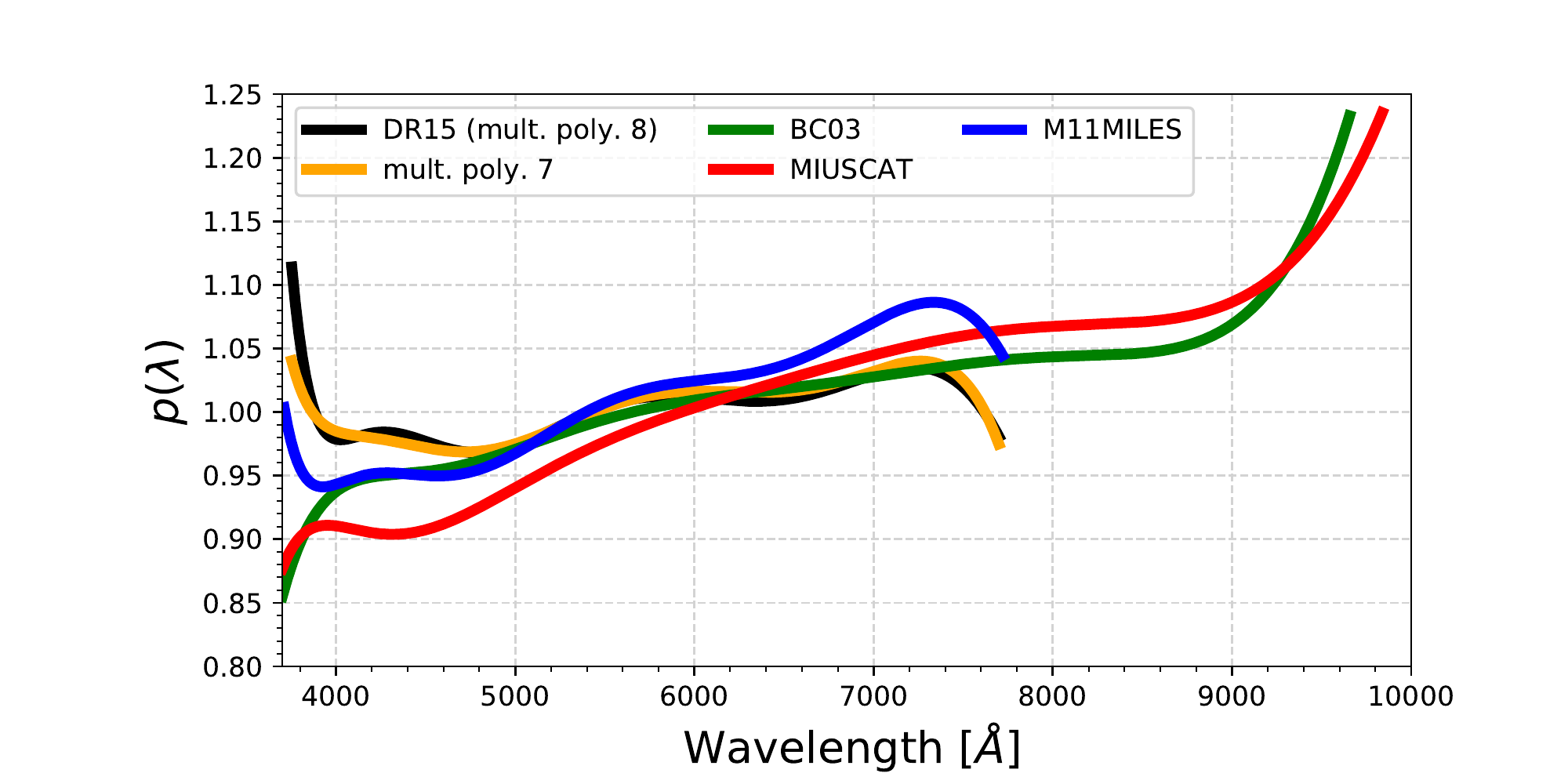}
    \caption{The median multiplicative polynomial correction for one
    massive elliptical test galaxy (8258-6102), fit using different
    stellar and SSP templates. An upturn in the polynomial correction is
    evident for $\rm \lambda < 4000~\AA$ for the DR15 (\mileshc, in
    black) and the M11 MILES (in blue) templates. We attribute this
    upturn to an issue in the flux calibration of the MILES stellar
    library. Interestingly the same effect is not as evident for the
    MIUSCAT (in red) templates, which are also based on MILES stars. The
    shape of the downturn is independent of the polynomial order used
    (in orange we show the result of using 7th-order multiplicative
    polynomials and \mileshc). In the far red ($\rm \lambda >9000~\AA$)
    we note a red upturn, picked up both by BC03 (in green) and MIUSCAT.
    For the moment we cannot exclude that this red upturn is an artifact
    of the MaNGA flux calibration.}
	\label{poly_all_tests}
\end{figure*}

\subsection{Multiplicative polynomials}
\label{sec5.1}

In this section we assess the role and importance of multiplicative
polynomials in the second fitting stage of the \DAP\ (the simultaneous
fitting of continuum and emission lines). A complementary discussion of
the role of additive polynomials during the first fitting stage of the
\DAP\ (stellar kinematics) is presented in Section 7.3.3 of \cite{Westfall2019_arxiv}.  We remind the reader that the multiplicative
polynomials are only applied to the stellar-continuum templates, and not
to the emission-line (Gaussian) templates. In this sense, their effect
on the line fluxes is only indirect. We nonetheless address this issue
here as a check on the quality of our continuum model and for its
relation to the overall flux calibration of the MaNGA survey. 

\subsubsection{The role of polynomials}

The inclusion of polynomials during the \pPXF\ fit may be advantageous
for several reasons.
\begin{enumerate}
\item Multiplicative polynomials can compensate for residual differences
in the relative flux calibration of the science data with respect to the
stellar templates. Assuming the spectral templates are perfectly
calibrated (and in presence of negligible extinction), one may use the
shape of the recovered polynomials to test the quality of the flux
calibration of the data.
\item Polynomials can mimic the shape of canonical extinction curves.
\item Polynomials can provide low-order corrections to the
stellar-population models, which may be especially valuable when
theoretical stellar spectra are used. In addition, they can help to
reproduce the shape of the spectra of stars that are not present or
under-represented in the library used (as we have discussed in Section
\ref{sec4.2}).
\end{enumerate}

\subsubsection{The typical shape of the polynomial correction in DR15}
In this section, therefore, we start by looking at the typical shapes of
the multiplicative polynomials used in the second fitting stage for the
DR15 \DAP\ run. To do so, we selected a random sample of 100 DR15
galaxies and reconstructed the multiplicative polynomials used in each
of their spaxels. In Figure \ref{mult_poly} we show the shape of the
median multiplicative correction applied as a function of rest-frame
wavelength.  The sample of galaxies is subdivided into two mass bins
(red for $\log(M_\star/M_\odot) >10$; blue for
$\log(M_\star/M_\odot) <10$) and the shaded areas
correspond to the 16th and 84th percentiles of the distribution. We also
show in black the expected multiplicative correction for a Calzetti
extinction curve and  two values of $E(B-V)$.\footnote{We recall here
that at this stage the data has already been corrected for Galactic
foreground extinction.} The extinction curves are scaled arbitrarily to
the median of the polynomial corrections to highlight the similarity in
relative shape.

The trends observed in the figure can be qualitatively interpreted as
follows. The massive bin contains a larger number of passive galaxies,
which are largely devoid of gas and thus suffer lower extinction.  The
shape of the polynomials are consistent with the values of $E(B-V)$
measured for the continuum by full spectral fitting in the outskirts of
MaNGA galaxies \citep{Goddard2017b}. 

\subsubsection{Deviations from smooth polynomial shapes and consequences
for flux calibration}

The key features in Figure \ref{mult_poly} are the deviations from the
expected smooth extinction curves, namely the upturn in the mean
correction at the blue end of the MaNGA wavelength range and a
similar downturn redder than 7000~\AA.  We determined that these deviations are 
not due to imperfections in the MaNGA flux calibration for the following reasons.

First, the upturn in the blue occurs at different observed wavelengths
for galaxies at different redshifts. For example, if one considers
massive galaxies in the MaNGA primary and secondary samples, which are
selected in the same fashion but separated by a small redshift interval,
the blue upturn moves to longer observed wavelength in the secondary
sample. If the upturn was due to imperfections in the MaNGA flux
calibration derived from standard-star spectra, it would always appear
at the same observed wavelength.  Secondly, the downturn observed redder
than 7000~\AA\ occurs in the middle of the MaNGA spectral coverage (but
at the edge of the spectral coverage of \mileshc) and is therefore more
likely to be originating from the \mileshc\ than the MaNGA data. 

In order to test whether imperfect relative flux calibration of the
\mileshc\ library is responsible for the deviations observed in Figure
\ref{mult_poly} we selected one test galaxy (a massive red galaxy,
8258-6102, with good S/N throughout) and examined the stacked polynomial
shapes obtained after fitting the galaxy with different template sets.
The results are presented in Figure \ref{poly_all_tests}. 
It is interesting to note that between 4000~\AA\ and 7000~\AA\
\mileshc\ (labeled DR15), M11-MILES, MIUSCAT (all based on MILES stars)
and BC03 (based on STELIB) agree to better than 10\%. Bluer than
4000~\AA\ BC03 presents a downturn, while both M11-MILES and \mileshc\
show an upturn. MIUSCAT, on the other hand, gives rise to a flattening.
It should also be noted that the downturn at 7000~\AA\ is present both in
\mileshc\ and M11-MILES, pointing towards a problem with the MILES
stars. 

To check the behaviour of the code at the edges of the wavelength range we changed
 the degree of multiplicative polynomials from 8 to 7 (i.e. from even to odd parity). 
If the behaviour of the polynomials at the edges was entirely dictated from the fit
within the central wavelength region, we would expect that a change of parity 
would lead to a change in symmetry of the recovered multiplicative correction, 
which is however not seen in Figure \ref{mult_poly}. We concluded, therefore, that
the offsets seen at the edges of the fitted MaNGA wavelength range are likely to
be real.
 
It is possible that \mileshc\ suffers from the lack of hot stars, including 
blue horizontal branch stars. These stellar types may have been accounted differently 
by different SSP models, generating the discrepancy observed between BC03, MIUSCAT and M11-MILES
at the blue edge of the optical wavelength range.

Interestingly the SSP templates that extend redder than 9000~\AA\ show
the need for an upward correction to match the MaNGA data. The presence
of this red upturn has been identified via visual inspection in some of
the MaNGA spectra.  Since this spectral range is not fit in DR15, we
postpone further study of this potential systematic effect.

\begin{figure*}
	\centering
	\includegraphics[width=\textwidth, trim=0 0 0 0, clip]{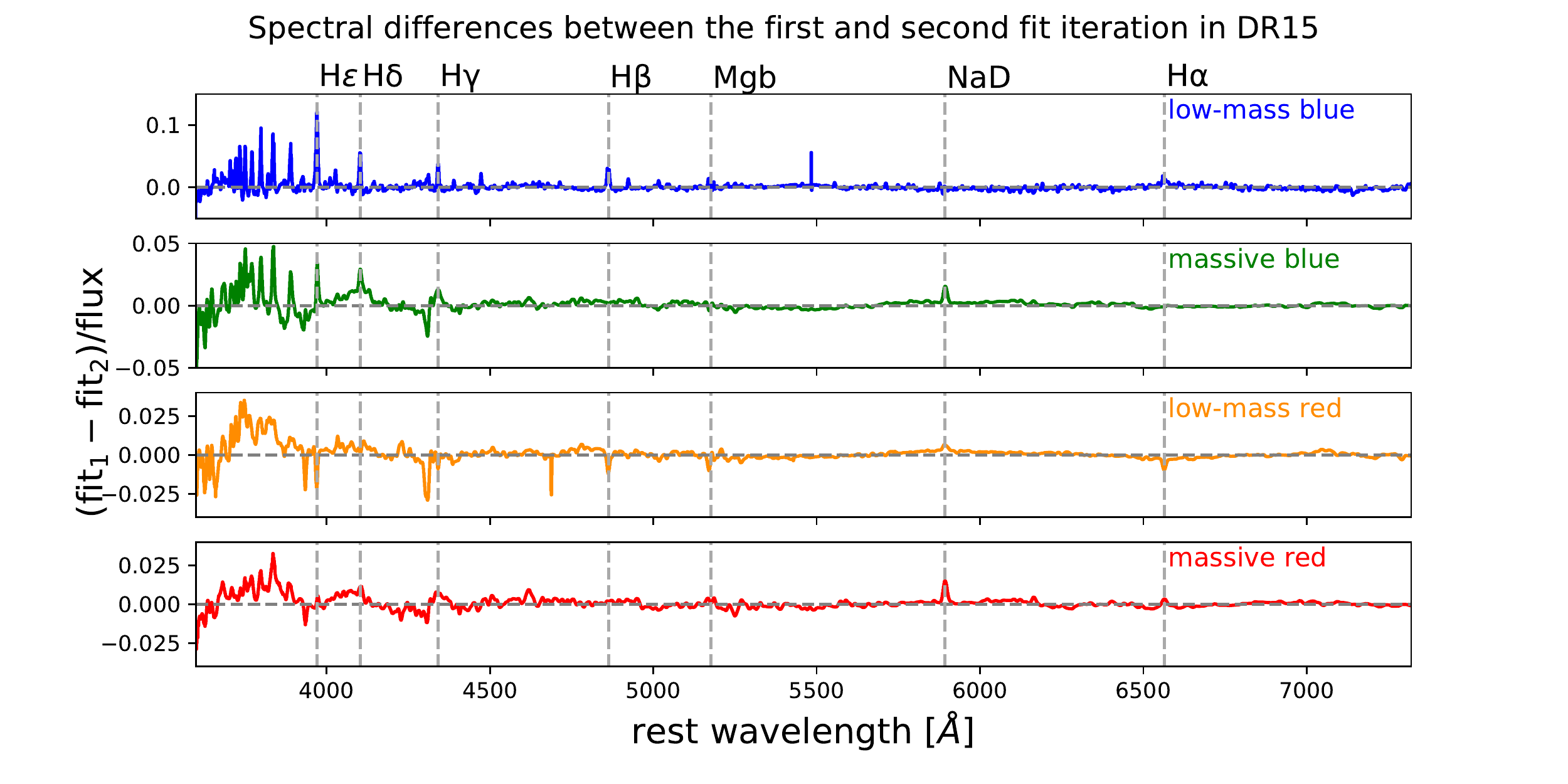}
    \caption{The spectral differences between the best-fit stellar
    continuum in the first fitting stage of the \DAP\ and that of the
    second fitting stage normalised by the input spectrum.. At the position of specific emission lines the y-axis value can be interpreted as the fractional error in the amplitude of the line due to the different continuum models. The four spectra shown are taken
    from the central regions of different test galaxies, spanning a
    range in stellar mass and color (low-mass blue: 7815-6101; massive
    blue: 8138-12704; low-mass red: 8329-1901; massive red: 8258-6102).
    The first fitting stage masks regions potentially
    contaminated by emission lines and allows the use of additive
    polynomials while the second fitting stage performs simultaneous
    fitting of gas and stellar templates and makes use of multiplicative
    polynomials. The main differences between the best fits from the two
    stages are seen in regions of strong Balmer and metal absorption.}
	\label{salerno1}
	
	\includegraphics[width=\textwidth, trim=0 0 0 0, clip]{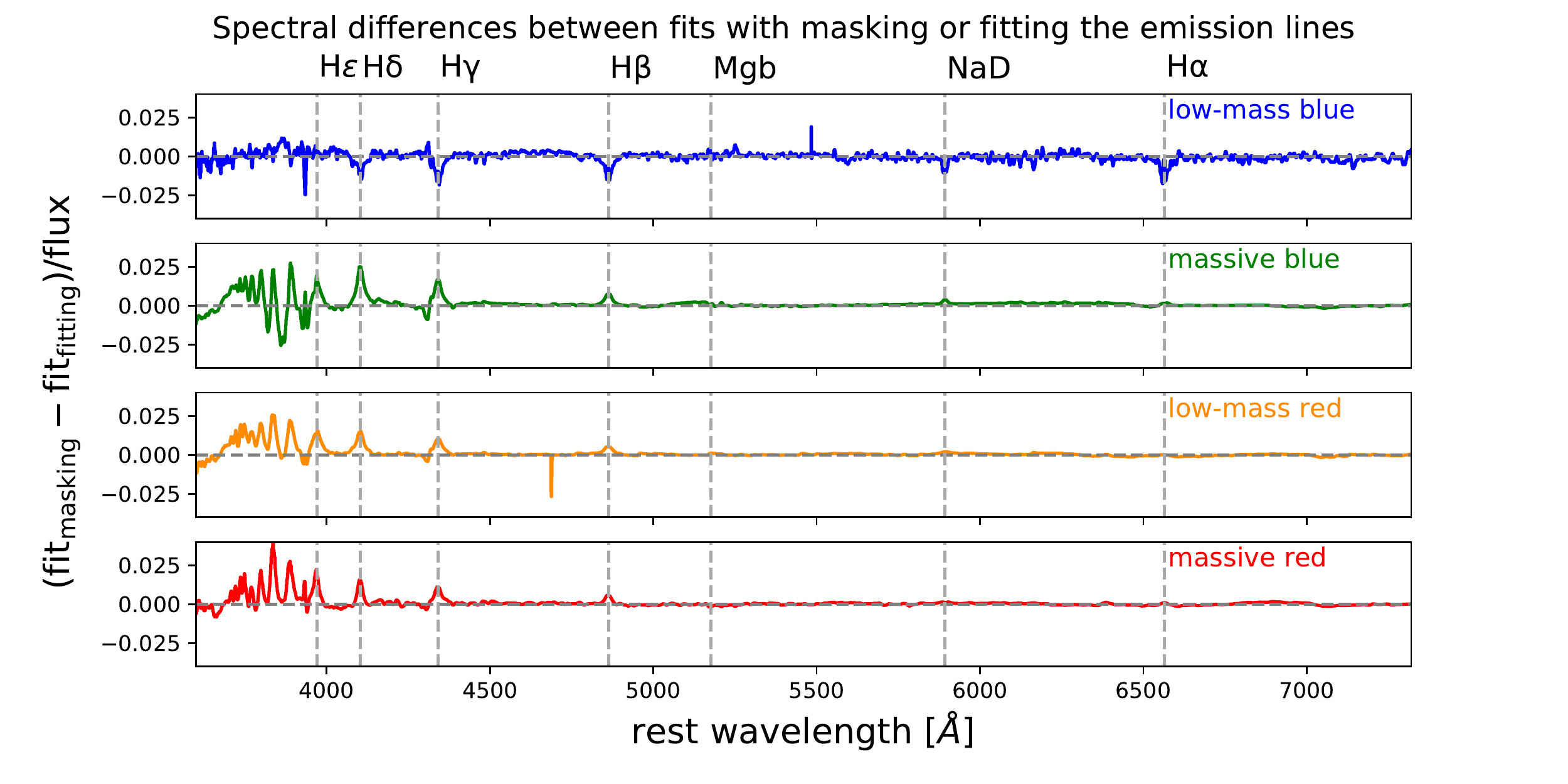}
    \caption{Same as Figure \protect\ref{salerno1}, but using 8th-order
    multiplicative polynomials in {\it both} fitting stages of the \DAP. The
    difference between the two best-fit models is only
    driven by the masking of the emission lines in the first stage (in
    the second stage emission lines are fit at the same time as the
    continuum). Small differences ($<$ 2 \%  of the H$\alpha$ flux) can
    be seen in the best-fit models at the position of the Balmer lines.}
	\label{salerno2}
	\label{Salerno_effect}
\end{figure*}

\subsection{The combined effect of masking and polynomials}
\label{sec5.2}

The MaNGA \DAP\ implements simultaneous fitting of emission-line and
continuum templates following the recommendation from previous work
\citep{Sarzi2005, Oh2011}. \cite{Sarzi2005}, in particular, demonstrated
the advantage of this algorithmic choice when dealing with the limited
wavelength range of the SAURON data, where the emission lines lie in
close vicinity to the key metallicity and age-sensitive features. In
Section \ref{sec4.3}, however, we have demonstrated that, even by
performing simultaneous fitting, residual degeneracies between Balmer
absorption and line emission are still present, leading to noticeably
different best-fit models when using different template libraries. In
this section, therefore, we perform some illustrative tests to evaluate
the impact of the masking on the recovered best-fit continuum under the
Balmer lines.

We first examine the difference between the best-fit continua obtained
by the first and the second fitting stages in the \DAP\ for the central
spaxels in four test galaxies (Figure \ref{salerno1}, same galaxies as
in Figure \ref{ha_models_new}).  A difference between the two best-fit
stellar-continuum models in this comparison may be due to either the
effect of masking (emission-line regions are masked in the first fit but
not in the second) or the difference in the use of polynomials (additive
polynomials in the first fit and multiplicative polynomials in the
second fit). In Figure \ref{salerno1} we plot the difference between the two best-fit models
normalised by the input spectrum. At the wavelength of a specific emission line, this can be interpreted as the fractional error in the amplitude of the line introduced by these different choices in continuum fitting.

The largest
deviations are seen in regions corresponding to strong
absorption lines (like the NaD doublet, evident in all the four examples
except the low-mass blue galaxy) and the Balmer absorption lines.
However, in the case of Balmer lines, the differences can be both
positive or negative. We attribute this behavior to the different
implementation of polynomials in the two fitting stages. In the case of
the low-mass blue galaxy, where the most prominent absorption features
are the Balmer lines, additive polynomials lead to shallower
absorption-line profiles, which result in positive residuals at the
positions of the Balmer lines in Figure \ref{salerno1}. For the
higher-mass galaxies other metal absorption lines dominate the spectrum,
and therefore determine the shape of the additive polynomials, causing both
positive, null or negative residuals around the Balmer lines.
We note that the deviations observed in the low-mass blue galaxy are much larger than those observed in the red galaxies, with significant changes already seen at H$\beta$ ($\sim 3 \%$) and increasing to $13 \%$ at H$\epsilon$.

A cleaner test to isolate the effect of masking is to apply the same
type of polynomials to both the first and second fitting stage. We have
therefore repeated the exercise just described by using 8th-order
\textit{multiplicative} polynomials for both fitting stages. The
resulting normalized flux differences are shown in Figure \ref{salerno2}
and look substantially different from Figure \ref{salerno1}. Now the
differences around metal absorption lines are reduced and the Balmer
lines correspond to the largest residuals (of the order
of $\sim 1-2\%$). These are again stronger for the high-order Balmer lines (in particular H$\delta$ and H$\gamma$). Integrating over the line profile the systematic differences in H$\alpha$ flux are less
than 2\%, which is negligible in most cases when compared to
the discrepancies caused by changes in the template library. However the changes 
in measured fluxes are more substantial for the high-order lines.

There is an interesting difference between the young spectrum of the
low-mass blue galaxy, which displays deeper Balmer absorption in the
second (unmasked) fit, and the other spectra characterized by older
stellar populations, which show shallower absorption in the second fit.
The reasons for this difference must be related to how the inclusion of
the masked regions affects the best-fit template mix. In the future, it
would be of interest to repeat the same exercise in the context of
stellar-population synthesis and assess the effect of masking on the
recovery of stellar-population parameters, which is likely more
significant than the effect on the emission lines.

\subsection{The choice of tying emission line kinematic parameters}
\label{sec5.3}

\begin{figure*}		
	\centering
	\includegraphics[width=0.9\textwidth, trim=0 0 0 0, clip]{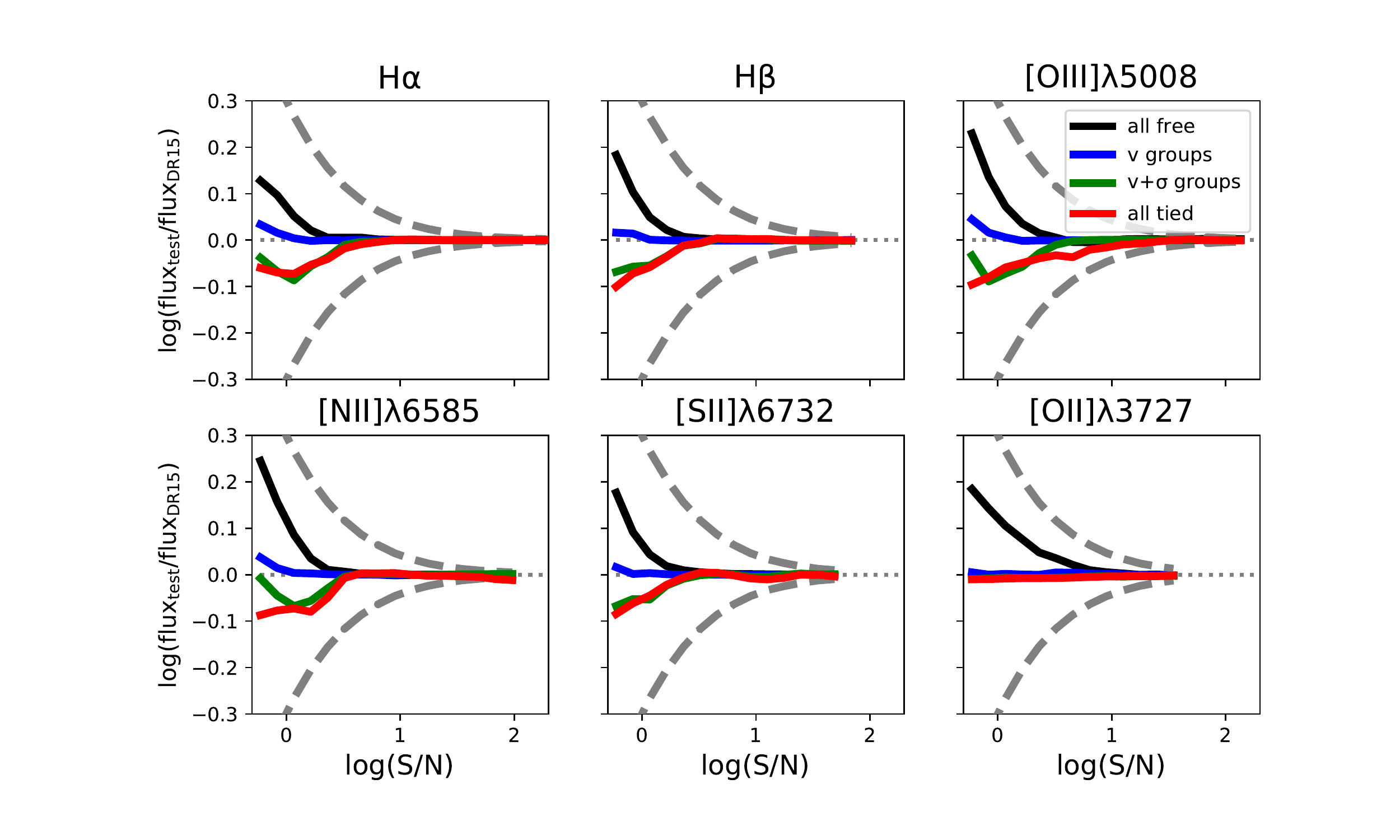}
    \caption{The median of the ratio of emission-line fluxes (in dex)
    obtained between runs with different tying prescriptions for the emission lines and DR15 as a function of S/N. The gray dashed lines
    correspond to the scatter expected considering the random errors.
    The largest systematic discrepancies between the tying strategy
    implemented in DR15 and other test runs are found for S/N < 6.}
	\label{tying1}
\end{figure*}

\begin{figure*}		
	\centering
	\includegraphics[width=0.9\textwidth, trim=0 0 0 0, clip]{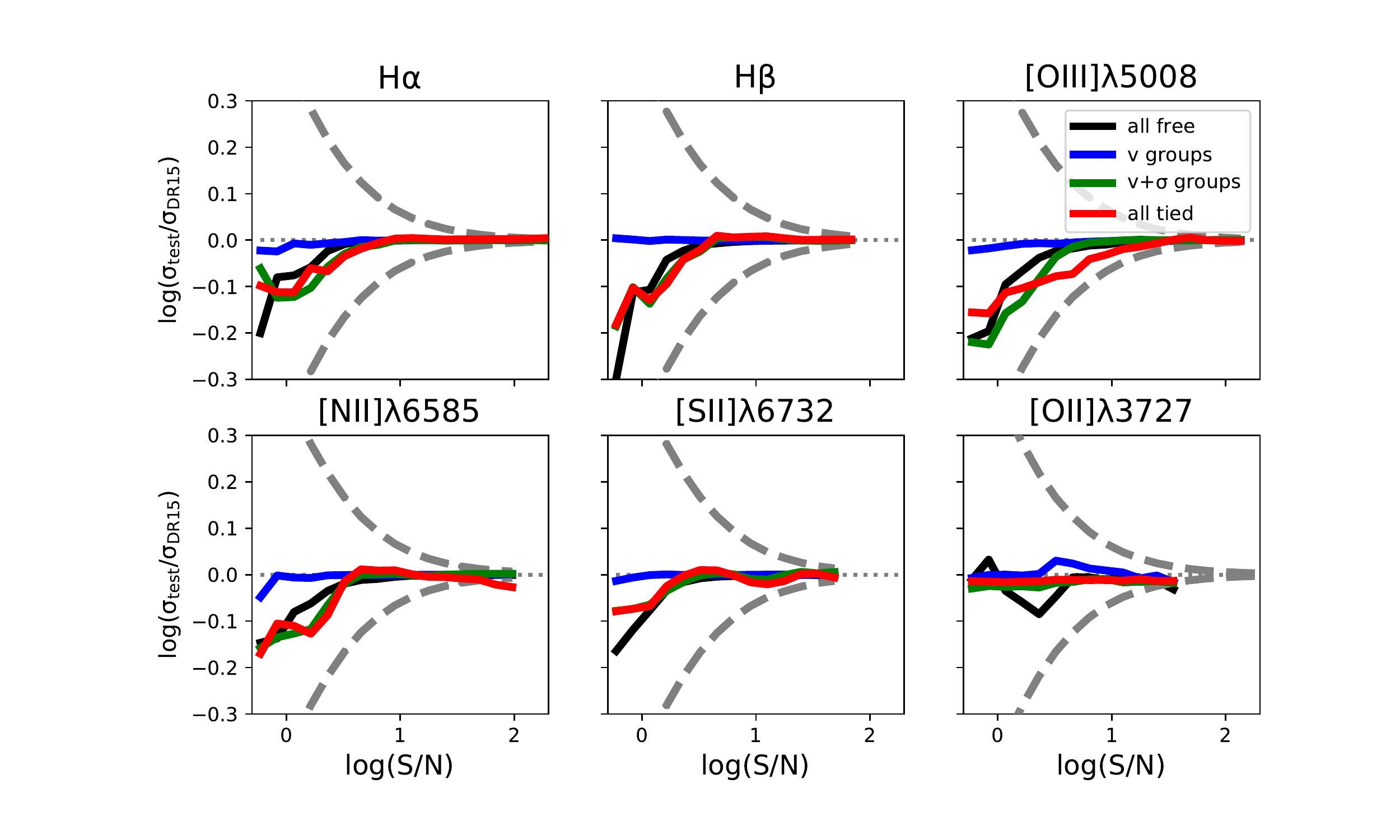}
    \caption{The median ratio of velocity dispersions (in dex) obtained
    in different test runs (as per legend) and
    DR15. The gray dashed lines correspond to the scatter expected
    considering the random errors.}
	\label{tying2}
\end{figure*}

When several emission lines are fit across a large wavelength range,
whether or not to tie the kinematic parameters for different lines
becomes a debatable problem. In general, tying the velocities and
velocity dispersions of different lines is an advantage in the low-S/N
regime, where stronger lines contribute much more to the overall
$\chi^2$, therefore effectively determining the kinematic parameters of
weaker ones. Several reason exist, however, to be skeptical of tying
kinematic parameters. First, without an accurate knowledge of the
LSF and its change with wavelength it is not
possible to correctly fix the astrophysical velocity dispersions of
widely separated lines. Likewise, small errors in the wavelength
calibration can induce problems when fitting all emission lines with a
common velocity.  Finally, there are astrophysical reasons to expect
emission lines emitted by different ionic species in different
ionization stages to have different kinematics. 

In this section we test the effect of making different assumptions
regarding the tying of kinematic parameters. We considered the sample of
15 galaxies described in Section \ref{sec4.3}, selected to evenly sample
the $NUV -r$ versus $\log(M_\star/M_\odot)$ plane.
We consider the schemes described below.

\paragraph{All parameters free (all free)} In this scheme, all
velocities, dispersions and amplitudes of the different emission lines
are fit individually as free parameters.

\paragraph{Tie velocities (DR15)} In this scheme, the
velocities of all lines are tied together, while the velocity
dispersions are fit independently. This tying scheme may be beneficial
when uncertainties in the LSF prevent the tying of the astrophysical
dispersions and is the scheme adopted in the DR15 run.

\paragraph{Tie velocities in three groups (v groups)} In this
scheme we define three groups of emission lines:

\begin{enumerate}
\item  Balmer lines: H$\alpha$, H$\beta$, H$\gamma$, H$\delta$,
H$\epsilon$, H$\zeta$, H$\eta$, H$\theta$
\item Low-ionization lines: [\oii]$\lambda\lambda$3727,29,
[\oi]$\lambda\lambda$6300,64, [\nii]$\lambda\lambda$6548,84,
[\sii]$\lambda\lambda$6717,31.
\item High-ionization lines: [\neiii]$\lambda\lambda$3869,3968,
\heii$\lambda$4687, [\oiii]$\lambda\lambda$4959,5007, \hei$\lambda$5876.
\end{enumerate}

The velocities of different lines are tied within the same group. None
of the velocity dispersions are tied. This scheme is a variant of the
DR15 run that allows for different astrophysical velocities for lines of
different species and ionization states.

\paragraph{Tie velocities and dispersions in three groups (v+$\sigma$ groups)} The scheme is the same as the previous one, but we tie both
velocity and velocity dispersions for the lines in the same group.

\paragraph{All parameters tied (all tied)} In this scheme we tie
the velocity and velocity dispersions of all lines together.

\begin{figure*}	[!htb]
	\centering
	\includegraphics[width=0.8\textwidth, trim=0 0 0 0, clip]{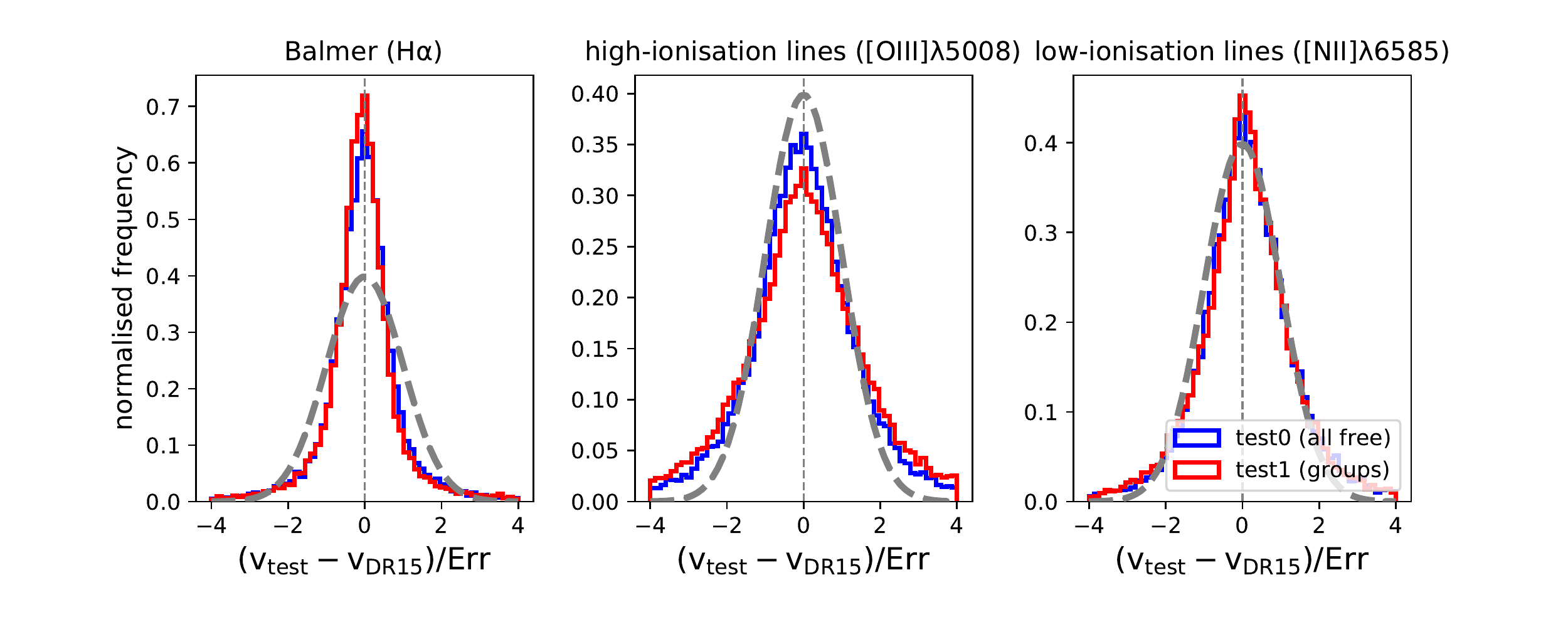}
    \caption{The velocity difference between test runs 0 and 1 and DR15
    normalized by the error. A normalized Gaussian is shown in dashed
    gray for comparison.  }
	\label{tying3}
\end{figure*}

\smallskip

In all schemes, except the first one (all free) we impose the flux ratios
set by atomic physics when fitting the line doublets of
[\oi]$\lambda\lambda$6300,64, [\nii]$\lambda\lambda$6548,84 and
[\oiii]$\lambda\lambda$4959,5007; see Table \ref{emission_line_groups}. 

In Figure \ref{tying1} we show the flux ratio (in dex) between each of
the test runs and DR15, as a function of line S/N for different emission
lines. The thick gray dashed lines correspond to the deviations expected
given the random errors in the flux measurements.

For S/N $<$ 2, the all free case gives
\textit{larger} fluxes than DR15 for all lines considered. Figure
\ref{tying2} shows a similar plot but for the velocity dispersions,
demonstrating that the deviations towards larger fluxes are accompanied
by a \textit{lower} sigmas. This may be due to the fact that, when
velocities are not tied, the algorithm may be fitting noise spikes at
low S/N. These spikes tend to have a width of one pixel, leading to
smaller dispersions and higher amplitudes.

The `v groups' case, where velocities are allowed to vary within emission line
groups, is indistinguishable from DR15 in terms of fluxes and
sigmas.

Finally, the `v+$\sigma$ groups' and `all tied' test runs produce
marginally lower sigmas and fluxes than DR15.

In Figure \ref{tying3} we show the velocity difference between DR15 and
different tests runs for H$\alpha$ [\nii]$\lambda\lambda$6584 and
[\oiii]$\lambda\lambda$5007. It is evident that there is no systematic
velocity shift if velocities are tied in groups rather than all
together. If the velocities are not tied (or tied in groups) the scatter
in the resulting velocities with respect to DR15 is comparable to the
random error (estimated in DR15) for the metal (both high and
low-ionization) lines, but smaller for the Balmer lines. Overall the
consistency of the velocities determined following different tying
prescriptions validates the DR15 approach and demonstrates that the
MaNGA data does not suffer from any detectable systematic in the
wavelength calibration \citep[see also Figure 19 of ][]{Law2016}, which
would necessarily invalidate some of our tying schemes.

\begin{figure*}	[htb]
	\centering
	\includegraphics[width=0.65\textwidth, trim=0 0 0 0, clip]{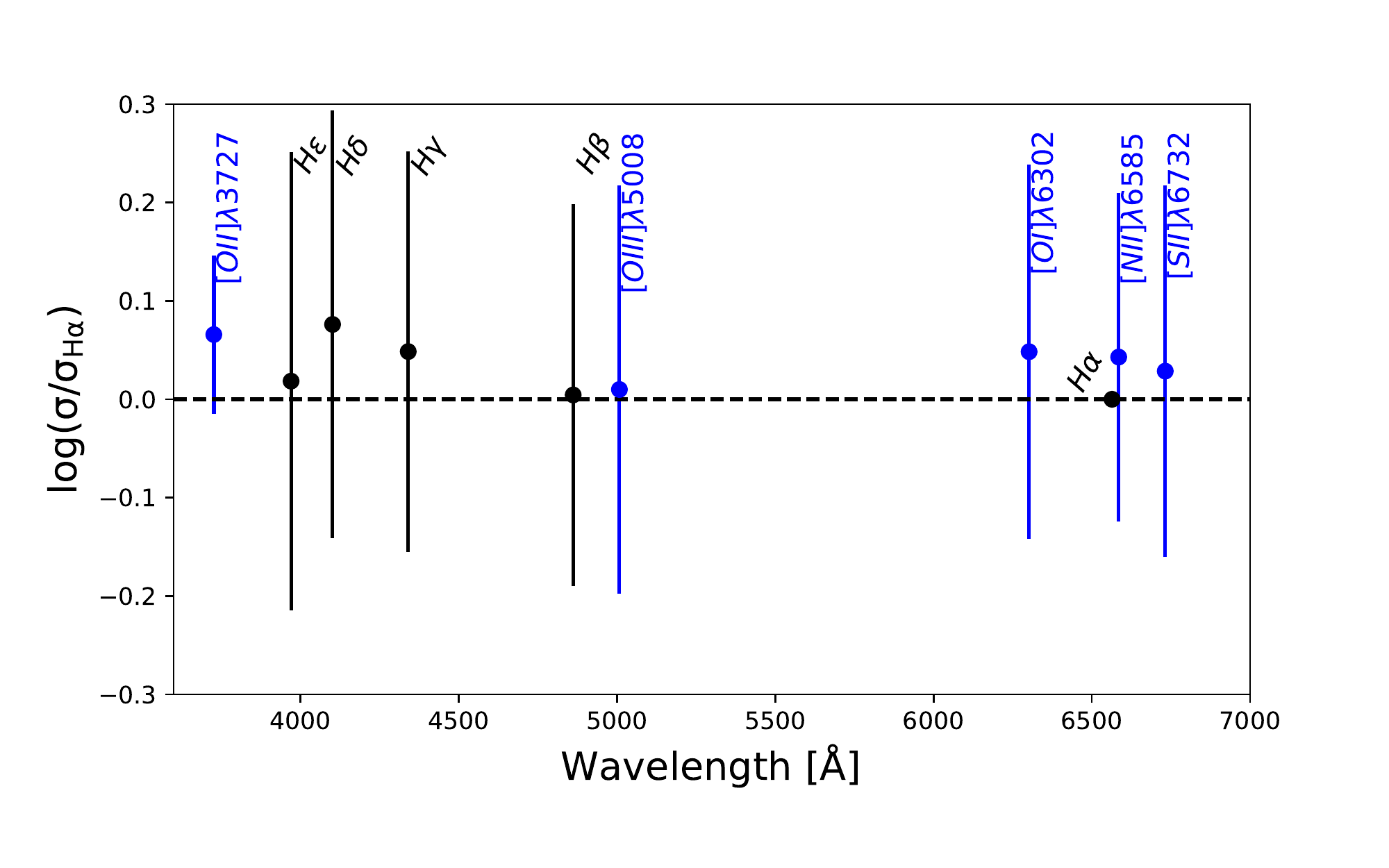}
    \caption{The ratio (in dex) between the intrinsic velocity
    dispersion of different emission lines with respect to H$\alpha$ in
    DR15 as a function of wavelength. The intrinsic (astrophysical)
    dispersion is measured by subtracting, in quadrature, our estimate
    of the instrumental velocity dispersion from the measured line
    velocity dispersion. Balmer lines are represented in black and metal
    lines in blue. Only spaxels with S/N $>$ 10 in the specific emission
    line are considered to generate the plot. The error bars represent
    the 1-$\sigma$ scatter.}
	\label{sigma_vs_wav}
\end{figure*}

Finally in Figure \ref{sigma_vs_wav} we show the ratio (in dex) between
the H$\alpha$ velocity dispersion and that of other Balmer (in black)
and metal (in blue) emission lines as a function of wavelength, after
subtracting in quadrature the DR15 estimate of the instrumental velocity
dispersion. Only spaxels with S/N $>$ 10 on the specific line are
considered. The figure demonstrates that in DR15 the velocity dispersion
of different emission lines within the MILES wavelength range are in
good agreement with each other, with no significant wavelength-dependent
systematic. The largest discrepancies are found in the blue end of the
wavelength range, where both [\oii] and  H$\delta$ are larger than
H$\alpha$ by 0.07 dex ($\sim$17\%) on average. 

\section{Recommendations and future work}
\label{sec6}

\subsection{Recommendations on the use of the DR15 \DAP\ data products}
\label{sec6.1}

Here we briefly summarize our recommendations for usage of the \DAP\
DR15 output regarding emission lines. 

Users whose science goal would benefit from the best spatial resolution
afforded by MaNGA should use the output from the hybrid binning scheme
contained in the {\tt HYB10-GAU-MILESHC} directory. If, on the other hand,
one requires the emission-line properties to be computed on the same
(Voronoi) binning scheme as the continuum the output in {\tt
VOR10-GAU-MILESHC} should be used instead.

All of the emission-line maps are included as extensions in the \DAP\
\MAPS\ files. Each extension corresponds to a 3D array, where two
dimensions correspond to the on-sky spatial pixels and the third
dimension allows the user to choose a specific emission line. These
extensions can be opened with a standard fits viewer (e.g. {\tt
DS9}\footnote{{\url http://ds9.si.edu/site/Home.html }}, {\tt
QFitsView}\footnote{{\url
http://www.mpe.mpg.de/{\textasciitilde}ott/QFitsView/ }}) and appear as
a datacube. The correspondence between the index in the \MAPS\ file
extension and the line name is given in the extension header. A
practical example of how to perform this association in an automatic
fashion in {\tt python} is given below.

\begin{lstlisting}
# Declare a function that creates a dictionary for the columns in the multi-channel extensions
def column_dictionary(hdu, ext):
    columndict = {}
    for k, v in hdu[ext].header.items():
        if k[0] == 'C':
        		try:
					# the -1 makes the indices zero-based as opposed to 1-based
		    		i = int(k[1:])-1
				except ValueError:
					continue
				columndict[v] = i
	return columndict
# end of the function declaration    
\end{lstlisting}

We strongly encourage users to take close consideration of the masks
provided in the \MAPS\ file. In the {\tt python} programming language
masked arrays, as implemented in the in {\tt numpy} package, are
particularly suitable for the task of manipulating data with associated
mask information. Masked arrays allow to perform arithmetic and other simple operations (like taking a median) while automatically ignoring the masked pixels. The code example below provides an example of how to
obtain the H$\alpha$ flux map from the \DAP\ \MAPS\ file and encode it
into a {\tt numpy} masked array.

\begin{lstlisting}
import numpy as np
from astropy.io import fits 

# open the maps file (hybrid binning scheme) for galaxy 8138-12704
hdu = fits.open('manga-8138-12704-MAPS-HYB10-GAU-MILESHC.fits.gz')

emline = column_dictionary(hdu, 'EMLINE_GFLUX')

#  get the Ha map as a masked array
flux_Ha = np.ma.MaskedArray(
hdu['EMLINE_GFLUX'].data[emline['Ha-6564'],:,:], mask=hdu['EMLINE_GFLUX_MASK'].data[emline['Ha-6564'],:,:] > 0)
\end{lstlisting}

Users interested in a comprehensive software framework in the {\tt
python} language to access and manipulate the MaNGA data are encouraged
to use our purpose-built {\tt Marvin} package
\citep{ArXiv_Cherinka2018}. Visualisation tools for the \DAP{} data
products are also directly available on the {\tt Marvin} web interface
at {\url{https://dr15.sdss.org/marvin/}}.

Finally we warn users of the following bugs which were discovered in the
DR15 \DAP\ output. These problems have already been resolved at the
software level, and the \DAP\ output will be corrected in the next data
release.

\begin{enumerate}
\item \textbf{The [\oii]$\lambda\lambda$3727,29 velocity-dispersion
masks and errors are incorrect in DR15.} In particular, the vast
majority of spaxels in the velocity-dispersion maps of [\oii] are
masked in DR15 because of a bug. The maps of the velocity dispersions
themselves are, however, correct. We therefore recommend to ignore the
mask extensions for the [\oii]$\lambda\lambda$3727,29 line velocity
dispersions and apply the H$\alpha$ velocity dispersion masks instead.
Unfortunately the same bug caused the inverse variance extensions for
the [\oii] lines to be filled with zeros. This has been fixed for future
releases, but for DR15 uncertainties for the [\oii] velocity dispersions
are not available.
\item \textbf{The $H\zeta$ line parameters are unreliable in DR15}. This
is due to a blend with the nearby \hei\ line at 3889.749~\AA\ (vacuum)
which was not included in the line list for DR15, but will be included
in future releases. We recommend against use of this line in DR15.
\end{enumerate}

\subsection{Future work}
\label{sec6.2}

There are many ways to improve on DR15 with respect to the emission-line
properties. First, we would like to transition away from \mileshc\
towards a stellar library derived from MaStar spectra
\citep{Arxiv_Yan2018}.  MaStar is a new library of stellar spectra
observed with the MaNGA instrument suite at APO, and offers several
advantages over MILES, in terms of its carefully controlled flux
calibration, wider coverage in terms of stellar parameters, wider
wavelength range and similar LSF to the MaNGA galaxy spectra. 

The generation of a hierarchically clustered set of MaStar spectra for
kinematics extraction and the production of a new generation of SSP
templates based on MaStar spectra are currently being pursued by the MaNGA team. We
anticipate incorporating a `MaStar-HC' library and/or a set of
MaStar-based SSP templates for the simultaneous fitting of continuum and
emission lines in the second fitting stage of the \DAP. The use of
continuum templates with wider wavelength coverage would also allow us
to provide accurate modeling of the continuum for the
[\siii]$\lambda$8831,9071 9533 lines in the near-IR, which are sensitive
tracers of the ionization parameter of the ISM \citep{Kewley2002}.

A detailed characterization of the MaNGA LSF is currently underway and
will be described in a forthcoming publication.
Nonetheless, the results presented in Section \ref{sec5.3}, and 
Figure \ref{sigma_vs_wav} in particular, demonstrate that the current estimate of the 
MaNGA LSF is sufficiently accurate to move towards a tying scheme 
where both velocity and velocity dispersion are tied for physically
motivated groups of lines. We anticipate that the next MaNGA data release
will adopt either the `all tied' or the `v+$\sigma$ groups' tying approach described in
Section \ref{sec5.3}.

Further improvements may include a treatment of multi-Gaussian kinematic
components and broad emission lines in AGN. While automatically detecting
very broad emission lines ($\sigma > 1000$ \kms) is a relatively simple
task, characterizing the significance of line asymmetries and/or
additional kinematic component for millions of low S/N,
medium-resolution MaNGA spaxels represents a significant task, likely to
remain outside the scope of a general-purpose data analysis pipeline
like the \DAP.  \cite{Gallagher2018} describes the first attempt at this
type of analysis on the MaNGA data.

Given the timescale of the SDSS-IV project, which is scheduled to terminate data collection in 2020,
we do not expect further substantial project-led developments of the MaNGA \DAP.  Members of the
astronomical community interested in adapting the MaNGA \DAP\ to their
own specific data format and scientific interests are welcome to make
use of the \DAP\ source code, which is publicly released on
GitHub.\footnote{\url{https://github.com/sdss/mangadap}}


\section{Summary and conclusions}
\label{sec7}

In this paper we have tested the algorithmic choices and output produced
by the MaNGA \DAP\ with regard to the quality of the stellar-continuum
modeling and the determination of emission-line fluxes and kinematics.
We have further described and assessed the choices made for SDSS DR15,
which corresponds to the first public release of \DAP\ data products.
We hope that the analysis presented in this paper will serve both as a
reference for the community interested in the intricacies of spectral
fitting and to those who wish to use the high-level data products from
the MaNGA survey released in DR15.

The main conclusions of this work are summarized below.

\begin{enumerate}
\item We derive a tight relationship between the amplitude-to-noise and
the ratio between measured flux and flux error (S/N). We therefore
consider the S/N as an appropriate metric for emission lines in MaNGA.
For S/N < 30 emission lines are statistically well-fit by the \DAP, with
a $\chi^2$/dof $\sim$ 0.8. An increase in $\chi^2$/dof is observed at
higher S/N, which we expect is associated with template mismatch (i.e.,
non-Gaussian line profiles are more well measured at high S/N). By comparing
Gaussian and non-parametric (summed) fluxes we conclude that, despite the increase in
$\chi^2$/dof at high S/N, our Gaussian line fluxes in that regime are accurate.
\item We generate mock datacubes with realistic error prescriptions and
demonstrate that the estimated errors for flux and velocity dispersion
behave in a statistically correct way down to S/N $\sim$ 1.5. Errors in
the velocity are underestimated for S/N $<$ 10, and we provide an
empirical formula (Equation \ref{correct_velocity}) to correct this
underestimation. We note that applying this correction is the responsibility of the user,
since the correction is not automatically applied to the \DAP\ output.
\item We analyze the error statistics from repeat observations. The
conclusions largely support what is observed for the idealized
simulations. In addition, repeat observations show an underestimation of
the errors in the high-S/N regime. We have demonstrated that this trend
can be entirely explained by small astrometric errors in individual
exposures, which are consistent with the uncertainties derived by the
MaNGA astrometry registration routine.  In light of this, we leave it up
to the user to consider whether adding this extra error contribution is
advisable for their specific science goals.
\item We tested how well the hierarchically clustered MILES library
(\mileshc), employed in DR15, can be used to fit very young stellar
populations (taken from the BC03 SSP library). \mileshc\ can reproduce the correct spectral shape even without the use of polynomials, and most
of the Balmer absorption lines for ages older than 25 Myr. Helium
absorption lines and Balmer lines in a 5 Myr old population are more
difficult to reproduce, even allowing the introduction of polynomials.
\item We have studied how the emission-line fluxes may differ if the
continuum is fit with a set of different SSP template libraries
(M11-MILES, BC03 and MIUSCAT). We find large discrepancies in the
recovered fluxes ($>$ 0.1 dex) for S/N $<$ 10. Metal lines are less
affected; however, [\nii] seems to be affected in a coherent fashion
with nearby H$\alpha$. These differences in flux can cause larger
discrepancies in derived line ratios, extinction correction and
metallicity-sensitive indicators. We find discrepancies of 0.1 dex for
the O3N2 and log(R23) metallicity-sensitive indices even for EW(H$\alpha$)
$>$ 6~\AA, where flux from H\textsc{ii} regions generally dominates over
diffuse ionized gas and LIER emission. The choice of template library
appears therefore to be the largest source of systematic error studied
in this paper.
\item By generating mock cubes with a particular template library and
using an alternative library to fit them, we demonstrate that the
derived emission-line errors remain statistically accurate even in
presence of template mismatch.
\item There is no evidence pointing towards inaccuracies in the MaNGA
flux calibration in the MILES wavelength range, although there is a hint
of a red upturn for $\rm \lambda > 9000$, which we have not investigated
further since this wavelength range is not fit in DR15. The small
deviations (< 10\%) from the smooth curves expected for a physical
extinction model present in the DR15 polynomials are probably due to
inaccuracies in the MILES library flux calibration. 
\item Simultaneously fitting the continuum and emission lines, as done
in the \DAP, has a minor effect on the Balmer line fluxes (<2 \% on H$\alpha$, although the effect can be larger on the higher order Balmer lines) and no
measurable effect on strong metal lines. The use of additive, rather that
multiplicative polynomials, leads to discrepancies in others areas of
the spectrum, but still at the few percent level.
\item We have investigated different tying strategies and compared them
with the approach followed in DR15, in which all lines were fit with a
common velocity and independent velocity dispersions. Large differences
($>$ 0.1 dex) are found comparing DR15 and the case where all kinematic
components are left free, but only for S/N $<$ 2. Treating all
velocities independently, tying velocities of all the lines, or groups
of lines with similar ionization potential does not lead to any
systematic changes in the best-fit velocity. Considering the DR15
determination of the instrumental dispersion at the position of
different emission lines, velocity dispersions of different lines agree
with H$\alpha$ on average to better than 0.07 dex across the full MILES
wavelength range. These facts demonstrate the accuracy of the MaNGA wavelength calibration and
LSF determination.
\end{enumerate}

The data products generated by the MaNGA \DAP\ are made publicly
available at {\tt http://www.sdss.org/dr15/manga/}, while the \DAP\
source code can be accessed via GitHub at
\url{https://github.com/sdss/mangadap}.

\acknowledgements

We thank the anonymous referee for the enlightening and supportive report. 
MAB acknowledges NSF Award AST-1517006.  CAT acknowledges NSF Award
AST-1554877. SFS acknowledges the following projects for their
support: CONACYT FC-2016-01-1916, CONACYT BC-285080 and PAPIIT IN100519.
MC acknowledges support from a Royal Society University Research
Fellowship. RY acknowledges support by NSF award AST-1715898.
This work makes use of data from SDSS-IV. Funding for SDSS
has been provided by the Alfred P.~Sloan Foundation and Participating
Institutions. Additional funding towards SDSS-IV has been provided by
the U.S. Department of Energy Office of Science. SDSS-IV acknowledges
support and resources from the Center for High-Performance Computing at
the University of Utah. The SDSS web site is {\tt www.sdss.org}. This
research made use of Marvin, a core Python package and web framework for
MaNGA data, developed by Brian Cherinka, Jos\'e S\'anchez-Gallego, and
Brett Andrews \citep{ArXiv_Cherinka2018}.
SDSS-IV is managed by the Astrophysical Research Consortium for the
Participating Institutions of the SDSS Collaboration including the
Brazilian Participation Group, the Carnegie Institution for Science,
Carnegie Mellon University, the Chilean Participation Group, the French
Participation Group, Harvard-Smithsonian Center for Astrophysics,
Instituto de Astrof\'isica de Canarias, The Johns Hopkins University,
Kavli Institute for the Physics and Mathematics of the Universe (IPMU) /
University of Tokyo, Lawrence Berkeley National Laboratory, Leibniz
Institut f\"ur Astrophysik Potsdam (AIP),  Max-Planck-Institut f\"ur
Astronomie (MPIA Heidelberg), Max-Planck-Institut f\"ur Astrophysik (MPA
Garching), Max-Planck-Institut f\"ur Extraterrestrische Physik (MPE),
National Astronomical Observatory of China, New Mexico State University,
New York University, University of Notre Dame, Observat\'ario Nacional /
MCTI, The Ohio State University, Pennsylvania State University, Shanghai
Astronomical Observatory, United Kingdom Participation Group,
Universidad Nacional Aut\'onoma de M\'exico, University of Arizona,
University of Colorado Boulder, University of Oxford, University of
Portsmouth, University of Utah, University of Virginia, University of
Washington, University of Wisconsin, Vanderbilt University, and Yale
University.
	
The MaNGA data used in this work is publicly available at {\tt
http://www.sdss.org/dr15/manga/manga-data/}.
	

\appendix
\section{Comparison with Pipe3D}

We have shown that the extraction of emission-line fluxes by the \DAP\
is statistically robust between repeat observations, and that the
estimated errors on the flux are well-determined down to $\mathrm{S/N} =
1.5$. However, as we have shown in Sections \ref{sec3} and \ref{sec4},
there may exist systematics in the recovered fluxes of lines based on
the choices of continuum model used and other aspects of the fitting
methodology. To explore the potential differences that may exist between
different fitting routines, we compare some of the derived emission-line
properties between the \DAP\ and {\tt Pipe3D}. 

The philosophy behind the fitting of the emission lines differs markedly
between the \DAP\ and {\tt Pipe3D}:  While the \DAP\ fits the emission
lines with (positive) Gaussian templates simultaneously with the stellar
continuum, {\tt Pipe3D} only fits Gaussians to the strong emission
lines. The strong lines are collected into four groups based on their
wavelengths, and the velocities of these lines within these groups are
kinematically tied. These groups are (i) the [\oii] $\lambda\lambda
3727,3729$ doublet, (ii) H$\beta$ and [\oiii]$\lambda\lambda$ 4959,
5007, (iii) [\nii]$\lambda\lambda$ 6548,84 and H$\alpha$; and (iv)
[\sii] $\lambda\lambda$ 6717,31. For the weaker emission lines, {\tt
Pipe3D} performs a moment-based analysis to numerically integrate the
weak emission lines, and uses a Monte Carlo method to estimate the
errors (\citealt{Sanchez2016b}, Sections 3.5-3.6). 
This fitting algorithm can return a negative value for
low S/N spectra.  

While the \DAP\ fits the stellar
component of the spectrum using the \mileshc\ library, {\tt Pipe3D}
models the stellar light with a set of of $156$ SSP
models which the authors refer to as the GSD156 library
\citep{CidFernandes2013a}. This set of templates includes the Granada models of
\cite{GonzalezDelgado2005} for stellar populations younger than $63 \,
\mathrm{Myr}$ and the \cite{Vazdekis2010} models for older stellar
populations. The GSD156 library covers a grid in age and
metallicity, with ages ranging from $ 1 \, \mathrm{Myr}$ to $14.1 \,
\mathrm{Gyr}$, and covering four values of metallicity ($Z/Z_{\odot} =
$0.2, 0.4,  1.0 and 1.5). 

\begin{figure}
\includegraphics[width=\textwidth, trim=40 0 40 0, clip]{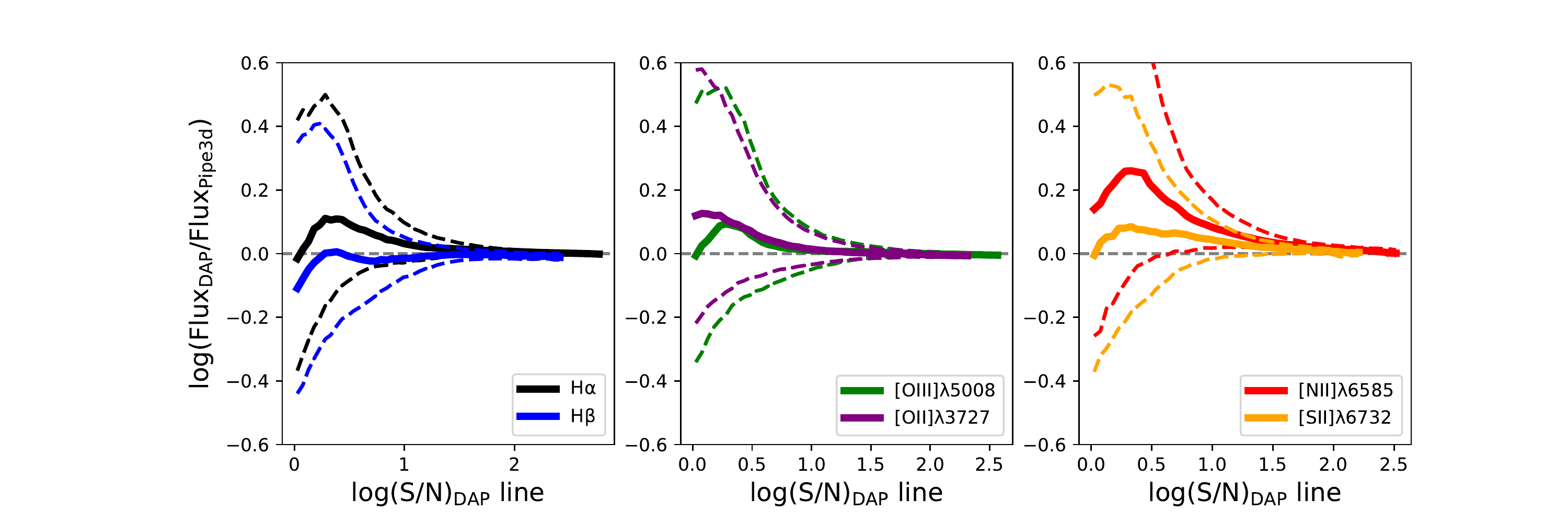}

\caption{The ratio of emission-line flux as measured by the \DAP\ to the
flux measured by {\tt Pipe3D} as a function of the emission-line
S/N (from the \DAP). The solid lines represent the median of the
distribution, while the dashed lines represent the 16th and 84th
percentiles of the distributions.}
\label{Flux_vs_SN}
\end{figure}

\subsection{Comparison of line fluxes}

We first compare the fluxes between {\tt Pipe3D} and the \DAP\ for a set of strong lines ([\oii]$\lambda\lambda 3727,3729$, H$\beta$,
[\oiii]$\lambda 5007$, H$\alpha$, [\nii]$\lambda 6584$, and
[\sii]$\lambda 6731$). We perform this comparison using a large sample of $3\ \times\ 10^6$ spaxels from $4565$ data cubes.
We compare the ratio of the fluxes from each suite pipeline as a
function of the \DAP\ S/N in Figure
\ref{Flux_vs_SN}. At high S/N, the line fluxes agree to
within $0.02 \, \mathrm{dex}$, or $\sim 5 \%$. For $\mathrm{S/N}<10$ the
\DAP\ consistently estimates emission lines to be brighter than {\tt
Pipe3D}. For H$\alpha$ the median of this effect is of order
0.05 dex (12\%) for 3< S/N< 10 and as much as 0.12 dex
(28\%) for [\nii]$\lambda 6584$ in the same S/N range. 
H$\beta$ behaves differently from the other lines tested, as the \DAP\ reports lower 
average fluxes compared to {\tt Pipe3D} at low S/N. For the Balmer lines, this may be
related to the differences in the stellar-absorption-line fits between
the two pipelines; however, this explanation seems less likely for the
forbidden lines. We note that for weak lines, the \DAP\ stipulation that
emission-line fluxes must be positive will introduce a positive bias,
which may partially explain the upward skew seen in the distributions
for low S/N lines in Figure \ref{Flux_vs_SN}.

\subsection{The effect on line ratios and metallicity}
\label{appendix_ratios}

The small systematic offsets observed in the line fluxes can be enhanced 
when one computes some commonly used line ratios. Given the importance 
of establishing the consistency of these higher level measurements, 
we compare the values of $\rm E(B-V)$, [\nii]/H$\alpha$, [\oiii]/H$\beta$ obtained by 
 \DAP\ and {\tt Pipe3D} as a function of EW(H$\alpha$). 

For  EW(H$\alpha$) $> 20$ \AA, the line ratios considered are in reasonable agreement between the 
two pipelines, although both for  $\rm E(B-V)$ and [\nii]/H$\alpha$ a roughly constant systematic offset ($\sim 0.02$ dex) is observed even at high EW. The agreement considerably worsens at low EW.
There exist two main differences between the methodologies of {\tt
	Pipe3D} and the \DAP\ that might influence the fluxes of the emission
lines. These two pipelines use different templates libraries to fit the
stellar continua. This difference can result in differences in line
fluxes, particularly at low EW. Since the [\nii]$\lambda 6584$
line is adjacent to $\mathrm{H\alpha}$, the features in the wings of the
absorption line can impact the measured flux of the [\nii]$\lambda 6584$
line, particularly in spectra with high stellar velocity dispersion.

\begin{figure}
\centering
\includegraphics[width=0.9\textwidth, trim=0 0 25 0, clip]{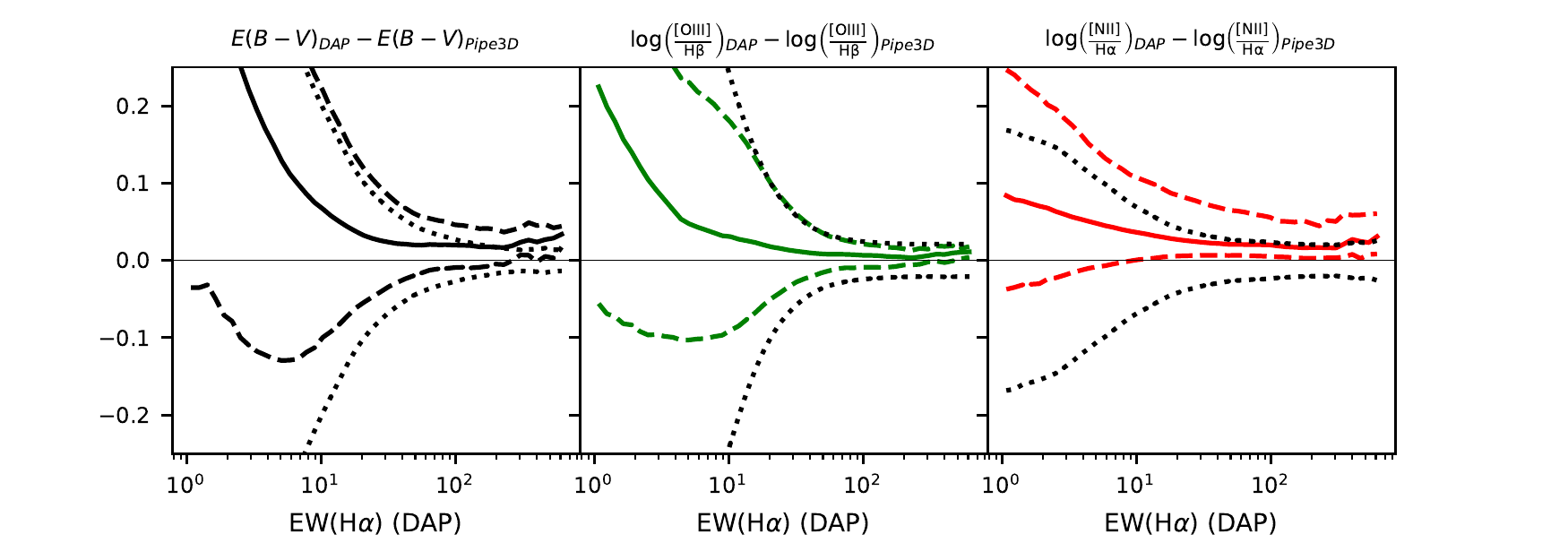}
\caption{The difference in emission-line ratios as measured by the \DAP\
to those measured by {\tt Pipe3D} as a function of EW(H$\alpha$). 
The solid lines are the median difference and the dashed lines are the 16th and 84th
percentiles of the distribution. While skewed upwards at low
EW, the distributions of $\mathrm{H\alpha/H\beta}$ and
[\oiii]/H$\beta$ are consistent within the measurement
uncertainties. This is not true for [\nii]/H$\alpha$, for which
the \DAP\ has measured a higher value, even at high H$\alpha$ EW.}
\label{Ratios_vs_EW}
\end{figure}

\begin{figure}
\centering
\includegraphics[width=0.51\textwidth, trim=0 0 40 0, clip]{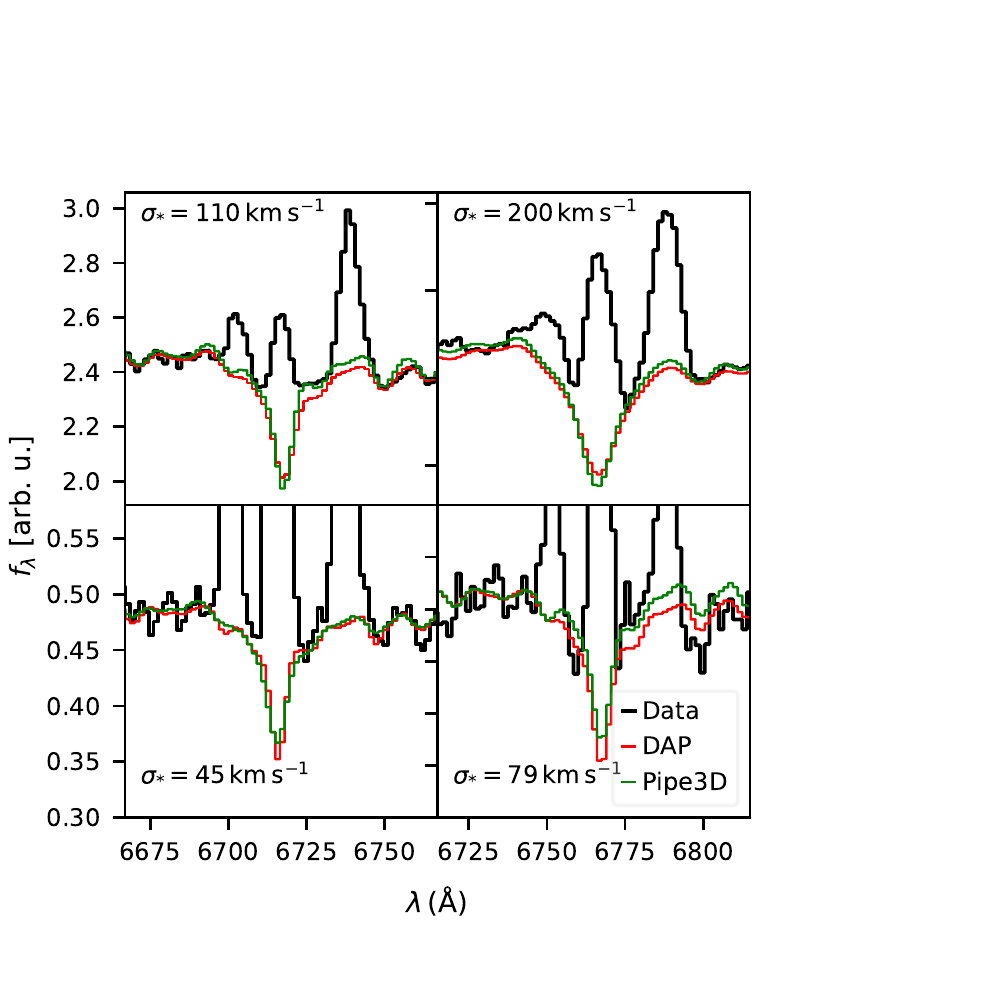}
\includegraphics[width=0.42\textwidth, trim=10 0 25 0, clip]{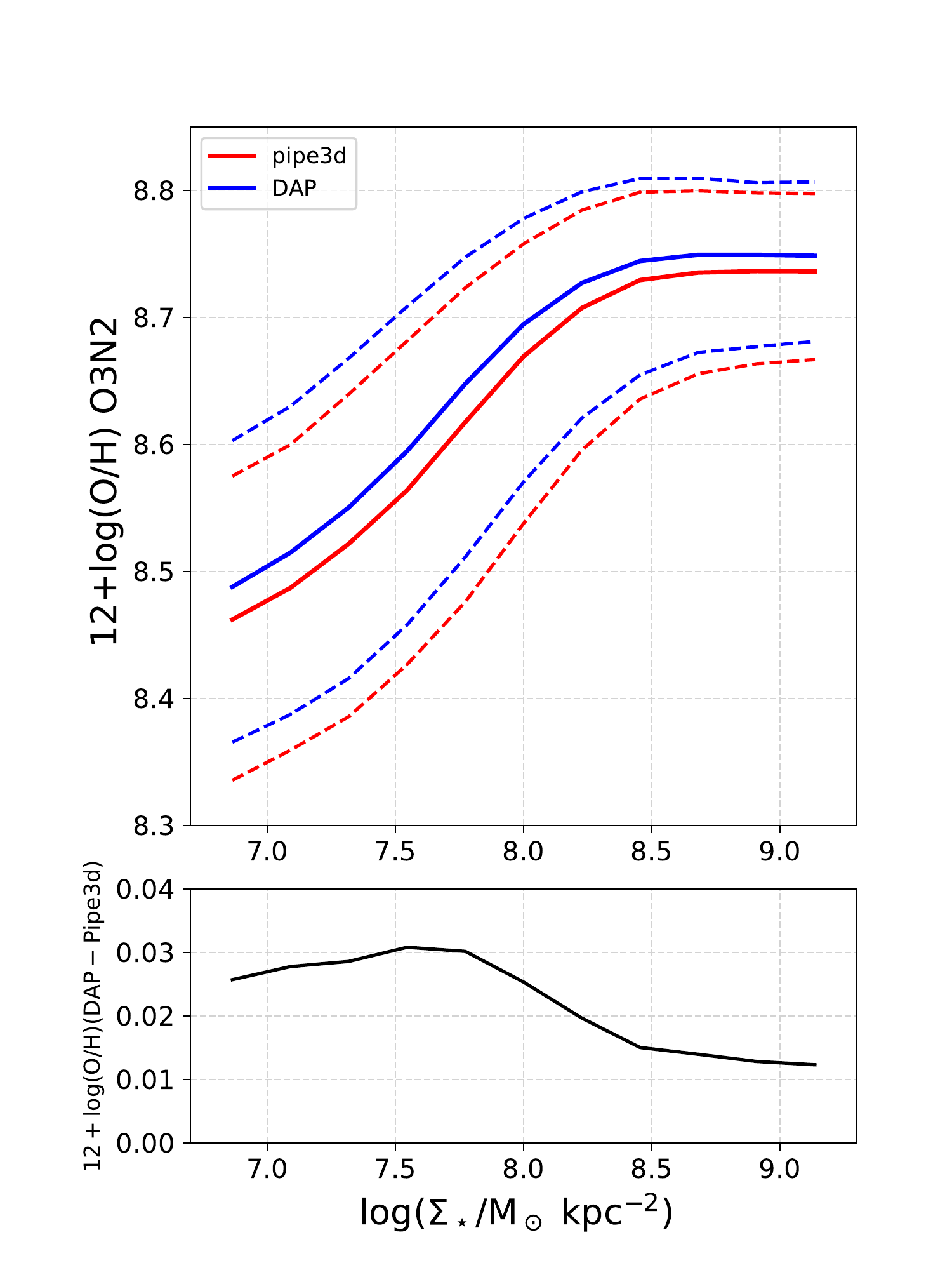}
\caption{
On the left we show example fits performed by the \DAP\ and {\tt Pipe3D}
around the [\nii]/H$\alpha$ region. Systematic differences in the absorption line profile can
extend underneath the [\nii] emission line. This may explain the
systematic differences in the [\nii] fluxes between pipelines. 
On the right, the resolved mass-metallicity relation, computed using the O3N2 diagnostic and the \protect\cite{Pettini2004} metallicity calibration, using the \DAP\ and {\tt Pipe3D} line fluxes for single spaxels in the MaNGA survey. The stellar mass surface density ($\Sigma_{\star}$) is always taken from the {\tt Pipe3D} VAC. Only spaxels with S/N $>$ 3 on the relevant line fluxes and classified as star-forming in the BPT diagram are plotted. The solid lines represent the median relations and the dashed lines are the 16th and 84th percentiles. The resolved mass-metallicity relations obtained from the two different pipelines are in good agreement with regards to the shape of the relation, despite showing a systematic offset in metallicity of the order of 0.01-0.03 dex, as shown in the bottom panel. }
\label{nii_discrepancy}
\end{figure}

The reasons for this behavior are explored in the left-hand panel of
Figure \ref{nii_discrepancy}, where we show four example spectra and the
stellar-continuum fits performed by the \DAP\ and {\tt Pipe3D}. These
spectra were chosen to cover a range of [\nii] EW and
$\sigma_\star$ values. In these examples we can see that differences in
the adopted continuum model can have an impact on the flux in the
emission lines, even in the spectral region around the forbidden lines.
For the highest stellar velocity dispersions, the H$\alpha$ absorption
can spread underneath the [\nii] lines, leading to an overestimate
of the [\nii] flux.

The potential effect on a specific scientific result of the discrepancy between the 
\DAP\ and {\tt Pipe3D} fluxes is illustrated in Fig. \ref{nii_discrepancy}, right panel, where we show the 
resolved mass-metallicity relation obtained from the two pipelines using the O3N2 metallicity calibrator and the  \cite{Pettini2004} calibration. 
We consider only spaxels
classified as star forming using the \cite{Kewley2001} line the [SII] BPT diagram.
All spaxels with S/N $<$ 3 for the emission lines required for the O3N2 diagnostic are excluded. 
The stellar mass surface density values for each spaxel are taken from the 
{\tt Pipe3D}  VAC. These values are corrected for the effect of dust extinction, but not for the
potential effect of galaxy inclination. The shape of the resulting resolved mass-metallicity relation agrees well 
with previous studies based on a smaller sample of MaNGA galaxies \citep{Ballesteros2016}. We note, 
moreover, that the determinations obtained from both pipelines are in very good agreement 
with regards to the shape of the relation, despite a systematic shift  in the median metallicity. This shift is of the order 
of 0.03 dex at low metallicity, decreasing to 0.01 dex at high metallicity. Since the systematics 
associated with metallicity calibrations are larger than this offset, we consider that the choice of 
pipeline does not significantly affect this science case.

The comparison performed in this Appendix highlights the difficulties in measuring the strengths
of emission lines in galaxy spectra. Systematic differences in the
derived fluxes for strong, high-EW lines are within a few per-cent for
the majority of the lines examined here. We have checked that the effect on the resolved mass-metallicity relation is small (0.03 dex at most) but clearly systematic.
However, caution should be taken
not to over-interpret fluxes and flux ratios of emission lines,
particularly in the regime of low line EW and for weak lines which are 
likely to be severely affected by the quality of the continuum subtraction.


\bibliography{library}

\begin{thebibliography}{}
\expandafter\ifx\csname natexlab\endcsname\relax\def\natexlab#1{#1}\fi
\providecommand{\url}[1]{\href{#1}{#1}}
\providecommand{\dodoi}[1]{doi:~\href{http://doi.org/#1}{\nolinkurl{#1}}}
\providecommand{\doeprint}[1]{\href{http://ascl.net/#1}{\nolinkurl{http://ascl.net/#1}}}
\providecommand{\doarXiv}[1]{\href{https://arxiv.org/abs/#1}{\nolinkurl{https://arxiv.org/abs/#1}}}

\bibitem[{{Aguado} {et~al.}(2018){Aguado}, {Ahumada}, {Almeida}, {Anderson},
  {Andrews}, {Anguiano}, {Aquino Ortiz}, {Aragon-Salamanca},
  {Argudo-Fernandez}, {Aubert}, {Avila- Reese}, {Badenes}, {Barboza Rembold},
  {Barger}, {Barrera-Ballesteros}, {Bates}, {Bautista}, {Beaton}, {Beers},
  {Belfiore}, {Bernardi}, {Bershady}, {Beutler}, {Bird}, {Bizyaev}, {Blanc},
  {Blanton}, {Blomqvist}, {Bolton}, {Boquien}, {Borissova}, {Bovy}, {Nielsen
  Brandt}, {Brinkmann}, {Brownstein}, {Bundy}, {Burgasser}, {Byler}, {Cano
  Diaz}, {Cappellari}, {Carrera}, {Cervantes Sodi}, {Chen}, {Cherinka},
  {Doohyun Choi}, {Chung}, {Coffey}, {Comerford}, {Comparat}, {Covey}, {da
  Silva Ilha}, {da Costa}, {Dai}, {Damke}, {Darling}, {Davies}, {Dawson}, {de
  Sainte Agathe}, {Deconto Machado}, {Del Moro}, {De Lee}, {Diamond-Stanic},
  {Dominguez Sanchez}, {Donor}, {Drory}, {du Mas des Bourboux}, {Duckworth},
  {Dwelly}, {Ebelke}, {Emsellem}, {Escoffier}, {Fernandez-Trincado},
  {Feuillet}, {Fischer}, {Fleming}, {Fraser-McKelvie}, {Freischlad},
  {Frinchaboy}, {Fu}, {Galbany}, {Garcia-Dias}, {Garcia-Hernandez}, {Garma
  Oehmichen}, {Geimba Maia}, {Gil-Marin}, {Grabowski}, {Gu}, {Guo}, {Ha},
  {Harrington}, {Hasselquist}, {Hayes}, {Hearty}, {Hernandez Toledo}, {Hicks},
  {Hogg}, {Holley-Bockelmann}, {Holtzman}, {Hsieh}, {Hunt}, {Hwang},
  {Ibarra-Medel}, {Jimenez Angel}, {Johnson}, {Jones}, {Jonsson}, {Kinemuchi},
  {Kollmeier}, {Krawczyk}, {Kreckel}, {Kruk}, {Lacerna}, {Lan}, {Lane}, {Law},
  {Lee}, {Li}, {Lian}, {Lin}, {Lin}, {Lintott}, {Long}, {Longa-Pena},
  {Mackereth}, {de la Macorra}, {Majewski}, {Malanushenko}, {Manchado},
  {Maraston}, {Mariappan}, {Marinelli}, {Marques-Chaves}, {Masseron},
  {Masters}, {McDermid}, {Medina Pena}, {Meneses-Goytia}, {Merloni},
  {Merrifield}, {Meszaros}, {Minniti}, {Minsley}, {Muna}, {Myers}, {Nair},
  {Correa do Nascimento}, {Newman}, {Nitschelm}, {Olmstead}, {Oravetz},
  {Oravetz}, {Ortega Minakata}, {Pace}, {Padilla}, {Palicio}, {Pan}, {Pan},
  {Parikh}, {Parker}, {Peirani}, {Penny}, {Percival}, {Perez-Fournon},
  {Peterken}, {Pinsonneault}, {Prakash}, {Raddick}, {Raichoor}, {Riffel},
  {Riffel}, {Rix}, {Robin}, {Roman-Lopes}, {Rose}, {Ross}, {Rossi}, {Rowlands},
  {Rubin}, {Sanchez}, {Sanchez- Gallego}, {Sayres}, {Schaefer}, {Schiavon},
  {Schimoia}, {Schlafly}, {Schlegel}, {Schneider}, {Schultheis}, {Seo},
  {Shamsi}, {Shao}, {Shen}, {Shetty}, {Simonian}, {Smethurst}, {Sobeck},
  {Souter}, {Spindler}, {Stark}, {Stassun}, {Steinmetz}, {Storchi-Bergmann},
  {Stringfellow}, {Suarez}, {Sun}, {Taghizadeh-Popp}, {Talbot}, {Tayar},
  {Thakar}, {Thomas}, {Tissera}, {Tojeiro}, {Troup}, {Unda-Sanzana},
  {Valenzuela}, {Vargas-Maga na}, {Vazquez Mata}, {Wake}, {Weaver}, {Weijmans},
  {Westfall}, {Wild}, {Wilson}, {Woods}, {Yan}, {Yang}, {Zamora}, {Zasowski},
  {Zhang}, {Zheng}, {Zheng}, {Zhu}, {Zinn}, \& {Zou}}]{Arxiv_Aguado2018}
{Aguado}, D.~S., {Ahumada}, R., {Almeida}, A., {et~al.} 2018, arXiv e-prints,
  arXiv:1812.02759.
\newblock \doarXiv{1812.02759}

\bibitem[{Aihara {et~al.}(2011)Aihara, {Allende Prieto}, An, Anderson, Aubourg,
  Balbinot, Beers, Berlind, Bickerton, Bizyaev, Blanton, Bochanski, Bolton,
  Bovy, Brandt, Brinkmann, Brown, Brownstein, Busca, Campbell, Carr, Chen,
  Chiappini, Comparat, Connolly, Cortes, Croft, Cuesta, {Da Costa}, Davenport,
  Dawson, Dhital, Ealet, Ebelke, Edmondson, Eisenstein, Escoffier, Esposito,
  Evans, Fan, {Femen{\'{i}}a Castell}, Font-Ribera, Frinchaboy, Ge, Gillespie,
  Gilmore, {Gonzlez Hernndez}, Gott, Gould, Grebel, Gunn, Hamilton, Harding,
  Harris, Hawley, Hearty, Ho, Hogg, Holtzman, Honscheid, Inada, Ivans, Jiang,
  Johnson, Jordan, Jordan, Kazin, Kirkby, Klaene, Knapp, Kneib, Kochanek,
  Koesterke, Kollmeier, Kron, Lampeitl, Lang, {Le Goff}, Lee, Lin, Long,
  Loomis, Lucatello, Lundgren, Lupton, Ma, MacDonald, Mahadevan, Maia, Makler,
  Malanushenko, Malanushenko, Mandelbaum, Maraston, Margala, Masters, McBride,
  McGehee, McGreer, M{\'{e}}nard, Miralda-Escud{\'{e}}, Morrison, Mullally,
  Muna, Munn, Murayama, Myers, Naugle, {Fausti Neto}, Nguyen, Nichol,
  O'Connell, Ogando, Olmstead, Oravetz, Padmanabhan, Palanque-Delabrouille,
  Pan, Pandey, Pris, Percival, Petitjean, Pfaffenberger, Pforr, Phleps, Pichon,
  Pieri, Prada, Price-Whelan, Raddick, Ramos, Reyl{\'{e}}, Rich, Richards, Rix,
  Robin, Rocha-Pinto, Rockosi, Roe, Rollinde, Ross, Ross, Rossetto, Snchez,
  Sayres, Schlegel, Schlesinger, Schmidt, Schneider, Sheldon, Shu, Simmerer,
  Simmons, Sivarani, Snedden, Sobeck, Steinmetz, Strauss, Szalay, Tanaka,
  Thakar, Thomas, Tinker, Tofflemire, Tojeiro, Tremonti, Vandenberg, {Vargas
  Mag{\~{a}}a}, Verde, Vogt, Wake, Wang, Weaver, Weinberg, White, White, Yanny,
  Yasuda, Yeche, \& Zehavi}]{Aihara2011}
Aihara, H., {Allende Prieto}, C., An, D., {et~al.} 2011, ApJS, 193, 29

\bibitem[{Baldwin {et~al.}(1981)Baldwin, Phillips, \& Terlevich}]{Baldwin1981}
Baldwin, J.~A., Phillips, M.~M., \& Terlevich, R. 1981, PASP, 93, 5

\bibitem[{Barrera-Ballesteros {et~al.}(2017)Barrera-Ballesteros, S{\'{a}}nchez,
  Heckman, \& Blanc}]{Barrera-Ballesteros2017}
Barrera-Ballesteros, J.~K., S{\'{a}}nchez, S.~F., Heckman, T., \& Blanc, G.~A.
  2017, ApJ, 844, 80

\bibitem[{Barrera-Ballesteros {et~al.}(2016)Barrera-Ballesteros, Heckman, Zhu,
  Zakamska, S{\'{a}}nchez, Law, Wake, Green, Bizyaev, Oravetz, Simmons,
  Malanushenko, Pan, Lopes, \& Lane}]{Ballesteros2016}
Barrera-Ballesteros, J.~K., Heckman, T.~M., Zhu, G.~B., {et~al.} 2016, MNRAS,
  463, 2513

\bibitem[{Beifiori {et~al.}(2011)Beifiori, Maraston, Thomas, \&
  Johanssin}]{Beifiori2011}
Beifiori, A., Maraston, C., Thomas, D., \& Johanssin, J. 2011, A{\&}A, 531,
  A109

\bibitem[{Belfiore {et~al.}(2016)Belfiore, Maiolino, \&
  Bothwell}]{Belfiore2016}
Belfiore, F., Maiolino, R., \& Bothwell, M. 2016, MNRAS, 455, 1218

\bibitem[{Belfiore {et~al.}(2017)Belfiore, Maiolino, Maraston, Emsellem,
  Bershady, Masters, Bizyaev, Boquien, Brownstein, Bundy, Diamond-Stanic,
  Drory, Heckman, Law, Malanushenko, Oravetz, Pan, Roman-Lopes, Thomas,
  Weijmans, Westfall, \& Yan}]{Belfiore2017}
Belfiore, F., Maiolino, R., Maraston, C., {et~al.} 2017, MNRAS, 466, 2570

\bibitem[{Blanton {et~al.}(2017)Blanton, Bershady, Abolfathi, Albareti, Prieto,
  Almeida, Alonso-garc{\'{i}}a, Anders, Anderson, Andrews, Aquino-ort{\'{i}}z,
  \& Al.}]{Blanton2017}
Blanton, M.~R., Bershady, M.~A., Abolfathi, B., {et~al.} 2017, AJ, 154, 28

\bibitem[{Brinchmann {et~al.}(2004)Brinchmann, Charlot, White, Tremonti,
  Kauffmann, Heckman, \& Brinkmann}]{Brinchmann2004}
Brinchmann, J., Charlot, S., White, S. D.~M., {et~al.} 2004, MNRAS, 351, 1151

\bibitem[{Bruzual \& Charlot(2003)}]{Bruzual2003}
Bruzual, G., \& Charlot, S. 2003, MNRAS, 344, 1000

\bibitem[{Bundy {et~al.}(2015)Bundy, Bershady, Law, Yan, Drory, MacDonald,
  Wake, Cherinka, S{\'{a}}nchez-Gallego, Weijmans, Thomas, Tremonti, Masters,
  Coccato, Diamond-Stanic, Arag{\'{o}}n-Salamanca, Avila-Reese, Badenes,
  Falc{\'{o}}n-Barroso, Belfiore, Bizyaev, Blanc, Bland-Hawthorn, Blanton,
  Brownstein, Byler, Cappellari, Conroy, Dutton, Emsellem, Etherington,
  Frinchaboy, Fu, Gunn, Harding, Johnston, Kauffmann, Kinemuchi, Klaene,
  Knapen, Leauthaud, Li, Lin, Maiolino, Malanushenko, Malanushenko, Mao,
  Maraston, McDermid, Merrifield, Nichol, Oravetz, Pan, Parejko, Sanchez,
  Schlegel, Simmons, Steele, Steinmetz, Thanjavur, Thompson, Tinker, van~den
  Bosch, Westfall, Wilkinson, Wright, Xiao, \& Zhang}]{Bundy2015}
Bundy, K., Bershady, M.~A., Law, D.~R., {et~al.} 2015, ApJ, 798, 7

\bibitem[{Calzetti(2001)}]{Calzetti2001}
Calzetti, D. 2001, PASP, 113, 1449

\bibitem[{Cappellari(2017)}]{Cappellari2017}
Cappellari, M. 2017, MNRAS, 466, 798

\bibitem[{Cappellari \& Copin(2003)}]{Cappellari2003}
Cappellari, M., \& Copin, Y. 2003, MNRAS, 342, 345

\bibitem[{Cappellari \& Emsellem(2004)}]{Cappellari2004}
Cappellari, M., \& Emsellem, E. 2004, PASP, 116, 138

\bibitem[{Cappellari {et~al.}(2011)Cappellari, Emsellem, Krajnovi{\'{c}},
  McDermid, Scott, {Verdoes Kleijn}, Young, Alatalo, Bacon, Blitz, Bois,
  Bournaud, Bureau, Davies, Davis, de~Zeeuw, Duc, Khochfar, Kuntschner,
  Lablanche, Morganti, Naab, Oosterloo, Sarzi, Serra, \&
  Weijmans}]{Cappellari2011}
Cappellari, M., Emsellem, E., Krajnovi{\'{c}}, D., {et~al.} 2011, MNRAS, 413,
  813

\bibitem[{Cenarro {et~al.}(2001)Cenarro, Cardiel, Gorgas, Peletier, Vazdekis,
  \& Prada}]{Cenarro2001}
Cenarro, A.~J., Cardiel, N., Gorgas, J., {et~al.} 2001, MNRAS, 326, 959

\bibitem[{Charlot \& Fall(2000)}]{Charlot2000}
Charlot, S., \& Fall, S.~M. 2000, ApJ, 539, 718

\bibitem[{{Cherinka} {et~al.}(2018){Cherinka}, {Andrews},
  {S{\'a}nchez-Gallego}, {Brownstein}, {Argudo-Fern{\'a}ndez}, {Blanton},
  {Bundy}, {Jones}, {Masters}, {Law}, {Rowlands}, {Weijmans}, {Westfall}, \&
  {Yan}}]{ArXiv_Cherinka2018}
{Cherinka}, B., {Andrews}, B.~H., {S{\'a}nchez-Gallego}, J., {et~al.} 2018,
  arXiv e-prints, arXiv:1812.03833.
\newblock \doarXiv{1812.03833}

\bibitem[{{Cid Fernandes} {et~al.}(2013){Cid Fernandes}, P{\'{e}}rez,
  {Garc{\'{i}}a Benito}, {Gonz{\'{a}}lez Delgado}, de~Amorim, S{\'{a}}nchez,
  Husemann, {Falc{\'{o}}n Barroso}, S{\'{a}}nchez-Bl{\'{a}}zquez, Walcher, \&
  Mast}]{CidFernandes2013a}
{Cid Fernandes}, R., P{\'{e}}rez, E., {Garc{\'{i}}a Benito}, R., {et~al.} 2013,
  A{\&}A, 557, A86

\bibitem[{Croom {et~al.}(2012)Croom, Lawrence, Bland-Hawthorn, Bryant, Fogarty,
  Richards, Goodwin, Farrell, Miziarski, Heald, Jones, Lee, Colless, Brough,
  Hopkins, Bauer, Birchall, Ellis, Horton, Leon-Saval, Lewis,
  L{\'{o}}pez-S{\'{a}}nchez, Min, Trinh, \& Trowland}]{Croom2012}
Croom, S.~M., Lawrence, J.~S., Bland-Hawthorn, J., {et~al.} 2012, MNRAS, 421,
  872

\bibitem[{Draine(2011)}]{Draine2011}
Draine, B. T.~T. 2011, {The physics of the insterstellar and intergalactic
  medium} (Princeton University Press)

\bibitem[{Drory {et~al.}(2015)Drory, MacDonald, Bershady, Bundy, Gunn, Law,
  Smith, Stoll, Tremonti, Wake, Yan, Weijmans, Byler, Cherinka, Cope,
  Eigenbrot, Harding, Holder, Huehnerhoff, Jaehnig, Jansen, Klaene, Paat,
  Percival, \& Sayres}]{Drory2015}
Drory, N., MacDonald, N., Bershady, M.~A., {et~al.} 2015, AJ, 149, 77

\bibitem[{Emsellem {et~al.}(2004)Emsellem, Cappellari, Peletier, McDermid,
  Bacon, Bureau, Copin, Davies, Krajnovic, Kuntschner, Miller, \&
  de~Zeeuw}]{Emsellem2004}
Emsellem, E., Cappellari, M., Peletier, R.~F., {et~al.} 2004, MNRAS, 352, 721

\bibitem[{Falc{\'{o}}n-Barroso {et~al.}(2011)Falc{\'{o}}n-Barroso,
  S{\'{a}}nchez-Bl{\'{a}}zquez, Vazdekis, Ricciardelli, Cardiel, Cenarro,
  Gorgas, \& Peletier}]{Falcon-Barroso2011}
Falc{\'{o}}n-Barroso, J., S{\'{a}}nchez-Bl{\'{a}}zquez, P., Vazdekis, A.,
  {et~al.} 2011, A{\&}A, 532, A95

\bibitem[{{Gallagher} {et~al.}(2018){Gallagher}, {Maiolino}, {Belfiore},
  {Drory}, {Riffel}, \& {Riffel}}]{Gallagher2018}
{Gallagher}, R., {Maiolino}, R., {Belfiore}, F., {et~al.} 2018, arXiv e-prints,
  arXiv:1806.03311.
\newblock \doarXiv{1806.03311}

\bibitem[{Goddard {et~al.}(2017)Goddard, Thomas, Maraston, Westfall,
  Etherington, Riffel, Mallmann, Zheng, Argudo-Fernandez, Lian, Bershady,
  Bundy, Drory, Law, Yan, Wake, Weijmans, Bizyaev, Brownstein, Lane, Maiolino,
  Masters, Merrifield, Nitschelm, Pan, Roman-Lopes, Storchi-Bergmann, \&
  Schneider}]{Goddard2017b}
Goddard, D., Thomas, D., Maraston, C., {et~al.} 2017, MNRAS, 466, 4731

\bibitem[{{Gonz{\'{a}}lez Delgado} {et~al.}(2005){Gonz{\'{a}}lez Delgado},
  Cervino, Martins, Leitherer, \& Hauschildt}]{GonzalezDelgado2005}
{Gonz{\'{a}}lez Delgado}, R.~M., Cervino, M., Martins, L.~P., Leitherer, C., \&
  Hauschildt, P.~H. 2005, MNRAS, 357, 945

\bibitem[{Green {et~al.}(2018)Green, Croom, Scott, Cortese, Medling, {Francesco
  D'Eugenio}, Bryant, Bland-Hawthorn, Allen, Sharp, Ho, Groves, Drinkwater,
  Mannering, Harischandra, van~de Sande, Thomas, {Simon O'Toole}, McDermid,
  Vuong, Sealey, Bauer, Brough, Catinella, Cecil, Colless, Couch, Driver,
  Federrath, Foster, Goodwin, Hampton, Hopkins, Jones, Konstantopoulos,
  Lawrence, Leon-Saval, Liske, Lopez-Sa{\'{n}}chez, Lorente, Mould, Obreschkow,
  Owers, Richards, Robotham, Schaefer, Sweet, Taranu, Tescari, Tonini, \&
  Zafar}]{Green2018a}
Green, A.~W., Croom, S.~M., Scott, N., {et~al.} 2018, MNRAS, 475, 716

\bibitem[{Groves {et~al.}(2012{\natexlab{a}})Groves, Brinchmann, \&
  Walcher}]{Groves2012a}
Groves, B., Brinchmann, J., \& Walcher, C.~J. 2012{\natexlab{a}}, MNRAS, 419,
  1402

\bibitem[{Groves {et~al.}(2012{\natexlab{b}})Groves, Krause, Sandstrom,
  Schmiedeke, Leroy, Linz, Kapala, Rix, Schinnerer, Tabatabaei, Walter, \&
  da~Cunha}]{Groves2012}
Groves, B., Krause, O., Sandstrom, K., {et~al.} 2012{\natexlab{b}}, MNRAS, 426,
  892

\bibitem[{Gunn {et~al.}(2006)Gunn, Siegmund, Mannery, Owen, Hull, Leger, Carey,
  Knapp, York, Boroski, Kent, Lupton, Rockosi, Evans, Waddell, Anderson, Annis,
  Barentine, Bartoszek, Bastian, Bracker, Brewington, Briegel, Brinkmann,
  Brown, Carr, Czarapata, Drennan, Dombeck, Federwitz, Gillespie, Gonzales,
  Hansen, Harvanek, Hayes, Jordan, Kinney, Klaene, Kleinman, Kron, Kresinski,
  Lee, Limmongkol, Lindenmeyer, Long, Loomis, McGehee, Mantsch, {Neilsen, Jr.},
  Neswold, Newman, Nitta, {Peoples, Jr.}, Pier, Prieto, Prosapio, Rivetta,
  Schneider, Snedden, \& Wang}]{Gunn2006}
Gunn, J.~E., Siegmund, W.~A., Mannery, E.~J., {et~al.} 2006, AJ, 131, 2332

\bibitem[{Hampton {et~al.}(2017)Hampton, Medling, Groves, Kewley, Dopita,
  Davies, Ho, Kaasinen, Leslie, Sharp, Sweet, Thomas, Allen, Bland-Hawthorn,
  Brough, Bryant, Croom, Goodwin, Green, Konstantantopoulos, Lawrence,
  Lopez-Sa{\'{n}}chez, Lorente, McElroy, Owers, Richards, \&
  Shastri}]{Hampton2017}
Hampton, E.~J., Medling, A.~M., Groves, B., {et~al.} 2017, MNRAS, 470, 3395

\bibitem[{Ho {et~al.}(2016)Ho, Medling, Groves, Rich, Rupke, Hampton, Kewley,
  Bland-Hawthorn, Croom, Richards, Schaefer, Sharp, \& Sweet}]{Ho2016a}
Ho, I.~T., Medling, A.~M., Groves, B., {et~al.} 2016, Astrophys. Space Sci.,
  361

\bibitem[{Jones {et~al.}(2017)Jones, Kauffmann, D'Souza, Bizyaev, Law, Haffner,
  Bah{\'{e}}, Andrews, Bershady, Brownstein, Bundy, Cherinka, Diamond-Stanic,
  Drory, Riffel, Thomas, Wake, Yan, \& Zhang}]{Jones2017}
Jones, A., Kauffmann, G., D'Souza, R., {et~al.} 2017, A{\&}A, 599, A141

\bibitem[{Kewley \& Dopita(2002)}]{Kewley2002}
Kewley, L.~J., \& Dopita, M.~a. 2002, ApJS, 141, 35

\bibitem[{Kewley {et~al.}(2001)Kewley, Dopita, Sutherland, Heisler, \&
  Trevena}]{Kewley2001}
Kewley, L.~J., Dopita, M.~A., Sutherland, R.~S., Heisler, C.~A., \& Trevena, J.
  2001, ApJ, 556, 121

\bibitem[{Law {et~al.}(2015)Law, Yan, Bershady, Bundy, Cherinka, Drory,
  MacDonald, S{\'{a}}nchez-Gallego, Wake, Weijmans, Blanton, Klaene, Moran,
  Sanchez, \& Zhang}]{Law2015}
Law, D.~R., Yan, R., Bershady, M.~a., {et~al.} 2015, AJ, 150, 19

\bibitem[{Law {et~al.}(2016)Law, Cherinka, Yan, Andrews, Bershady, Bizyaev,
  Blanc, Blanton, Bolton, Brownstein, Bundy, Chen, Drory, D'Souza, Fu, Jones,
  Kauffmann, MacDonald, Masters, Newman, Parejko, S{\'{a}}nchez-Gallego,
  S{\'{a}}nchez, Schlegel, Thomas, Wake, Weijmans, Westfall, \&
  Zhang}]{Law2016}
Law, D.~R., Cherinka, B., Yan, R., {et~al.} 2016, AJ, 152, 83

\bibitem[{{Le Borgne} {et~al.}(2003){Le Borgne}, Sanahuja, \&
  Schaerer}]{LeBorgne2003a}
{Le Borgne}, J.~F., Sanahuja, B., \& Schaerer, D. 2003, A{\&}A, 402, 433

\bibitem[{Maraston \& Str{\"{o}}mb{\"{a}}ck(2011)}]{Maraston2011}
Maraston, C., \& Str{\"{o}}mb{\"{a}}ck, G. 2011, MNRAS, 418, 2785

\bibitem[{McCall {et~al.}(1985)McCall, Rybski, \& Shields}]{Mccall1985}
McCall, M.~L., Rybski, P.~M., \& Shields, G.~A. 1985, ApJS, 57, 1

\bibitem[{O'Donnell(1994)}]{O'Donnell1994}
O'Donnell, J.~E. 1994, ApJ, 422, 158

\bibitem[{Oh {et~al.}(2011)Oh, Sarzi, Schawinski, \& Yi}]{Oh2011}
Oh, K., Sarzi, M., Schawinski, K., \& Yi, S.~K. 2011, ApJS, 195, 13

\bibitem[{Osterbrock \& Ferland(2006)}]{Osterbrock2006}
Osterbrock, D.~E., \& Ferland, G.~J. 2006, {Astrophysics of gaseous nebulae and
  active galactic nuclei} (Mill Valley, CA: University Science Books)

\bibitem[{Pagel {et~al.}(1979)Pagel, Edmunds, Blackwell, Chun, \&
  Smith}]{Pagel1979}
Pagel, B. E.~J., Edmunds, M.~G., Blackwell, D.~E., Chun, M.~S., \& Smith, G.
  1979, MNRAS, 189, 95

\bibitem[{Pettini \& Pagel(2004)}]{Pettini2004}
Pettini, M., \& Pagel, B. E.~J. 2004, MNRAS, 348, L59

\bibitem[{Poetrodjojo {et~al.}(2018)Poetrodjojo, Groves, Kewley, Medling,
  Sweet, van~de Sande, Sanchez, Bland-Hawthorn, Brough, Bryant, Cortese, Croom,
  Lopez-Sanchez, Richards, Zafar, Lawrence, Lorente, Owers, \&
  Scott}]{Poetrodjojo2018a}
Poetrodjojo, H., Groves, B., Kewley, L.~J., {et~al.} 2018, MNRAS, 479, 5235

\bibitem[{S{\'{a}}nchez(2006)}]{Sanchez2006}
S{\'{a}}nchez, S.~F. 2006, Astron. Nachrichten, 327, 850

\bibitem[{S{\'{a}}nchez {et~al.}(2012)S{\'{a}}nchez, Rosales-Ortega, Marino,
  Iglesias-Paramo, Vilchez, Kennicutt, Diaz, Mast, Monreal-Ibero,
  Garcia-Benito, Bland-Hawthorn, Perez, Delgado, Husemann, Lopez-Sanchez,
  Fernandes, Kehrig, Walcher, de~Paz, \& Ellis}]{Sanchez2012}
S{\'{a}}nchez, S.~F., Rosales-Ortega, F.~F., Marino, R.~A., {et~al.} 2012,
  A{\&}A, 546, 2

\bibitem[{S{\'{a}}nchez {et~al.}(2014)S{\'{a}}nchez, Rosales-Ortega,
  Iglesias-P{\'{a}}ramo, Moll{\'{a}}, Barrera-Ballesteros, Marino, P{\'{e}}rez,
  S{\'{a}}nchez-Blazquez, {Gonz{\'{a}}lez Delgado}, {Cid Fernandes},
  de~Lorenzo-C{\'{a}}ceres, Mendez-Abreu, Galbany, Falcon-Barroso,
  Miralles-Caballero, Husemann, Garc{\'{i}}a-Benito, Mast, Walcher, {Gil de
  Paz}, Garc{\'{i}}a-Lorenzo, Jungwiert, V{\'{i}}lchez, J{\'{i}}lkov{\'{a}},
  Lyubenova, Cortijo-Ferrero, D{\'{i}}az, Wisotzki, M{\'{a}}rquez,
  Bland-Hawthorn, Ellis, van~de Ven, Jahnke, Papaderos, Gomes, Mendoza, \&
  L{\'{o}}pez-S{\'{a}}nchez}]{Sanchez2014}
S{\'{a}}nchez, S.~F., Rosales-Ortega, F.~F., Iglesias-P{\'{a}}ramo, J.,
  {et~al.} 2014, A{\&}A, 563, A49

\bibitem[{S{\'{a}}nchez {et~al.}(2016{\natexlab{a}})S{\'{a}}nchez, Perez,
  S{\'{a}}nchez-Blazquez, Gonz{\'{a}}lez, Ros{\'{a}}lez-Ortega,
  Cano-D{\'{i}}az, L{\'{o}}pez-Cob{\'{a}}, Marino, {De Paz}, Moll{\'{a}},
  L{\'{a}}pez-S{\'{a}}nchez, Ascasibar, \& Barrera-Ballesteros}]{Sanchez2016a}
S{\'{a}}nchez, S.~F., Perez, E., S{\'{a}}nchez-Blazquez, P., {et~al.}
  2016{\natexlab{a}}, RMxAA, 52, 21

\bibitem[{S{\'{a}}nchez {et~al.}(2016{\natexlab{b}})S{\'{a}}nchez, E.Perez,
  S{\'{a}}nchez-Blazquez, R.Garcia-Benito, Ibarra-Mede, J.J.Gonzalez,
  F.F.Rosales-Ortega, L.Sanchez-Menguiano, Ascasibar, Bitsakis, Law,
  M.Cano-D´ıaz, A, R.A.Marino, {Gil de Paz}, R., Barrera-Ballesteros,
  Galbany, D.Mast, V.Abril-Melgarejo, \& Roman-Lopes}]{Sanchez2016b}
S{\'{a}}nchez, S.~F., E.Perez, S{\'{a}}nchez-Blazquez, P., {et~al.}
  2016{\natexlab{b}}, RMxAA, 52, 171

\bibitem[{S{\'{a}}nchez {et~al.}(2016{\natexlab{c}})S{\'{a}}nchez, E.Perez,
  Sanchez-Blazquez, P., Ibarra-Mede, J.J.Gonzalez, F.F.Rosales-Ortega,
  L.Sanchez-Menguiano, Ascasibar, Bitsakis, Law, M.Cano-D´ıaz, A, R.A.Marino,
  {Gil de Paz}, R., Barrera-Ballesteros, Galbany, D.Mast, V.Abril-Melgarejo, \&
  Roman-Lopes}]{Sanchez2016}
S{\'{a}}nchez, S.~F., E.Perez, Sanchez-Blazquez, {et~al.} 2016{\natexlab{c}},
  RMxAA, 52, 171

\bibitem[{S{\'{a}}nchez {et~al.}(2017)S{\'{a}}nchez, Avila-Reese,
  Hernandez-Toledo, Cortes-Suarez, Rodriguez-Puebla, Ibarra-Medel, Cano-Diaz,
  Barrera-Ballesteros, Negrete, Calette, de~Lorenzo-Caceres, Ortega-Minakata,
  Aquino, Valenzuela, Clemente, Storchi-Bergmann, Riffel, Schimoia, Riffel,
  Rembold, Brownstein, Pan, Yates, Mallmann, \& Bitsakis}]{Sanchez2017}
S{\'{a}}nchez, S.~F., Avila-Reese, V., Hernandez-Toledo, H., {et~al.} 2017,
  RMxAA, 54, 217

\bibitem[{S{\'{a}}nchez {et~al.}(2019)S{\'{a}}nchez, Avila-Reese,
  Rodr{\'{i}}guez-Puebla, Ibarra-Medel, Calette, Bershady,
  Hern{\'{a}}ndez-Toledo, Pan, \& Bizyaev}]{Sanchez2018}
S{\'{a}}nchez, S.~F., Avila-Reese, V., Rodr{\'{i}}guez-Puebla, A., {et~al.}
  2019, MNRAS, 482, 1557

\bibitem[{Sanchez-Blazquez {et~al.}(2006)Sanchez-Blazquez, Peletier,
  Jimenez-Vicente, Cardiel, Cenarro, Falcon-Barroso, Gorgas, Selam, \&
  Vazdekis}]{Sanchez-Blazquez2006}
Sanchez-Blazquez, P., Peletier, R., Jimenez-Vicente, J., {et~al.} 2006, MNRAS,
  371, 703

\bibitem[{Sarzi {et~al.}(2005)Sarzi, Rix, Shields, Ho, Barth, Rudnick,
  Filippenko, \& Sargent}]{Sarzi2005}
Sarzi, M., Rix, H., Shields, J.~C., {et~al.} 2005, ApJ, 628, 169

\bibitem[{Sarzi {et~al.}(2006)Sarzi, Falcon-Barroso, Davies, Bacon, Bureau,
  Cappellari, {Tim de Zeeuw}, Emsellem, Fathi, Krajnovic, Kuntschner, McDermid,
  \& Peletier}]{Sarzi2006}
Sarzi, M., Falcon-Barroso, J., Davies, R.~L., {et~al.} 2006, MNRAS, 366, 1151

\bibitem[{Sarzi {et~al.}(2010)Sarzi, Shields, Schawinski, Jeong, Shapiro,
  Bacon, Bureau, Cappellari, Davies, {Tim de Zeeuw}, Emsellem,
  Falc{\'{o}}n-Barroso, Krajnovi{\'{c}}, Kuntschner, McDermid, Peletier,
  van~den Bosch, van~de Ven, \& Yi}]{Sarzi2010}
Sarzi, M., Shields, J.~C., Schawinski, K., {et~al.} 2010, MNRAS, 402, 2187

\bibitem[{Schlegel {et~al.}(1998)Schlegel, Finkbeiner, \& Davis}]{Schlegel1998}
Schlegel, D. J.~D., Finkbeiner, D. P.~D., \& Davis, M. 1998, ApJ, 500, 525

\bibitem[{Smee {et~al.}(2013)Smee, Gunn, Uomoto, Roe, Schlegel, Rockosi, Carr,
  Leger, Dawson, Olmstead, Brinkmann, Owen, Barkhouser, Honscheid, Harding,
  Long, Lupton, Loomis, Anderson, Annis, Bernardi, Bhardwaj, Bizyaev, Bolton,
  Brewington, Briggs, Burles, Burns, Castander, Connolly, Davenport, Ebelke,
  Epps, Feldman, Friedman, Frieman, Heckman, Hull, Knapp, Lawrence, Loveday,
  Mannery, Malanushenko, Malanushenko, Merrelli, Muna, Newman, Nichol, Oravetz,
  Pan, Pope, Ricketts, Shelden, Sandford, Siegmund, Simmons, Smith, Snedden,
  Schneider, SubbaRao, Tremonti, Waddell, \& York}]{Smee2013}
Smee, S.~A., Gunn, J.~E., Uomoto, A., {et~al.} 2013, AJ, 146, 32

\bibitem[{Tremonti {et~al.}(2004)Tremonti, Heckman, Kauffmann, Brinchmann,
  White, Seibert, Peng, \& Schlegel}]{Tremonti2004}
Tremonti, C.~A., Heckman, T.~M., Kauffmann, G., {et~al.} 2004, ApJ, 613, 898

\bibitem[{Vazdekis {et~al.}(2012)Vazdekis, Ricciardelli, Cenarro,
  Rivero-Gonz{\'{a}}lez, D{\'{i}}az-Garc{\'{i}}a, \&
  Falc{\'{o}}n-Barroso}]{Vazdekis2012}
Vazdekis, A., Ricciardelli, E., Cenarro, a.~J., {et~al.} 2012, MNRAS, 424, 157

\bibitem[{Vazdekis {et~al.}(2010)Vazdekis, S{\'{a}}nchez-Bl{\'{a}}zquez,
  Falc{\'{o}}n-Barroso, Cenarro, Beasley, Cardiel, Gorgas, \&
  Peletier}]{Vazdekis2010}
Vazdekis, A., S{\'{a}}nchez-Bl{\'{a}}zquez, P., Falc{\'{o}}n-Barroso, J.,
  {et~al.} 2010, MNRAS, 1671, 1639

\bibitem[{Wake {et~al.}(2017)Wake, Bundy, Diamond-stanic, Yan, Blanton,
  Bershady, S{\'{a}}nchez-gallego, Drory, Jones, \& Kauffmann}]{Wake2017}
Wake, D.~A., Bundy, K., Diamond-stanic, A.~M., {et~al.} 2017, AJ, 154, 86

\bibitem[{{Westfall} {et~al.}(2019){Westfall}, {Cappellari}, {Bershady},
  {Bundy}, {Belfiore}, {Ji}, {Law}, {Schaefer}, {Shetty}, {Tremonti}, {Yan},
  {Andrews}, {Brownstein}, {Cherinka}, {Coccato}, {Drory}, {Maraston},
  {Parikh}, {S{\'a}nchez-Gallego}, {Thomas}, {Weijmans}, {Barrera-Ballesteros},
  {Du}, {Goddard}, {Li}, {Masters}, {Ibarra Medel}, {S{\'a}nchez}, {Yang},
  {Zheng}, \& {Zhou}}]{Westfall2019_arxiv}
{Westfall}, K.~B., {Cappellari}, M., {Bershady}, M.~A., {et~al.} 2019, arXiv
  e-prints, arXiv:1901.00856.
\newblock \doarXiv{1901.00856}

\bibitem[{Wilkinson {et~al.}(2017)Wilkinson, Maraston, Goddard, Thomas, \&
  Parikh}]{Wilkinson2017}
Wilkinson, D.~M., Maraston, C., Goddard, D., Thomas, D., \& Parikh, T. 2017,
  MNRAS, 472, 4297

\bibitem[{Wilkinson {et~al.}(2015)Wilkinson, Maraston, Thomas, Coccato,
  Tojeiro, Cappellari, Belfiore, Bershady, Blanton, Bundy, Cales, Cherinka,
  Drory, Emsellem, Fu, Law, Li, Maiolino, Masters, Tremonti, Wake, Wang,
  Weijmans, Xiao, Yan, Zhang, Bizyaev, Brinkmann, Kinemuchi, Malanushenko,
  Malanushenko, Oravetz, Pan, \& Simmons}]{Wilkinson2015}
Wilkinson, D.~M., Maraston, C., Thomas, D., {et~al.} 2015, MNRAS, 449, 328

\bibitem[{Yan {et~al.}(2016{\natexlab{a}})Yan, Tremonti, Bershady, Law,
  Schlegel, Bundy, Macdonald, Bizyaev, Blanc, Blanton, Cherinka, Sebastian,
  Eigenbrot, Gunn, Harding, Hogg, Wake, Weijmans, Xiao, Zhang, Drory,
  Macdonald, Bizyaev, Blanc, Blanton, Cherinka, Eigenbrot, Gunn, Harding, Hogg,
  S{\'{a}}nchez-Gallego, S{\'{a}}nchez, Wake, Weijmans, Xiao, \&
  Zhang}]{Yan2016}
Yan, R., Tremonti, C., Bershady, M.~A., {et~al.} 2016{\natexlab{a}}, AJ, 151, 8

\bibitem[{Yan {et~al.}(2016{\natexlab{b}})Yan, Bundy, Law, Bershady, Andrews,
  Cherinka, Diamond-Stanic, Drory, MacDonald, S{\'{a}}nchez-Gallego, Thomas,
  Wake, Weijmans, Westfall, Zhang, Arag{\'{o}}n-Salamanca, Belfiore, Bizyaev,
  Blanc, Blanton, Brownstein, Cappellari, D'Souza, Emsellem, Fu, Gaulme,
  Graham, Goddard, Gunn, Harding, Jones, Kinemuchi, Li, Li, Maiolino, Mao,
  Maraston, Masters, Merrifield, Oravetz, Pan, Parejko, Sanchez, Schlegel,
  Simmons, Thanjavur, Tinker, Tremonti, {Van Den Bosch}, Zheng, D'Souza,
  Emsellem, Fu, Gaulme, Graham, Goddard, Gunn, Harding, Jones, Kinemuchi, Li,
  Li, Maiolino, Mao, Maraston, Masters, Merrifield, Oravetz, Pan, Parejko,
  Sanchez, Schlegel, Simmons, Thanjavur, Tinker, Tremonti, {Van Den Bosch}, \&
  Zheng}]{Yan2016a}
Yan, R., Bundy, K., Law, D. D.~R., {et~al.} 2016{\natexlab{b}}, AJ, 152, 197

\bibitem[{{Yan} {et~al.}(2018){Yan}, {Chen}, {Lazarz}, {Bizyaev}, {Maraston},
  {Stringfellow}, {McCarthy}, {Meneses-Goytia}, {Law}, {Thomas}, {Falcon
  Barroso}, {S{\'a }nchez-Gallego}, {Schlafly}, {Zheng},
  {Argudo-Fern{\'a}ndez}, {Beaton}, {Beers}, {Bershady}, {Blanton},
  {Brownstein}, {Bundy}, {Chambers}, {Cherinka}, {De Lee}, {Drory}, {Galbany},
  {Holtzman}, {Imig}, {Kaiser}, {Kinemuchi}, {Liu}, {Luo}, {Magnier},
  {Majewski}, {Nair}, {Oravetz}, {Oravetz}, {Pan}, {Sobeck}, {Stassun},
  {Talbot}, {Tremonti}, {Waters}, {Weijmans}, {Wilhelm}, {Zasowski}, {Zhao}, \&
  {Zhao}}]{Arxiv_Yan2018}
{Yan}, R., {Chen}, Y., {Lazarz}, D., {et~al.} 2018, arXiv e-prints,
  arXiv:1812.02745.
\newblock \doarXiv{1812.02745}

\bibitem[{Zhang {et~al.}(2017)Zhang, Yan, Bundy, Bershady, Haffner, Walterbos,
  Maiolino, Tremonti, Thomas, Drory, Jones, Belfiore, S{\'{a}}nchez,
  Diamond-stanic, Bizyaev, Nitschelm, Andrews, Brinkmann, Brownstein, Cheung,
  Li, Law, {Roman Lopes}, Oravetz, Pan, {Storchi Bergmann}, Simmons, Sebastian,
  Diamond-stanic, Bizyaev, Nitschelm, Andrews, Brinkmann, Brownstein, Cheung,
  Li, Law, Lopes, \& Oravetz}]{Zhang2017}
Zhang, K., Yan, R., Bundy, K., {et~al.} 2017, MNRAS, 466, 3217

\end{thebibliography}

\label{lastpage}

\end{document}